\documentclass[superscriptaddress,reprint,citeautoscript,prx, longbibliography,floatfix]{revtex4-2} 
\usepackage{graphicx}
\usepackage[margin=0.7in]{geometry}
\usepackage{amsmath}
\usepackage{hyperref}
\usepackage{multirow}
\usepackage{newtxtext}                          
\usepackage{braket}
\usepackage[ruled,vlined,linesnumbered]{algorithm2e}
\usepackage{booktabs}
\usepackage[dvipsnames]{xcolor}
\usepackage{hyperref}
\usepackage[normalem]{ulem}
\usepackage[textwidth=1.5cm,textsize=footnotesize]{todonotes}
\hypersetup{colorlinks, 
	linkcolor={blue!75!black!80!yellow},
	citecolor={blue!75!black!80!yellow}, 
	urlcolor={blue!75!black!80!yellow}
	}
\usepackage[cmintegrals]{newtxmath}
\usepackage{lipsum}
\usepackage{float}

\newcommand{\ECEPrinceton}{Department of Electrical and Computer Engineering, Princeton University, Princeton, New Jersey 08540, USA}
\newcommand{\PQIPrinceton}{Princeton Quantum Initiative, Princeton University, Princeton, New Jersey 08540, USA}
\newcommand{\PhysPrinceton}{Department of Physics, Princeton University, Princeton, New Jersey 08540, USA}
\newcommand{\ChemPrinceton}{Department of Chemistry, Princeton University, Princeton, New Jersey 08540, USA}

\begin{document}
\author{Ambrose Yang}\affiliation{\ECEPrinceton}
\author{Alexander M. Young}\affiliation{\PQIPrinceton}
\author{Paal H. Prestegaard}\affiliation{\ECEPrinceton}
\author{Atharv Joshi}\affiliation{\ECEPrinceton}
\author{Matthew S. Ai}\affiliation{\PQIPrinceton}
\author{Q. Rumman Rahman}\affiliation{\PQIPrinceton}
\author{Faranak Bahrami}\affiliation{\ECEPrinceton}
\author{Matthew P. Bland}\affiliation{\ECEPrinceton}
\author{Jui-Lin Niu}\affiliation{\ECEPrinceton}
\author{Ourania-Maria Glezakou-Elbert}\affiliation{\PhysPrinceton}
\author{Ray D. Chang}\affiliation{\ECEPrinceton}
\author{Jocelyn Liu}\affiliation{\ECEPrinceton}
\author{Zhengyi Lu}\affiliation{\ECEPrinceton}
\author{Robert J. Cava}\affiliation{\ChemPrinceton}
\author{Nathalie P. de Leon}\email{npdeleon@princeton.edu} \affiliation{\ECEPrinceton}\affiliation{\PQIPrinceton}
\author{Andrew A. Houck}\email{aahouck@princeton.edu}\affiliation{\ECEPrinceton}\affiliation{\PQIPrinceton}

\title{Real-Time Correlation of Relaxation, Dephasing, and Two-level System Dynamics in Transmon Qubits}

\begin{abstract}
Materials improvements in transmon qubits have yielded remarkable advances in qubit lifetimes $(T_1)$ and coherence times $(T_2)$. However, state-of-the-art superconducting qubits experience temporal fluctuations in their performance, which limit gate fidelities and quantum processor performance. We observe correlated increases in qubit relaxation and dephasing rates that last for minutes to hours, which we call $T_1$ and $T_2$ excursions. Using interleaved lifetime and coherence time measurements, noise spectroscopy, and two-level-system (TLS) spectroscopy, we find that these excursions are caused by single or few strongly-coupled TLS defects that spectrally diffuse into resonance with the qubit frequency. Our measurement protocol provides a powerful tool for understanding the noise sources that cause temporal fluctuations in superconducting qubit performance.

\end{abstract}

\maketitle

\section{Introduction}
Superconducting qubits are a promising physical platform to realize large-scale quantum processors, with numerous recent demonstrations in quantum error correction \cite{google_surface_2025, krinner2022realizing, Gong2021Experimental, sivak2023realtime, ofek2016extending}, quantum simulation \cite{barends2015digital, kandala2017hardware, marcos2013superconducting}, and quantum many-body physics \cite{zhang2022synthesizing, mi2022time, kollar2019hyperbolic, andersen2025thermalization, google_otoc_2025}. Recently, improving materials and device fabrication have enabled substantial advances in superconducting qubit lifetimes ($T_1$) and Hahn-echo coherence times ($T_{2\text{E}}$); in particular, transmons fabricated with tantalum on a high-purity silicon substrate (Ta-on-Si) have achieved $T_1$ and $T_{2\text{E}}$ at the millisecond scale \cite{bland_millisecond_2025, olszewski_krypton_2026, acharya_ultrahigh_2026}. However, superconducting qubits still suffer from temporal fluctuations in their lifetimes and coherence times. Temporal fluctuations in qubit lifetimes and coherence times are widely reported for qubits fabricated with various material stacks, including tantalum \cite{bland_millisecond_2025, olszewski_krypton_2026}, niobium \cite{carroll_dynamics_2022}, and aluminum \cite{burnett_decoherence_2019, berritta_real-time_2026, klimov_fluctuations_2018}. 

Temporal fluctuations in superconducting qubit $T_1$ and $T_{2\text{E}}$ have been studied extensively both experimentally \cite{carroll_dynamics_2022, burnett_decoherence_2019, klimov_fluctuations_2018, schlor_correlating_2019, zhao_combating_2022, zhu_disentangling_2025, berritta_real-time_2026} and theoretically \cite{muller_interacting_2015, you_stabilizing_2022, etxezarreta_martinez_time-varying_2021}. These fluctuations have most commonly been attributed to fluctuations of TLS defects that couple resonantly with the qubit \cite{burnett_decoherence_2019, klimov_fluctuations_2018, schlor_correlating_2019, muller_interacting_2015, zhao_combating_2022, weeden_statistics_2026, berritta_real-time_2026, you_stabilizing_2022, zhu_disentangling_2025}, but other mechanisms such as quasiparticles \cite{zhu_disentangling_2025, gustavsson_suppressing_2016} have also been proposed. Prior experimental investigations of temporal fluctuations in superconducting qubits have involved data analysis on repeated measurements of $T_1$ and $T_{2\text{E}}$ \cite{burnett_decoherence_2019, schlor_correlating_2019} or probing TLS spectral diffusion by tuning the qubit frequency \cite{klimov_fluctuations_2018, carroll_dynamics_2022}. Such experiments were performed on qubits with $T_1$ and $T_{2\text{E}}$ around $100 \; \mu$s \cite{burnett_decoherence_2019, schlor_correlating_2019, klimov_fluctuations_2018, carroll_dynamics_2022}, and have not yet been performed on state-of-the-art Ta-on-Si qubits, which have significantly reduced TLS losses and order-of-magnitude improvements in lifetimes and coherence times. 

\begin{figure*}
  \includegraphics{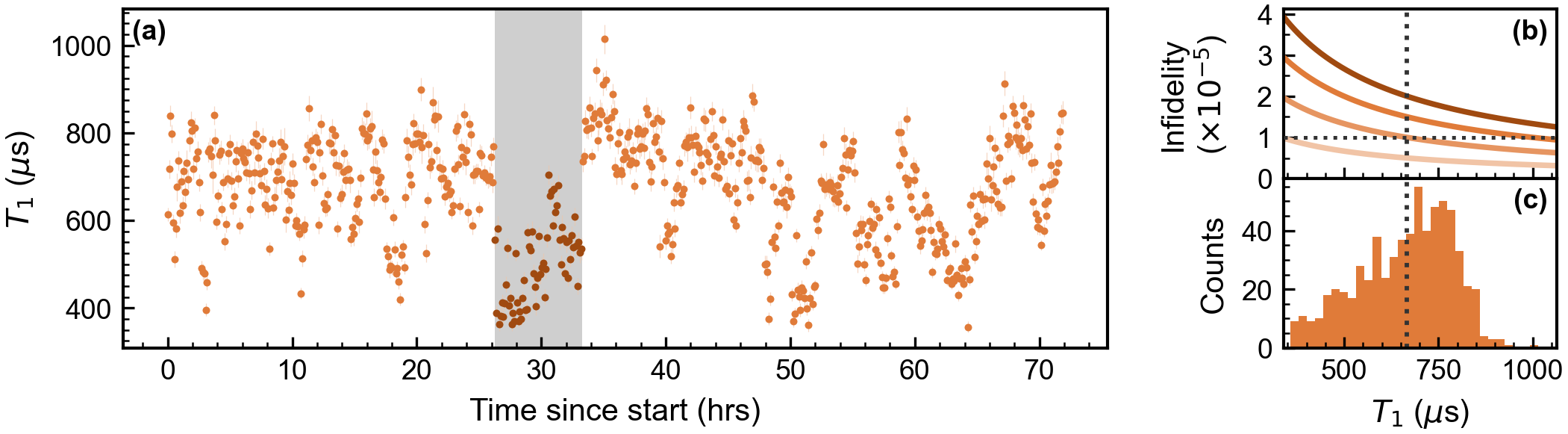}
  \caption{\textbf{Excursions in Qubit $T_1$:} 
  (a) Temporal fluctuations in qubit $T_1$ from dataset 3, measured repeatedly over 72 hours. Points in the gray shaded region show an excursion lasting 7 hours where the $T_1$ is notably reduced from its steady-state value.
  (b) Theoretical single-qubit gate fidelities following \cite{bland_millisecond_2025} assuming $T_1$ relaxation is the dominant source of gate errors for a 10 ns, 20 ns, 30 ns, and 40 ns $\pi/2$-pulse (light to dark orange). The horizontal dotted line denotes the 99.999\% gate fidelity threshold, and the vertical dotted line indicates the $T_1$ time required to reach this threshold assuming a 20 ns $\pi/2$-pulse. 
  (c) Histogram of $T_1$ values from (a). Threshold from (b) intersects at $T_1 = 667 \textrm{ $\mu$s}$. With a 20 ns $\pi/2$-pulse, $T_1$ temporal fluctuations prevent gate fidelity from reaching the 99.999\% milestone 45.5\% of the time.}
   \label{fig:motivation}
\end{figure*}

In this work we perform repeated measurements of $T_1$ ($T_{2\text{E}}$) in state-of-the-art, fixed-frequency, Ta-on-Si qubits over multiple days, in which we observe hours-long periods of decreased $T_1$ ($T_{2\text{E}}$) bookended by sudden jumps, which we call $T_1$ ($T_{2\text{E}}$) \textit{excursions} (Fig. \ref{fig:motivation}(a)). Such temporal fluctuations in qubit performance can cause detrimental downstream effects for large-scale quantum processors. $T_1$ excursions degrade gate operations because qubit relaxation during the gate pulse is a significant source of gate infidelity \cite{li_gate_2023}. For example, assuming a 20 ns $\pi/2$-pulse \cite{li_gate_2023}, temporal fluctuations in $T_1$ as in Fig. \ref{fig:motivation}(a) preclude single-qubit gate fidelities from reaching the 99.999\% fidelity milestone 45.5\% of the time (Fig. \ref{fig:motivation}(b), (c)). These temporally fluctuating error rates in turn also demand more complex quantum error correction protocols \cite{etxezarreta_martinez_time-varying_2021}.

We benchmark temporal fluctuations and investigate the underlying noise sources by measuring correlations between different measurements of qubit dynamics. We simultaneously perform measurements of $T_1$ and $T_{2\text{E}}$ \cite{schlor_correlating_2019}, noise spectroscopy using Carr-Purcell-Meiboom-Gill (CPMG) dynamical decoupling pulses \cite{bylander_noise_2011}, and TLS spectroscopy \cite{carroll_dynamics_2022}, combined into a single interleaved measurement protocol. We find that the distribution of $T_1$ is non-Gaussian (Appendix \ref{CoherenceDistr}), and that it corresponds to a high average value with excursions to lower values that persist for minutes to hours. Correlating $T_1$ and $T_{2\text{E}}$ measurements reveal that excursions in $T_1$ coincide with excursions in the dephasing rate. Real-time noise spectroscopy reveals increased noise amplitude in the kHz range in the qubit environment during $T_1$ and $T_{2\text{E}}$ excursions. We confirm that the increased noise amplitude and corresponding increased relaxation rate arise from the spectral diffusion of individual, strongly-coupled TLSs by performing simultaneous TLS spectroscopy. These measurements give rise to a model for qubit loss dynamics wherein the steady-state lifetime and coherence time are determined by the TLS bath in the capacitors, while excursions are driven by strongly-coupled TLSs in the near-junction region, where the electric fields are strongest.

\section{Correlated Relaxation and Dephasing}
\begin{figure*}
   \centering
  \includegraphics[width=\textwidth]{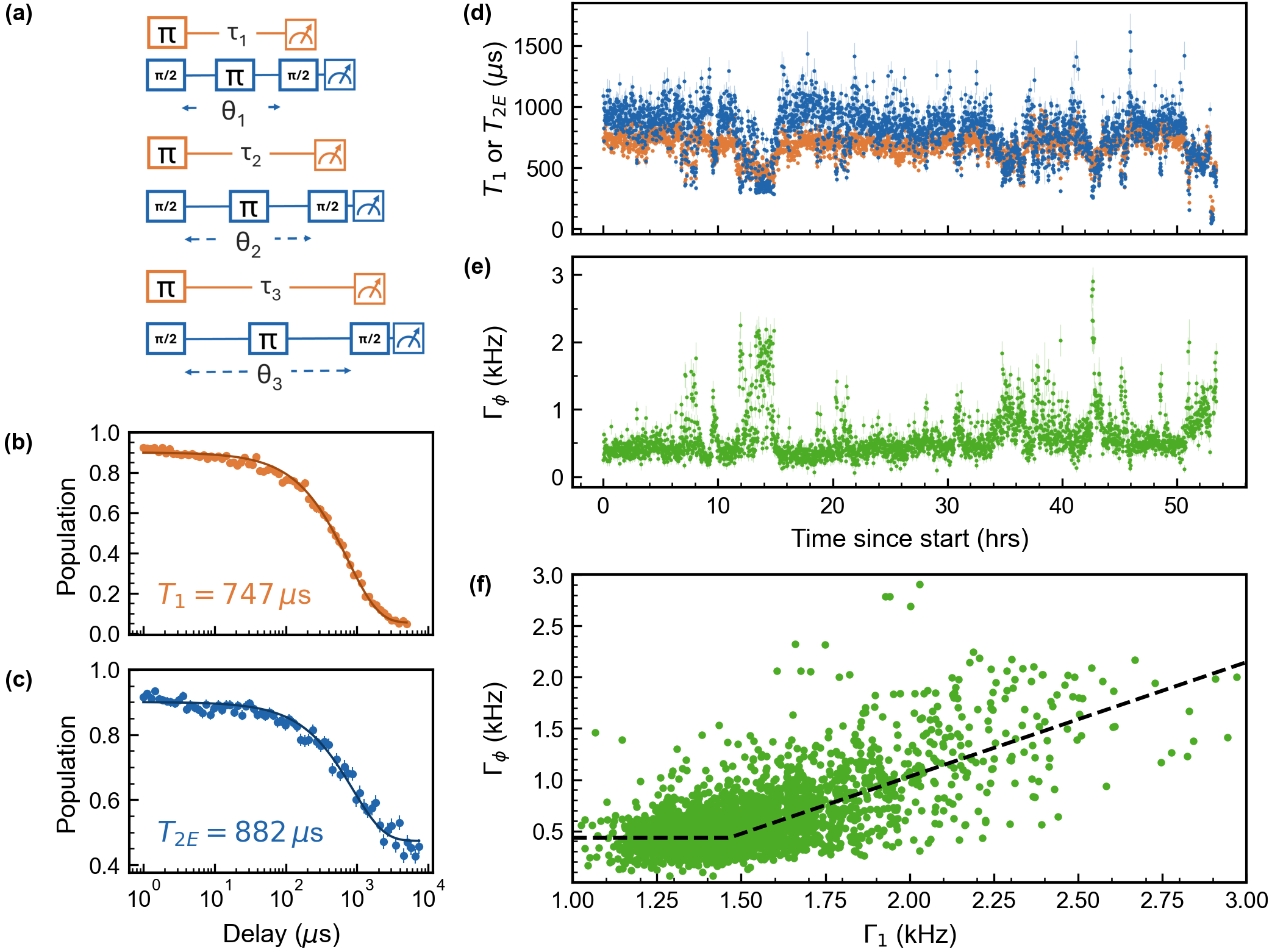}
  \caption{\textbf{Correlated relaxation and dephasing:} 
  (a) Shot-to-shot interleaving of a $T_1$ (orange pulse sequence) and $T_{2\text{E}}$ (blue pulse sequence) measurement. Each round consists of a single shot measurement at each delay time $(\tau_1, \tau_2, \dots, \tau_n)$ and $(\theta_1, \theta_2, \dots, \theta_n)$ for $T_1$ and $T_{2\text{E}}$ respectively. 
  (b, c) Qubit populations extracted by averaging across 500 rounds. Population versus delay time is fitted to an exponential decay to extract $T_1$ and $T_{2\text{E}}$. 
  (d) Repeated shot-to-shot interleaved measurement of $T_1$ and $T_{2\text{E}}$ over 54 hours from dataset 1, which show $T_1$ and $T_{2\text{E}}$ excursions that last for multiple minutes to hours. The y-axis range is upper-bounded by 1.7 ms to more clearly show trends in the data.
  (e) Calculated $\Gamma_\phi = 1/T_{2\text{E}} - 1/(2T_1)$ over the same time as (d). $\Gamma_\phi$ increases when $T_1$ and $T_{2\text{E}}$ are in an excursion, which suggests a common relaxation and dephasing mechanism. The y-axis range is limited to $\Gamma_\phi < 3 \textrm{ kHz}$ to more clearly show trends in the data.
  (f) Scatter plot of dephasing rate $\Gamma_\phi$ and relaxation rate $\Gamma_1$. Dashed line shows a piecewise linear fit, indicating uncorrelated relaxation and dephasing when $T_1 < 683 \textrm{ $\mu$s}$ and correlated relaxation and dephasing when $T_1 > 683 \textrm{ $\mu$s}$ with a fitted slope of $1.12$. The axis ranges are limited to $\Gamma_1 < 3 \textrm{ kHz}$ and $\Gamma_\phi < 3 \textrm{ kHz}$ to more clearly show trends in the data.
  }
   \label{fig:t1_t2e_corr}
\end{figure*}

To understand the noise sources causing temporal fluctuations in $T_1$ and $T_{2\text{E}}$, we first determine whether fluctuations in relaxation and dephasing rates are correlated in time. To do so, we repeatedly measure $T_1$ and $T_{2\text{E}}$ over time to extract the qubit relaxation rate $\Gamma_1 = 1 / T_1$ and dephasing rate $\Gamma_\phi = 1 / T_{2\text{E}} - 1 / (2T_1)$. To calculate $\Gamma_\phi$ accurately, we use a shot-to-shot interleaved measurement protocol to measure $T_1$ and $T_{2\text{E}}$ nearly simultaneously. Given a series of delay times $(\tau_1, \tau_2, \dots \tau_n)$ for a $T_1$ measurement and a series of delay times $(\theta_1, \theta_2, \dots \theta_n)$ for a $T_{2\text{E}}$ measurement, the interleaved measurement protocol alternates between $T_1$ and $T_{2\text{E}}$ pulse sequences with delay times $(\tau_1, \theta_1, \tau_2, \theta_2, \dots \tau_n, \theta_n)$ (Fig. \ref{fig:t1_t2e_corr}(a)).

$T_1$ and $T_{2\text{E}}$ measurements require averaging over many shots to extract populations over many delay times $\tau$. The populations in Fig. \ref{fig:t1_t2e_corr}(b) and \ref{fig:t1_t2e_corr}(c) are averaged over 500 shots for each of the 71 delay times to extract a qubit $T_1$ or $T_{2\text{E}}$. Without interleaving, each $T_1$ or $T_{2\text{E}}$ measurement separately takes about 1 minute, and therefore dynamics faster than a timescale of several minutes disproportionately affect either the $T_1$ or $T_{2\text{E}}$ measurement and disrupt the simultaneity of the experiment. In the interleaved measurement protocol, only dynamics faster than millisecond timescales disproportionately affect either the $T_1$ or $T_{2\text{E}}$ measurement (Appendix \ref{TimeResMeas}). We note that since the interleaved measurement protocol only permutes the order of the pulse sequences, it improves the simultaneity but not the temporal resolution of repeated $T_1$ and $T_{2\text{E}}$ measurements. To improve the temporal resolution of our measurement to around one minute, we implement an active reset protocol (Appendix \ref{ActiveReset}).

With this interleaved protocol, we measure $T_1$ and $T_{2\text{E}}$ for 54 hours (Fig. \ref{fig:t1_t2e_corr}(d)).  The integrated autocorrelation time of the $T_1$ data is 34 minutes (Appendix \ref{CorrTime}). This dataset shows multiple, minutes- to hours-long excursions where both $T_1$ and $T_{2\text{E}}$ decrease. Calculating $\Gamma_\phi$ (Fig. \ref{fig:t1_t2e_corr}(e)) reveals that these excursions exhibit not only increased relaxation but are correlated with increased dephasing as well. We quantify this correlated behavior with a scatter plot of $\Gamma_1$ and $\Gamma_\phi$ (Fig. \ref{fig:t1_t2e_corr}(f)). Excluding outliers (Appendix \ref{OutlierRemoval}), the relaxation and dephasing rates are correlated with a Spearman correlation coefficient of 0.51 and fitted linear slope of $\Gamma_\phi / \Gamma_1 = 0.89 \pm 0.02$. We have performed interleaved $T_1$ and $T_{2\text{E}}$ measurements on multiple qubits, which show similar correlated dynamics for qubits where the average $T_1$ and $T_{2\text{E}}$ are comparable (Appendix \ref{CorrRD}).

The relationship between $\Gamma_\phi$ and $\Gamma_1$ does not appear to be simply linear (Appendix \ref{ReLUFit}). To model the nonlinear relationship, we perform model selection with piecewise linear fitting. We fit to a linear combination of rectified linear units $\textrm{ReLU}(x) = \max(0, x)$ and use the Bayes Information Criterion (BIC) to prevent overfitting (Appendix \ref{ReLUFit}). The piecewise linear fit yields a zero slope for $\Gamma_1 < 1.47 \textrm{ kHz}$ ($T_1 > 683 \textrm{ $\mu$s}$) and a slope of 1.12 for $\Gamma_1 > 1.47 \textrm{ kHz}$ ($T_1 < 683 \textrm{ $\mu$s}$). This suggests that relaxation and dephasing are uncorrelated in the high-$T_1$ regime where $T_1 > 683 \textrm{ $\mu$s}$, while relaxation and dephasing are correlated in the low-$T_1$ regime where $T_1 < 683 \textrm{ $\mu$s}$, indicating that there is a common noise source causing correlated relaxation and dephasing during $T_1$ and $T_{2\text{E}}$ excursions.


\section{Fluctuations in Noise Spectral Density}
\begin{figure}
   \centering
  \includegraphics{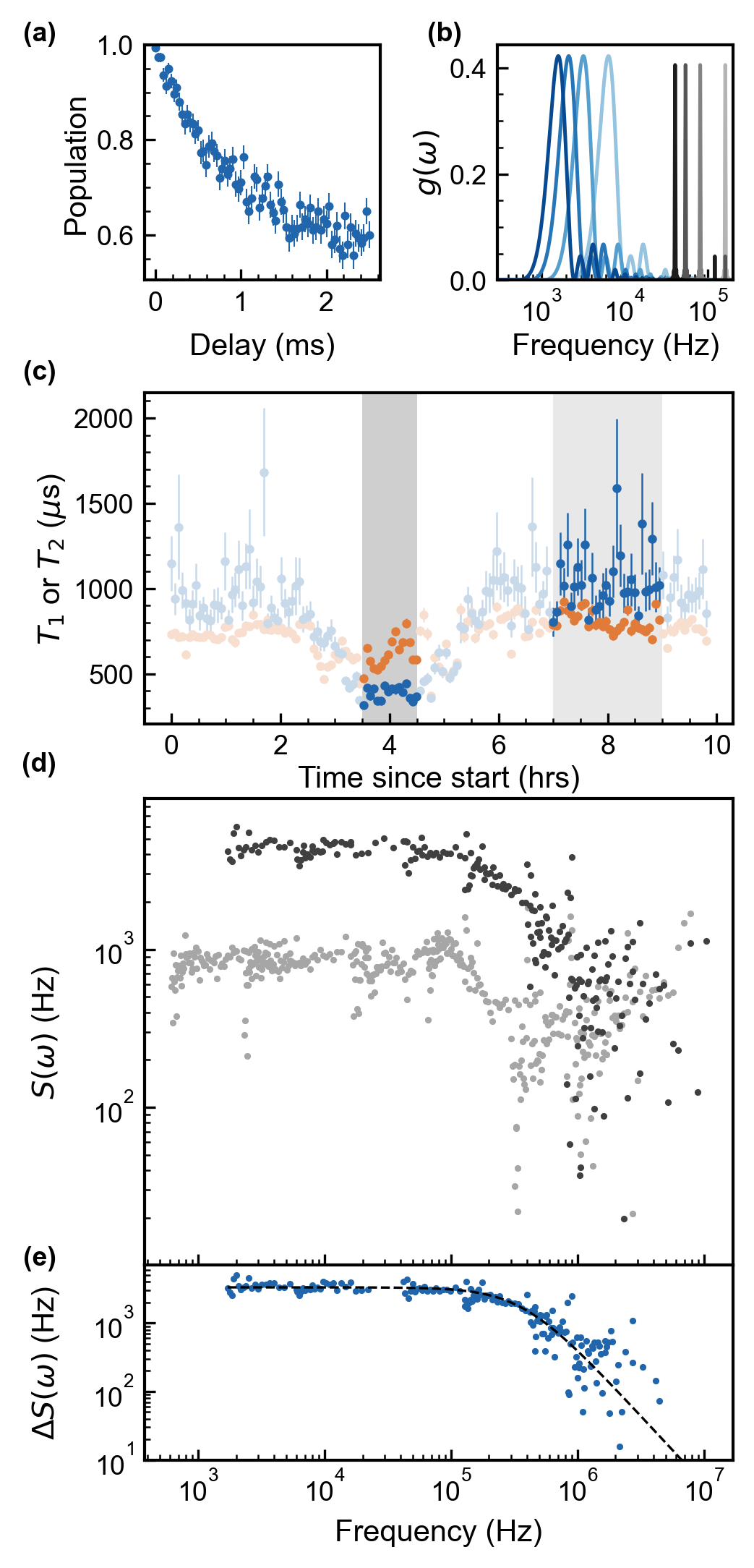}
  \caption{\textbf{Noise spectroscopy:} 
  (a) Sample decay trace showing the qubit population as a function of the total delay time $\tau$ for $N=3$ $\pi$-pulses. Each point on the decay curve can be used to estimate the noise spectral density at a different frequency (Appendix \ref{SwCalc}).
  (b) Filter function $g_N(\omega, \tau)$ for CPMG pulse sequence with $N = 3$ $\pi$-pulses for $\tau = 0.25, 0.5, 0.75 \textrm{ and } 1 \textrm{ ms}$ (lighter to darker blue). Filter function for $N = 81$ $\pi$-pulses for the same $\tau$ values (lighter to darker gray).
  (c) Repeated shot-to-shot interleaved measurement of $T_1$ (orange) and $T_{2}$ (blue) measured with an $N = 3$ CPMG sequence over 10 hours from dataset 2. The dark gray shaded data points from $t = 3.5$ to $4.5$ hours are inside an excursion. The light gray shaded data points from $t = 7$ to $9$ hours are outside an excursion.
  (d) $S(\omega)$ averaged across the shaded points in (e)  inside (dark grey) and outside (light grey) an excursion.
  (e) Noise difference between the two noise spectra in (d). The additional noise is fit to a Lorentzian curve with a roll-off at $365 \pm 18$ kHz inside the excursion.}
   \label{fig:cpmg}
\end{figure}

To further characterize the noise source causing increased relaxation and dephasing during an excursion, we examine the qubit environmental noise spectral density $S(\omega)$ using CPMG dynamical decoupling pulses. CPMG dynamical decoupling generalizes the Hahn echo pulse sequence with $N \geq 2$ evenly-spaced refocusing $\pi$-pulses. A pulse sequence with $N$ $\pi$-pulses and total free evolution time $\tau$ filters the qubit environmental noise with a filter function given by the Fourier transform of the pulse sequence \cite{bylander_noise_2011, viola_dynamical_1998}:
\begin{equation}
\begin{split}
g_N(\omega, \tau) &= \frac{1}{(\omega \tau)^2} \Big|1 + (-1)^{1 + N} \exp(i \omega \tau) \\
  &+ 2 \sum_{j=1}^N (-1)^j \exp(i \omega \delta_j \tau) \cos(\omega t_\pi / 2) \Big|^2
\end{split}
\label{eq:filter}
\end{equation}
where $\delta_j$ is the normalized position of the $j$th $\pi$-pulse and $t_\pi$ is the $\pi$-pulse time. This filter function is strongly peaked at $\omega^\prime = N / (2 \tau)$ and thus can be approximated as a Dirac $\delta$-function at $\omega^\prime$. Therefore, measurement of the qubit population using this pulse sequence probes $S(\omega)$ around $\omega^\prime$ (Fig. \ref{fig:cpmg}(a), Appendix \ref{SwCalc}). Sweeping both $N$ and $\tau$ allows us to reconstruct $S(\omega)$ from the kHz to the MHz range, where the frequency range is limited by the $\pi$-pulse fidelity and the qubit coherence time $T_2$ (Fig. \ref{fig:cpmg}(b), Appendix \ref{SwCutoff}).

Similar to Fig. \ref{fig:t1_t2e_corr}(a), we interleave CPMG measurements with various $N$ and $\tau$ with $T_1$ and $T_{2\text{E}}$ measurements, allowing for real-time correlation of $S(\omega)$ with qubit dynamics. This measurement technique allows us to compare $S(\omega)$ inside and outside of an excursion (Fig. \ref{fig:cpmg}(c) and Fig. \ref{fig:cpmg}(d)). Outside of an excursion, we observe a flat spectral noise density with a roll-off corresponding to the readout resonator bandwidth, indicating that noise is dominated by thermal photons in the readout resonator, as previously reported \cite{bland_millisecond_2025}. Inside an excursion, the noise amplitude in the kHz range increases. The excess noise exhibits a roll-off frequency around 365 kHz (Fig. \ref{fig:cpmg}(e)). Other excursions exhibit excess noise different roll-off frequencies (Appendix \ref{SwExamples}), indicating that they arise from a noise source with variable correlation times.

The most likely candidate causing coherence excursions are individual TLSs. When a TLS couples resonantly with a qubit, the TLS can exchange energy with the qubit and cause relaxation \cite{carroll_dynamics_2022, klimov_fluctuations_2018, burnett_decoherence_2019, chen_scalable_2025}. In addition, the TLS switching between its two low-lying energy states induces changes to the qubit frequency which causes dephasing \cite{schlor_correlating_2019}. TLS switching can be modeled as random telegraph noise, which has a Lorentzian noise spectral density \cite{rower_evolution_2023}:
\begin{equation}
    S(\omega) \propto \frac{1}{\omega^2 + \gamma^2}
    \label{sw_lorentzian}
\end{equation}
where the linewidth $\gamma$ represents the finite correlation time of the TLS switching. In addition, we observe non-Gaussian or non-stationary effects in the qubit noise environment \cite{etxezarreta_martinez_time-varying_2021} (Appendix \ref{NonStandardSw}). This suggests that the added noise during an excursion is caused by a small number of individual fluctuators rather than a statistical average of many fluctuators \cite{schlor_correlating_2019, burnett_decoherence_2019, muller_interacting_2015, lisenfeld_electric_2019}.

Correlating noise spectroscopy measurements with lifetime and coherence measurements allows us to rule out other possible physical mechanisms that may be responsible for $T_1$ and $T_{2\text{E}}$ excursions. Notably, the excess noise does not exhibit $1/f$ scaling, as would be expected for a TLS bath, charge noise, critical current noise, or other sources of low frequency electronic noise \cite{koch_model_2007, schreier_suppressing_2008, christensen_anomalous_2019, yan_spectroscopy_2012}. Quasiparticle bursts are also an unlikely candidate, since they occur on the millisecond timescale \cite{mcewen_gapEngineering_2024, kurilovich_correlated_2026, riste_millisecond_2013} while the excursions we observe are on the minutes- to hours-long timescale. Photon shot noise is also an unlikely candidate since the dispersive interaction in the qubit-resonator Hamiltonian only causes qubit dephasing but not relaxation \cite{blais_circuit_2021, krantz_quantum_2019}.

\begin{figure*}
   \centering
  \includegraphics[width=\textwidth]{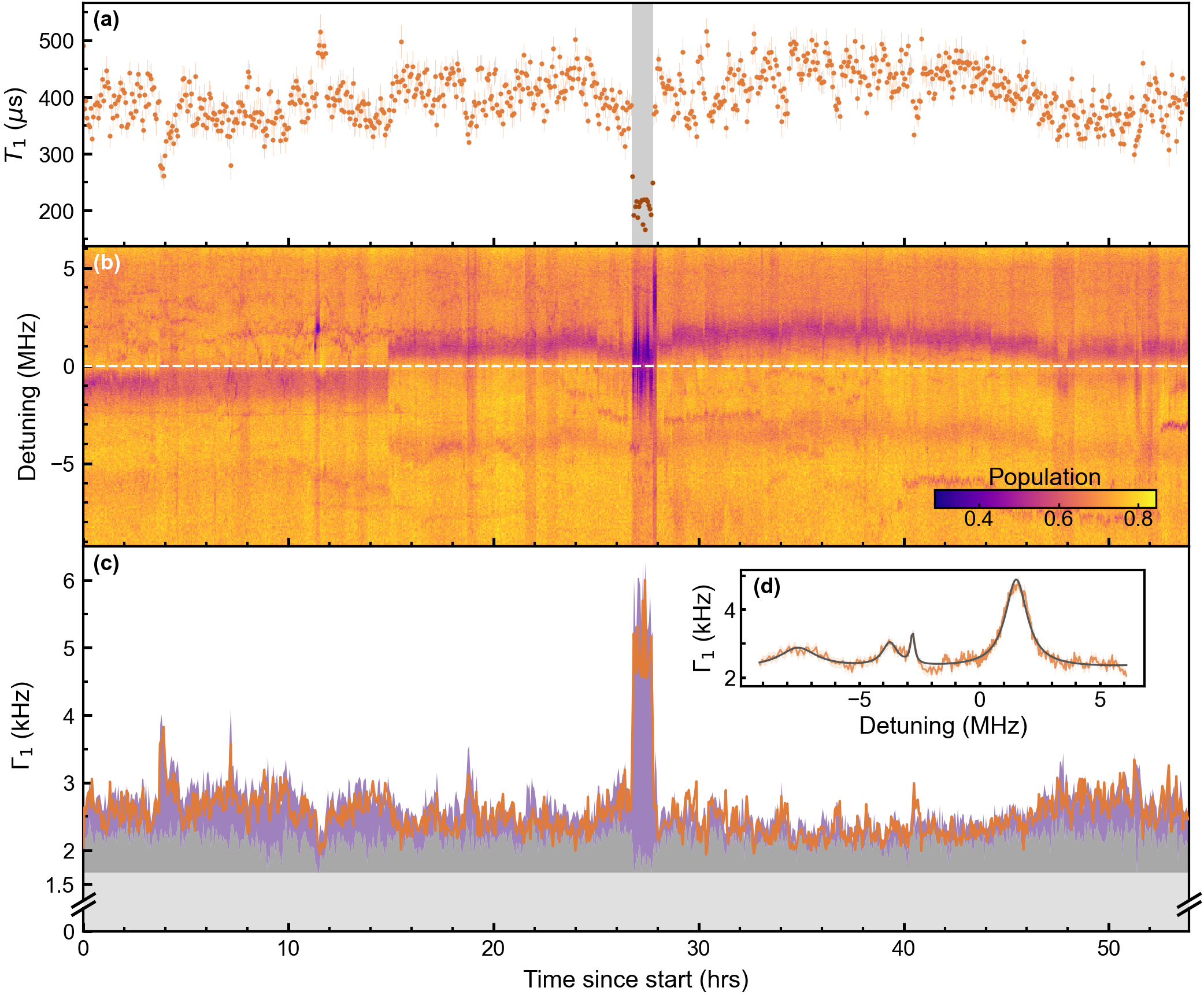}
  \caption{\textbf{TLS spectroscopy:} 
  (a) Measurement of $T_1$ over 54 hours from interleaved $T_1$ and AC Stark pulse sequences from dataset 7. A clear excursion (gray shaded region) is visible around hour 27. 
  (b) Qubit excited state population at $\tau = 100 \textrm{ $\mu$s}$ at various qubit frequencies over the same time window as (a). Purple bands represent frequencies with reduced population due to a resonant strongly coupled TLS. Excursion in (b) at hour 27 coincides with the spectral diffusion of a TLS into the qubit frequency (white dashed line).
  (c) Measured $\Gamma_1(t)$ (orange) and fitted $\hat{\Gamma}_1(f = 0, t)$, which is the sum of three terms: TLS contribution $\sum_i \Gamma_{1, \; \text{TLS}}^{(i)}(f = 0, t)$ (purple), time-dependent broadband fluctuations $\Gamma_{1, \; \text{Broadband}}(t)$ (dark grey), and time-independent background $\Gamma_{1, \; \text{Background}}(f = 0)$ (light grey). $\Gamma_1$ and $\hat{\Gamma}_1$ show good agreement and the time-variation of $\hat{\Gamma}_1$ is dominated by variation in the TLS contribution. 
  (d) $\Gamma_1(f,t)$ calculated from a vertical linecut of (b) averaged from $t = 28.90$ to $t = 29.50$ hours (orange) and $\hat{\Gamma}_{1}(f,t)$ fitted from Eq. \eqref{eq:TLSbudget} (black). Peaks correspond to individually resolvable TLSs. 
  }
   \label{fig:stark}
\end{figure*}

\section{TLS Spectroscopy}
The spectral noise density suggests that $T_1$ and $T_{2\text{E}}$ excursions are caused by a few individual strongly-coupled TLSs. To test this hypothesis, we perform direct spectroscopy of the TLSs by measuring the qubit population after applying an AC Stark shift to the qubit. Applying an off-resonant microwave tone with amplitude $\Omega$ and detuning $\Delta$ shifts the qubit's frequency by \cite{zhao_combating_2022, carroll_dynamics_2022}
\begin{equation} 
\delta \omega \approx \frac{\alpha \Omega^2}{2 \Delta (\Delta + \alpha)}
\end{equation}
where $\alpha$ is the transmon anharmonicity (Appendix \ref{ACStarkCal}) \cite{carroll_dynamics_2022}. 

Similar to previous work on TLS spectroscopy \cite{carroll_dynamics_2022, colao_zanuz_mitigating_2025, klimov_fluctuations_2018, weeden_statistics_2026}, we measure the relaxation rate as a function of qubit frequency over a 15 MHz bandwidth. To perform rapid measurements with high time resolution we estimate $T_1$ through repeated measurement of the qubit excited state population at $\tau = 100 \; \mu$s. These measurements are interleaved with existing $T_1$, $T_{2\text{E}}$, and CPMG pulse sequences (Table \ref{table:datasets}). A drop in $T_1$ over a narrow frequency band corresponds to energy exchange with an individual TLS, thus reporting on the spectral density of TLSs in the vicinity of the qubit.

Repeated TLS spectroscopy measurements over time reveal that the TLS resonant frequencies exhibit diffusive drifts and abrupt jumps (Fig. \ref{fig:stark}(b)), consistent with previous literature \cite{klimov_fluctuations_2018, carroll_dynamics_2022, muller_interacting_2015, thorbeck_two-level-system_2023-2}. Importantly, interleaving TLS spectroscopy and $T_1$ measurements shows that excursions in $T_1$ occur when a single TLS spectrally diffuses to be on resonance with the qubit frequency (Fig. \ref{fig:stark}(a)--(b)). Excursions in $T_1$ often occur abruptly (Fig. \ref{fig:motivation}(a), \ref{fig:t1_t2e_corr}(d), \ref{fig:stark}(a)); these abrupt changes in $T_1$ are due to the spectral jump of a TLS into resonance with the qubit.

We fit the data in Fig. \ref{fig:stark}(b) to the following model (Fig. \ref{fig:stark}(d)):
\begin{equation}
\label{eq:TLSbudget}
\begin{aligned}
    \hat{\Gamma}_{1}(f, t) &= \sum_i \Gamma_{1, \; \text{TLS}}^{(i)}(f, t) \\
    &+ \Gamma_{1, \; \text{Broadband}}(t) + \Gamma_{1, \; \text{Background}}(f) \\
    \Gamma_{1, \; \text{TLS} }^{(i)}(f, t) &= \frac{2g_i^2 \Gamma_{d, i}}{\Gamma_{d, i}^2 + 4\pi^2(f-f_{d, i})^2}
\end{aligned}
\end{equation}
The terms in the first sum $\Gamma_{1, \; \text{TLS}}^{(i)}(f, t)$ give the loss contribution of the individually resolvable TLSs indexed by $i$, where $g$ is the TLS coupling strength, $\Gamma_d$ is the TLS relaxation rate, $f_d$ is the TLS resonant frequency, and $f$ is the Stark-shifted qubit frequency \cite{weeden_statistics_2026}. The second term $\Gamma_{1, \; \text{Broadband}}(t)$ encompasses all time-dependent, frequency-independent noise sources, such as quasiparticles or a bath of weakly-coupled TLSs. This term is modeled as a constant vertical shift which is fit simultaneously with the Lorentzian peaks. The last term $\Gamma_{1, \text{Background}}(f)$ includes all time-independent losses, which may include Stark shift drive induced losses and TLSs that do not spectrally diffuse across the entire measurement. Details of the model and fitting procedure are provided in Appendix \ref{LorentzianFit}.

We study the contributions of TLSs to the temporal fluctuations of the qubit by computing $\hat{\Gamma}_1(f = 0, t)$. We use $\hat{\Gamma}_1(f = 0, t)$ to denote the predicted relaxation rate from Eq. \eqref{eq:TLSbudget} and $\Gamma_1(t)$ to denote the measured qubit relaxation rate from the interleaved full $T_1$ sweeps. To quantify the agreement between $\hat{\Gamma}_1(f = 0, t)$ and $\Gamma_1(t)$, we use Lin's Concordance Correlation Coefficient ($r_c$) \cite{lin_concordance_1989}, which accounts for both Pearson correlation and systematic error. We find good agreement between $\hat{\Gamma}_1(f = 0, t)$ and $\Gamma_1(t)$ with $r_c = 0.831$ (Appendix \ref{LinsCCC}).

We can quantify the relative contribution to the temporal fluctuations of $\Gamma_1$ from the two time-dependent terms in the model: individually resolvable TLSs and broadband fluctuations (Fig. \ref{fig:stark}(c)). The qubit $\Gamma_1$ is strongly correlated with the TLS term with Pearson correlation coefficient $0.902$ but is uncorrelated with the broadband term with Pearson correlation coefficient $-0.056$. Furthermore, the variance in the TLS term is 10.8 times larger than the variance in the broadband term. Together these show that the temporal fluctuations in qubit $\Gamma_1$ are dominated by the spectral diffusion of individual strongly-coupled TLSs.

The TLS-qubit coupling strength is proportional to $\mathbf{p} \cdot \mathbf{E}$, where $\mathbf{p}$ is the dipole moment of the TLS and $\mathbf{E}$ is the electric field at the location of the TLS. Since the electric fields in the transmon circuit are strongest near the Josephson junction, the strongly-coupled TLSs that cause $T_1$ excursions likely reside in the amorphous oxides along the Josephson junction leads or the surfaces near the junction. Prior literature reports experiments that map the location of TLSs, which also suggest that most of the strongly-coupled TLSs are near the Josephson junction \cite{lisenfeld_mappingTLS_2026}. Our observations are consistent with a model wherein the TLSs in the capacitor pads and bulk substrate limit the average qubit $T_1$, while TLSs near the junction limit the stability of qubit $T_1$ and cause $T_1$ excursions. 

\section{Conclusion}
In this work we measure real-time correlations between qubit lifetime and coherence times, noise environment, and TLS spectroscopy. We implement these measurements on state-of-the-art Ta-on-Si transmons and find that $T_1$ and $T_{2\text{E}}$ excursions are caused by the spectral diffusion of single, strongly-coupled TLSs. Characterizing and mitigating the TLSs causing excursions in qubit $T_1$ will become an increasingly important problem as qubit lifetimes continue to improve and as larger scale quantum processors are implemented.

The goal of developing high-stability superconducting qubits motivates several avenues for further exploration, such as further study into TLS spectral diffusion dynamics. While TLS spectroscopy via AC Stark shift of the qubit only probes a spectral range of $\sim 10 \textrm{ -- } 20 \textrm{ MHz}$, a natural next step is to use flux-tunable qubits, which can probe TLS dynamics over $\sim$ GHz of spectral range. In addition, since these strongly-coupled TLSs likely reside near the Josephson junction, alternative junction materials can improve the stability of superconducting qubit lifetimes and coherence times. Comparative study of TLS coupling strengths, linewidths, and spectral diffusion dynamics in different junction materials will enable the selection of superior junction materials for high-performance, high-stability superconducting quantum devices with reliably low gate errors.


\section*{Acknowledgements}
We thank Michel Devoret, David Schuster, Alex Opremcak, Xiao Mi, Vladislav Kurilovich, Jeff Thompson, Zeeshawn Kazi, Artur Lozovoi, and Lila Nassar for fruitful discussions. This work was primarily supported by the Air Force Office of Scientific Research/Laboratory of Physical Sciences (AFOSR/LPS) under Award No. FA9550-25-1-0172 and by Google Quantum AI under SOW No. 89201. Materials characterization and device fabrication was supported by the US Department of Energy, Office of Science, National Quantum Information Science Research Centers, Co-design Center for Quantum Advantage (C2QA) under Contract No. DESC0012704. A.M.Y. and M.P.B. were also supported by the National Science Foundation Graduate Research Fellowship Program (NSF GRFP) under Grant No. DGE-2444107. The authors acknowledge the use of Princeton’s Imaging and Analysis Center (IAC), which is partially supported by the Princeton Center for Complex Materials (PCCM), a National Science Foundation (NSF) Materials Research Science and Engineering Center (MRSEC; DMR-2011750), as well as the Princeton Micro/Nano Fabrication Laboratory. We also acknowledge MIT Lincoln Labs for supplying a traveling-wave parametric amplifier.

Princeton University Professor Nathalie de Leon is a Senior Staff Research Scientist with Google Quantum AI. Due to her income from Google, Princeton University has a management plan in place to mitigate a potential conflict of interest that could affect the design, conduct, and reporting of this research. Her academic group also has a sponsored research contract with Google Quantum AI. 

Princeton University Professor Andrew Houck was also a consultant for Quantum Circuits Incorporated (QCI)/DWave. Due to his income from QCI/DWave, Princeton University has a management plan in place to mitigate a potential conflict of interest that could affect the design, conduct and reporting of this research.


\appendix
\setcounter{figure}{0}
\renewcommand{\thefigure}{S\arabic{figure}}
\renewcommand{\theHfigure}{S\arabic{figure}}

\section{Additional Experimental Details}
\subsection{Device Fabrication and Measurement Apparatus}
\label{Methods}

\begin{figure}
   \centering
  \includegraphics[width=\columnwidth]{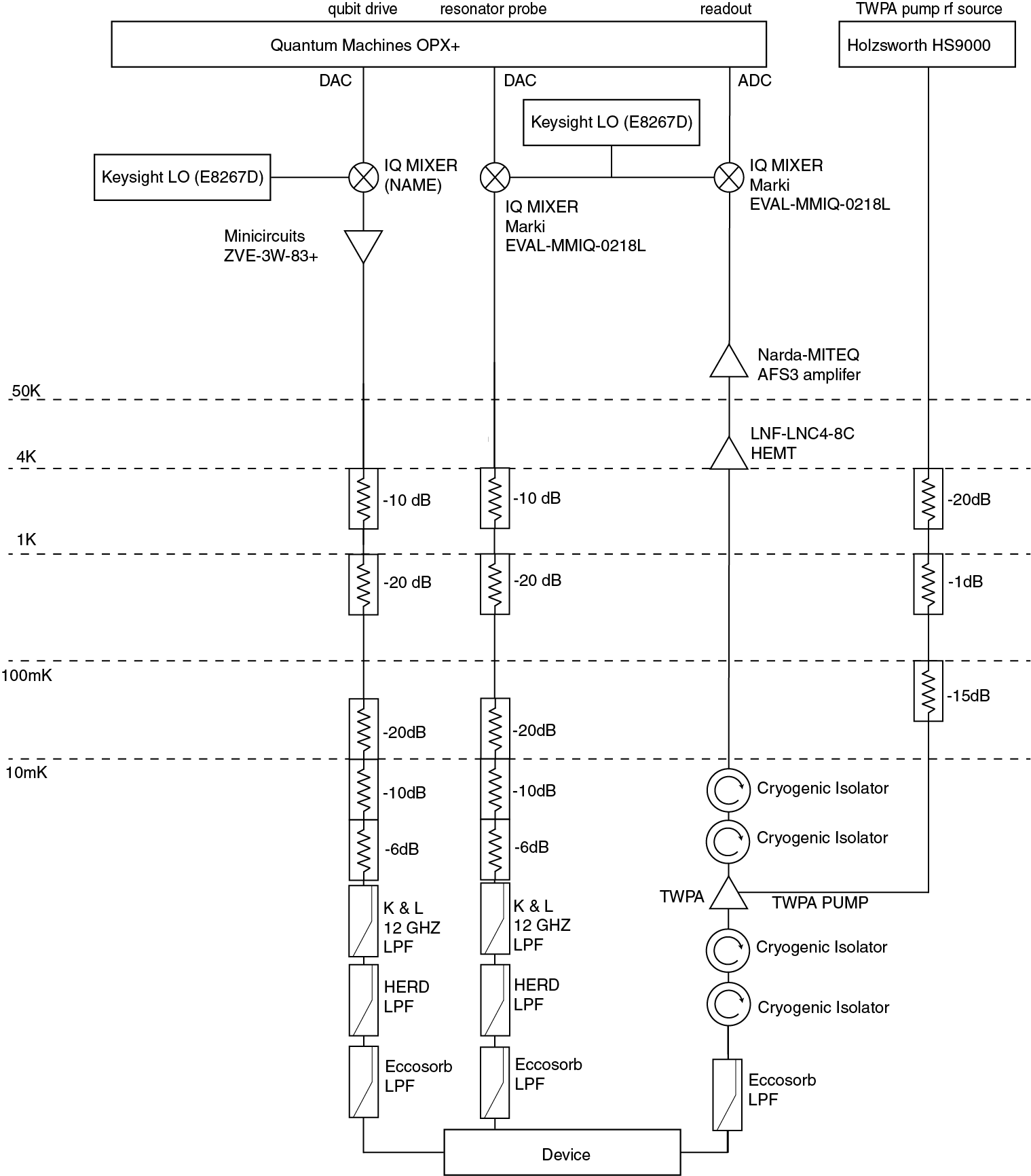}
  \caption{Fridge wiring diagram for qubit measurements} 
  \label{fig_si:fridge_wiring}
\end{figure}

The device fabrication and packaging procedures are identical to those in \cite{bland_millisecond_2025}. All aluminum/aluminum oxide evaporation for Josephson junction fabrication was performed in a Plassys UHV electron-beam evaporation system. The devices were measured in a BlueFors XLD400 dilution refrigerator with wiring configuration shown in Fig. \ref{fig_si:fridge_wiring}. All qubit measurements, including the interleaved measurement protocol, were implemented and performed on a Quantum Machines Operator X (OPX). 

\label{Measurement}

\subsection{Interleaved Measurement Protocol}
Each round of the interleaved measurement protocol consists of a subset of the following pulse sequences: $T_1$ pulse sequences with various delay times, $T_{2\text{E}}$ pulse sequences with various delay times, Ramsey pulse sequences with various delay times, CPMG pulse sequences with various number of $\pi$-pulses and delay times, and AC Stark shift pulses of a fixed delay time with various drive amplitudes. After each pulse sequence we use an active reset protocol as detailed in Appendix \ref{ActiveReset} to reset the qubit in its ground state, which improves the temporal resolution of the measurement. The pulses sequences in each round follow the same fixed order, unless they are \textit{shotgun sampled}, in which case the order of the pulse sequences is permuted each round. We calibrate the qubit readout using single-shot calibration every 30 minutes; the details of this calibration are provided in Appendix \ref{SSCal}.

We have taken a total of 9 datasets across 5 qubits. Table \ref{table:datasets} lists the measurements performed for each dataset, its measurement duration, the pulse sequences which were interleaved, and whether the pulse sequences were shotgun sampled. 

\begin{table*}[t]
\centering
\caption{\label{table:datasets}List of datasets taken with interleaved measurement protocol}
\begin{ruledtabular}
\begin{tabular}{c c c c c c c c c c}
     Dataset \# & Qubit \# & Duration (hours) & $T_1$ & $T_{2\text{E}}$ & Ramsey & $T_{2 \text{, CPMG}}$ & AC Stark & Shotgun Sampled? \\ \hline
       1 & 1 & 53 & \checkmark & \checkmark &  &  &  & No \\
       2 & 1 & 10 & \checkmark &  &  & \checkmark &  & No \\
       3 & 1 & 74 & \checkmark &  \checkmark  & \checkmark & \checkmark &  & No \\
       4 & 2 & 54 & \checkmark & \checkmark & \checkmark & \checkmark & & No\\ 
       5 & 3 & 91 & \checkmark &  \checkmark  & \checkmark & \checkmark &  & No \\
       6 & 4 & 91 & \checkmark &  \checkmark  & \checkmark & \checkmark &  & No \\
       7 & 5 & 53 & \checkmark &  \checkmark  & \checkmark & \checkmark & \checkmark & Yes \\
       8 & 1 & 83 & \checkmark &  \checkmark  & \checkmark & \checkmark & \checkmark & Yes \\
       9 & 1 & 61 & \checkmark & \checkmark & & & \checkmark & Yes\\
   \end{tabular}
   \end{ruledtabular}
\end{table*}

\subsection{Single-Shot Calibration}
\label{SSCal}

\begin{figure*}[t]
    \centering
    \includegraphics[width=1\linewidth]{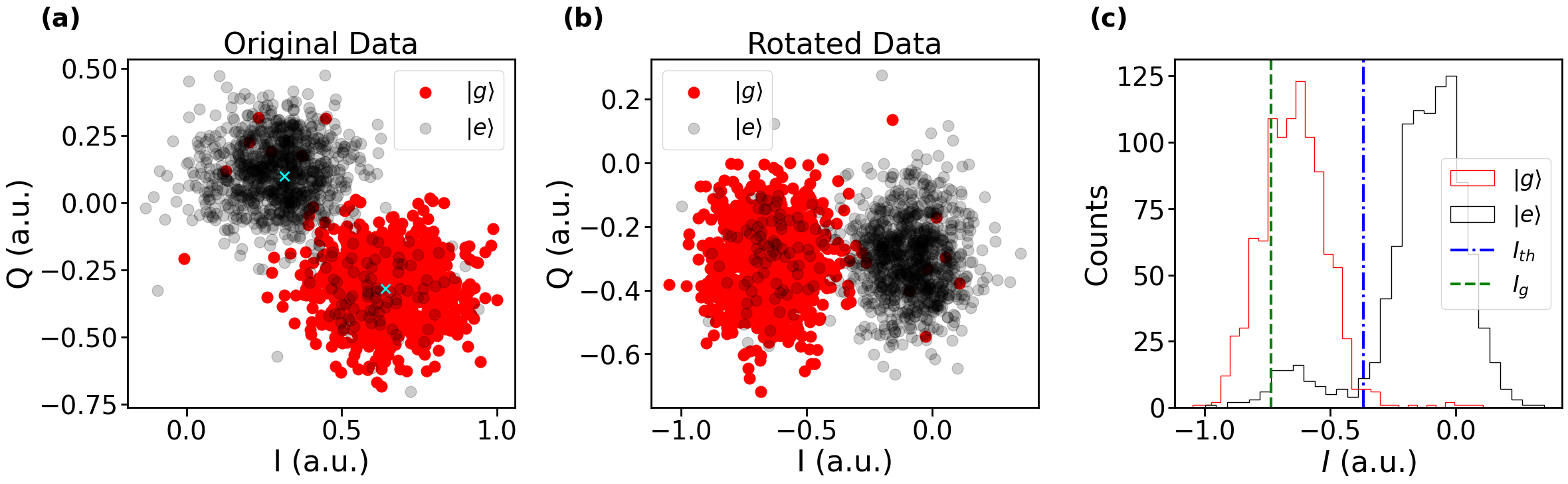}
    \caption{\textbf{Single shot calibration example}: (a) Resonator response $z = I + iQ$ of states prepared in $\ket{g}$ (red) and $\ket{e}$ (black) plotted on the complex plane. (b) Rotated data so that the maximum state discrimination is contained in the $I$ quadrature. (c) Histogram of $I$ quadrature component of the rotated single-shot data. Histogram bins have width $\sqrt{N}$. Vertical lines show the $\ket{g}$-$\ket{e}$ discrimination threshold (blue) and the 95$\%$ posterior confidence threshold (green).}
    \label{fig:SI_SS_cal}
\end{figure*}

We calibrate qubit readout with a single-shot calibration protocol consisting of measuring the readout resonator response for $N = 1000$ shots where the qubit is prepared in $\ket{g}$ and also $N = 1000$ shots where the qubit is prepared in $\ket{e}$ (Fig. \ref{fig:SI_SS_cal}(a)). For a qubit prepared in state $s \in \{g, e\}$, the amplitude and phase of the resonator response for shot $k$ ($1 \leq k \leq N$) can be written in complex form as $z_k^{(s)} = I_k^{(s)} + i Q_k^{(s)}$ with centroid $\bar{z}_s$. We rotate the data as $z_k \to z_k e^{-i\phi}$ with $\phi = \arg(\bar{z}_e - \bar{z}_g)$. This aligns the
vector connecting $\overline{z}_g$ and $\overline{z}_e$ with the positive $I$ axis so that there is maximum contrast between the two qubit states along the $I$ axis. We adopt the convention that we always rotate the data so that the shots corresponding to the qubit prepared in $\ket{e}$ lie at greater $I$ than those corresponding to the qubit prepared in $\ket{g}$ (Fig. \ref{fig:SI_SS_cal}(b)). We discriminate the rotated data using only the $I$ quadrature. Taking $\mathcal{I}$ to be the set of all $2N$ rotated $I$ values, we choose the discrimination threshold $I_{\textrm{th}}$ which maximizes the readout fidelity $\mathcal{F}(t)$ for a given threshold $t$:
\begin{equation}
    I_{\rm th} = \operatorname*{arg\,max}_{t \in \mathcal{I}} \mathcal{F}(t)
    \qquad
    \mathcal{F}(t) = \tfrac{1}{2}\left[ P(g|g;t) + P(e|e;t) \right]
\end{equation}
where $P(g|g;t)$ is the fraction of $\ket{g}$-prepared shots correctly assigned as $\ket{g}$ given a $I$-quadrature threshold $I < t$ and $P(e|e;t)$ is defined analogously for $\ket{e}$-prepared shots) (Fig. \ref{fig:SI_SS_cal}(c)). The rotation angle $\phi$ and threshold $I_{\rm th}$ are used to readout the qubit state for subsequent qubit measurements.

We also determine a posterior confidence threshold $I_g$, the
largest value in $\mathcal{I}$ for which the posterior probability that a shot with $I < I_g$ was prepared in $\ket{g}$ exceeds $95\%$ (Fig. \ref{fig:SI_SS_cal}(c)). This threshold is used in the active reset protocol (Appendix \ref{ActiveReset}).

\subsection{Active Reset}
\label{ActiveReset}

We perform measurement-based active reset by first measuring the qubit state and then rotating the resonator response by $\phi$ as specified in the most recent single-shot calibration. Thus the qubit state can be discriminated entirely from the $I$ quadrature of the rotated data. Reset is preformed with FPGA logic at runtime via the pseudocode in Algorithm \ref{ActiveResetAlgo}.

\begin{algorithm}[h]
\label{ActiveResetAlgo}
\caption{Active reset}
\KwIn{Thresholds $I_{\rm th}$, $I_g$, resonator linewidth $\kappa$}
\KwOut{Initial state assignment $s_0$}
Measure qubit, obtain rotated $I$\;
$s_0 \leftarrow \ket{e}$ \textbf{if} $I > I_{\rm th}$ \textbf{else} $\ket{g}$\;
\While{$I \geq I_g$}{
    \If{$I > I_{\rm th}$}{
        Apply $\pi$ pulse\;
    }
    Measure qubit, obtain rotated $I$\;
}
Wait $10/\kappa$ for resonator to depopulate\;
\Return $s_0$\;
\end{algorithm}

\section{Temporal Resolution of Qubit Measurements}
\label{TimeResMeas}

\begin{table}[!htbp]
\centering
\caption{\label{table:temporal-budget}Measurement time budgeting for various components of a single-shot qubit measurement}
\begin{ruledtabular}
\begin{tabular}{c c c c}
     Pulses & Delay Time & Readout & Reset  \\ \hline
     $2 \times 10^{-7}$ s  &  $1 \times 10^{-3}$ s &    $2 \times 10^{-5}$ s  &    $2 \times 10^{-5}$ s
   \end{tabular}
   \end{ruledtabular}
\end{table}

We use interleaved measurements to more accurately correlate $T_1$ and $T_{2\text{E}}$ measurements as well as CPMG noise spectroscopy and TLS spectroscopy. Without interleaved measurements, a typical $T_1$ or $T_{2\text{E}}$ measurement takes around 1 -- 2 minutes using the active reset protocol in Appendix \ref{ActiveReset} and averaging around 500 shots for each population measurement. If we use a measurement protocol alternating between full $T_1$ and $T_{2\text{E}}$ measurements, dynamics which are faster than minutes-long timescales can disproportionately affect a $T_1$ (or $T_{2\text{E}}$) measurement, despite the calculation of $\Gamma_\phi$ assuming that they are measured effectively simultaneously.

To analyze the timescale of dynamics required in an interleaved measurement to disproportionately affect one of the $T_1$ and $T_{2\text{E}}$ measurements,  Table \ref{table:temporal-budget} provides a temporal budget of the wall clock time required to perform a single-shot qubit measurement. We note that the wall clock time is dominated by the delay time, which is on the order of milliseconds. Thus the interleaved measurement protocol provides four orders of magnitude improvement in the simultaneity of $T_1$ and $T_{2\text{E}}$ measurements.

The temporal budgeting also reveals fundamental limits to the temporal resolution of our measurements. The time between single-shot qubit measurements is dominated by the delay time. Thus the fastest $T_1$ measurement under our current protocol (consisting of measuring the population at a single delay time) would be on the order of seconds (assuming averaging over $\sim 1000$ shots). The temporal budgeting also shows that faster pulses, readout time, or reset schemes offer little improvement to the temporal resolution of our measurements. In addition, this highlights the importance of an active reset scheme: a passive reset scheme which waits $\sim 3 T_1$ for the qubit to relax to the ground state would cause the reset time to be similar to the delay time.

\section{Statistics of $T_1$ and $T_{2\text{E}}$ Fluctuations}
\label{ExtraData}

\subsection{Distributions of $T_1$ and $T_{2\text{E}}$}
\label{CoherenceDistr}

\begin{figure*}
   \centering
  \includegraphics[width=\textwidth]{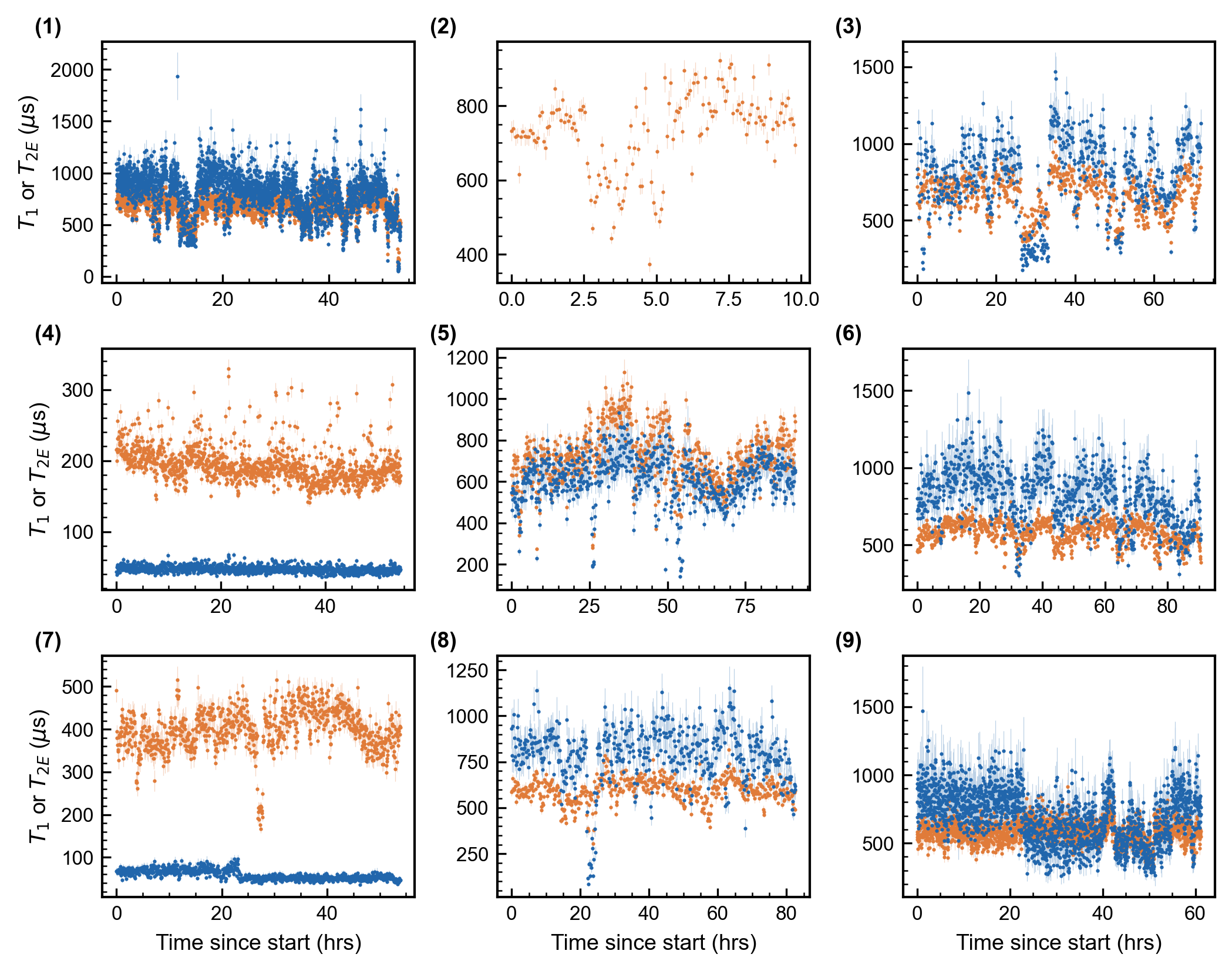}
  \caption{\textbf{Temporal Fluctuations of $T_1$ and $T_{2\text{E}}$}: repeated $T_1$ (orange) and $T_{2\text{E}}$ (blue) measurements over time for each dataset, which is indicated in parentheses at the top left of each plot. $T_1$ and $T_{\textrm{2E}}$ are measured by fitting to exponential decay of populations averaged over 500 shots for datasets 1, 7, 8, and 9, while averaging over 300 shots for datasets 2 through 6.
  }
   \label{fig_si:time_traces}
\end{figure*}

The temporal fluctuations in $T_1$ and $T_{2\text{E}}$ for each of the datasets listed in Table \ref{table:datasets} are shown in Fig. \ref{fig_si:time_traces}. Histograms of the measured $T_1$ and $T_{2\text{E}}$ values for each dataset are shown in Fig. \ref{fig_si:histograms}. As in the Supplementary Information of \cite{bland_millisecond_2025}, the Shapiro-Wilk and D'Agostino-Pearson tests of Gaussianity were performed on the distribution of $T_1$ and $T_{2\text{E}}$ for each dataset. Assuming a significance level of $\alpha = 0.01$, we find that the Shapiro-Wilk test classifies all the $T_1$ and $T_{2\text{E}}$ distributions as non-Gaussian except for the $T_{2\text{E}}$ distribution of dataset 6. The D'Agostino-Pearson test classifies all the $T_1$ and $T_{2\text{E}}$ distributions as non-Gaussian except for the $T_1$ distribution of dataset 5 and the $T_{2\text{E}}$ distribution of dataset 6.

\begin{figure*}
   \centering
  \includegraphics[width=\textwidth]{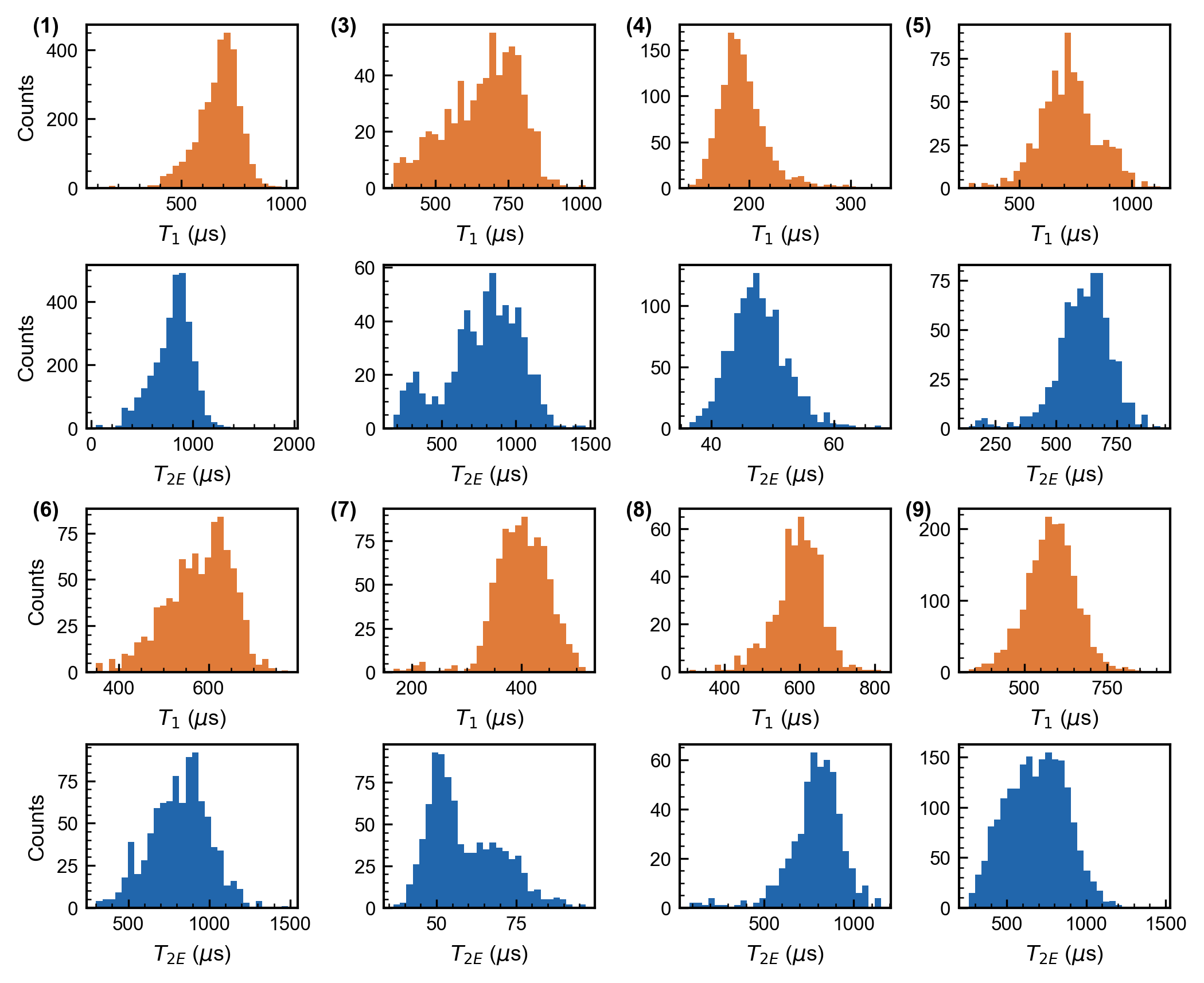}
  \caption{\textbf{Histograms of $T_1$ and $T_{2\text{E}}$}: Distributions of $T_1$ and $T_{2\text{E}}$ from measurements shown in Fig. \ref{fig_si:time_traces}. The distributions are notably non-Gaussian, which is quantified with Gaussianity statistical tests in Table \ref{table:gaussian-pvals}. $T_1$ and $T_{2\text{E}}$ values are from the dataset indicated in parentheses; dataset 2 is omitted since it has much fewer measurements than all other datasets.
  }
   \label{fig_si:histograms}
\end{figure*}

\begin{table}[H]
\centering
\caption{\label{table:gaussian-pvals}Gaussianity $p$-values for $T_1$ and $T_{2\text{E}}$ measurements}
\begin{ruledtabular}
\begin{tabular}{c c c c c}
             & \multicolumn{2}{c}{Shapiro-Wilk} & \multicolumn{2}{c}{D'Agostino-Pearson}\\
     Dataset & $T_1$ & $T_{2\text{E}}$ & $T_1$ & $T_{2\text{E}}$  \\ \hline
     1 & $3.42 \times 10^{-29}$ & $3.69 \times 10^{-22}$ & $1.47 \times 10^{-108}$ & $1.46 \times 10^{-31}$\\
     3 & $1.04 \times 10^{-9}$ & $3.48 \times 10^{-10}$ & $1.89 \times 10^{-7}$ & $4.33 \times 10^{-6}$\\
     4 & $1.84 \times 10^{-25}$ & $6.12 \times 10^{-10}$ & $1.47 \times 10^{-76}$ & $8.19 \times 10^{-16}$\\
     5 & $1.02 \times 10^{-3}$ & $1.26 \times 10^{-13}$ & $7.47 \times 10^{-2}$ & $7.70 \times 10^{-25}$\\
     6 & $3.96 \times 10^{-10}$ & $1.88 \times 10^{-2}$ & $2.94 \times 10^{-8}$ & $3.14 \times 10^{-1}$\\
     7 & $4.76 \times 10^{-17}$ & $2.23 \times 10^{-17}$ & $3.91 \times 10^{-35}$ & $4.84 \times 10^{-14}$ \\
     8 & $3.63 \times 10^{-7}$ & $1.90 \times 10^{-16}$ & $2.89 \times 10^{-9}$ & $4.08 \times 10^{-33}$ \\
     9 & $3.40 \times 10^{-5}$ & $6.56 \times 10^{-9}$ & $5.93 \times 10^{-4}$ & $7.73 \times 10^{-8}$\\
   \end{tabular}
   \end{ruledtabular}
\end{table}

\subsection{Correlation Time}
\label{CorrTime}

We calculate the correlation time of repeated qubit $T_1$ and $T_{2\text{E}}$ measurements using the \textit{integrated autocorrelation time} \cite{MadrasSokal1988}. Let $\rho_k$ be the autocorrelation of lag $k$ in a time-series dataset. The integrated autocorrelation time $\tau_{int}$ is defined as:
\begin{equation}
\tau_{int} \equiv \frac12 \sum_{k = -\infty}^\infty \rho_k = \frac12 + \sum_{k = 1}^\infty \rho_{k}
\label{eq:int_corr_time}
\end{equation}

This is a commonly used metric in Markov Chain Monte Carlo (MCMC) sampling as an estimate of the effective number of the independent measurements in a series of correlated measurements. Given $N$ time-series samples, the effective number of independent samples is given by $N / 2 \tau_{int}$. The factor of 2 is included in \eqref{eq:int_corr_time} so that for large $\tau \gg 1$, given an exponential decaying autocorrelation $\rho_k = e^{-|k| / \tau}$, the integrated autocorrelation time $\tau_{int} \approx \tau$. We report the integrated autocorrelation times as $\tau_{int} \Delta t$, where $\Delta t$ is the average time between successive $T_1$ or $T_{2\text{E}}$ measurements. 

In practice, summing $\rho_k$ for $k \rightarrow \infty$ gives an inaccurate estimate of $\tau_{int}$, since the values of $\rho_k$ for large $k$ are noisy. Instead, we sum autocorrelations $\rho_k$ up to a finite cutoff $M$:
\begin{equation}
    \tau_{int} \equiv \frac12 + \sum_{k=1}^M \rho_k
\end{equation}
This cutoff $M$ is determined as the minimum $M$ such that $M \geq c \tau_{int}$, where $c$ is a pre-selected hyperparameter. We choose $c = 5$ and show that the correlation time is robust against choices of $c$ between 4 and 6 (Table \ref{table:corr-time-sensitivity}). The cutoff $M$ is then found by finding the minimum $M$ satisfying the self-consistent equation \cite{MadrasSokal1988}:
\[M \geq c \left(\frac12 + \sum_{k = 1}^M \rho_{k}\right)\]
Intuitively, this is setting the cutoff to be $M = c \tau_{int}$, where $\tau_{int}$ is estimated by summing autocorrelations up to cutoff $M$.

\begin{figure*}
   \centering
  \includegraphics[width=\textwidth]{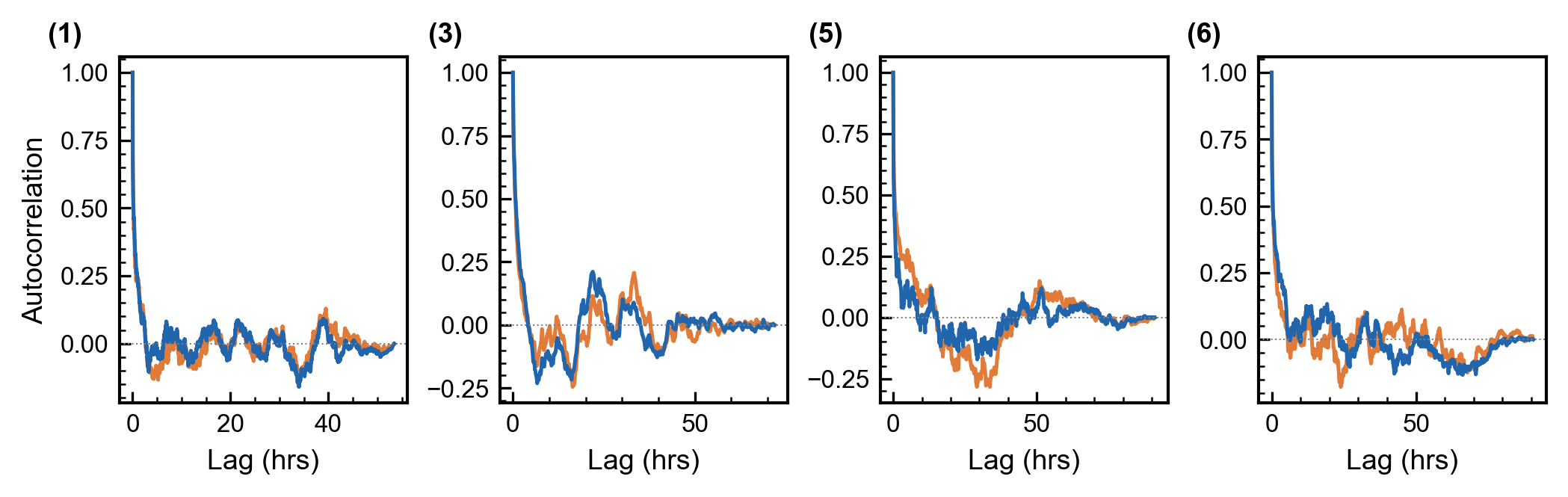}
  \caption{\textbf{Autocorrelation function for $T_1$ and $T_{2\text{E}}$ measurements}: Autocorrelation as a function of lag for repeated $T_1$ (orange) and $T_{2\text{E}}$ (blue) measurements over time. Autocorrelations are calculated from the dataset specified in parentheses on the top left of each plot.
  }
   \label{fig_si:autocorrelations}
\end{figure*}

\begin{table}[H]
\centering
\caption{\label{table:corr-time-sensitivity}Sensitivity analysis of hyperparameter $c$ on integrated autocorrelation time (listed in minutes)}
\begin{ruledtabular}
\begin{tabular}{c c c c c c c}
             & & \multicolumn{5}{c}{$c$ value} \\
     \multicolumn{2}{c}{Dataset} & 4.0 & 4.5 & 5.0 & 5.5 & 6.0  \\ \hline
     1 & $T_1$    & 32.4 & 34.2 & 33.9 & 33.2 & 32.3 \\
     1 & $T_{2\text{E}}$ & 32.8 & 34.4 & 34.4 & 33.7 & 32.7 \\
     3 & $T_1$    & 60.5 & 60.3 & 59.0 & 57.9 & 56.3 \\
     3 & $T_{2\text{E}}$ & 71.4 & 69.7 & 66.3 & 63.2 & 60.8 \\
     5 & $T_1$    & 145.3 & 152.9 & 168.7 & 169.0 & 164.4 \\
     5 & $T_{2\text{E}}$ & 46.2 & 48.0 & 51.3 & 59.1 & 65.4 \\
     6 & $T_1$    & 63.2 & 66.0 & 67.1 & 66.0 & 64.7 \\
     6 & $T_{2\text{E}}$ & 92.2 & 93.5 & 95.2 & 99.5 & 100.2 \\
   \end{tabular}
   \end{ruledtabular}
\end{table}

\section{Correlations Between $\Gamma_\phi$ and $\Gamma_1$}
\label{CorrRD}

\subsection{Outlier Removal}
\label{OutlierRemoval}
\begin{figure*}
   \centering
  \includegraphics{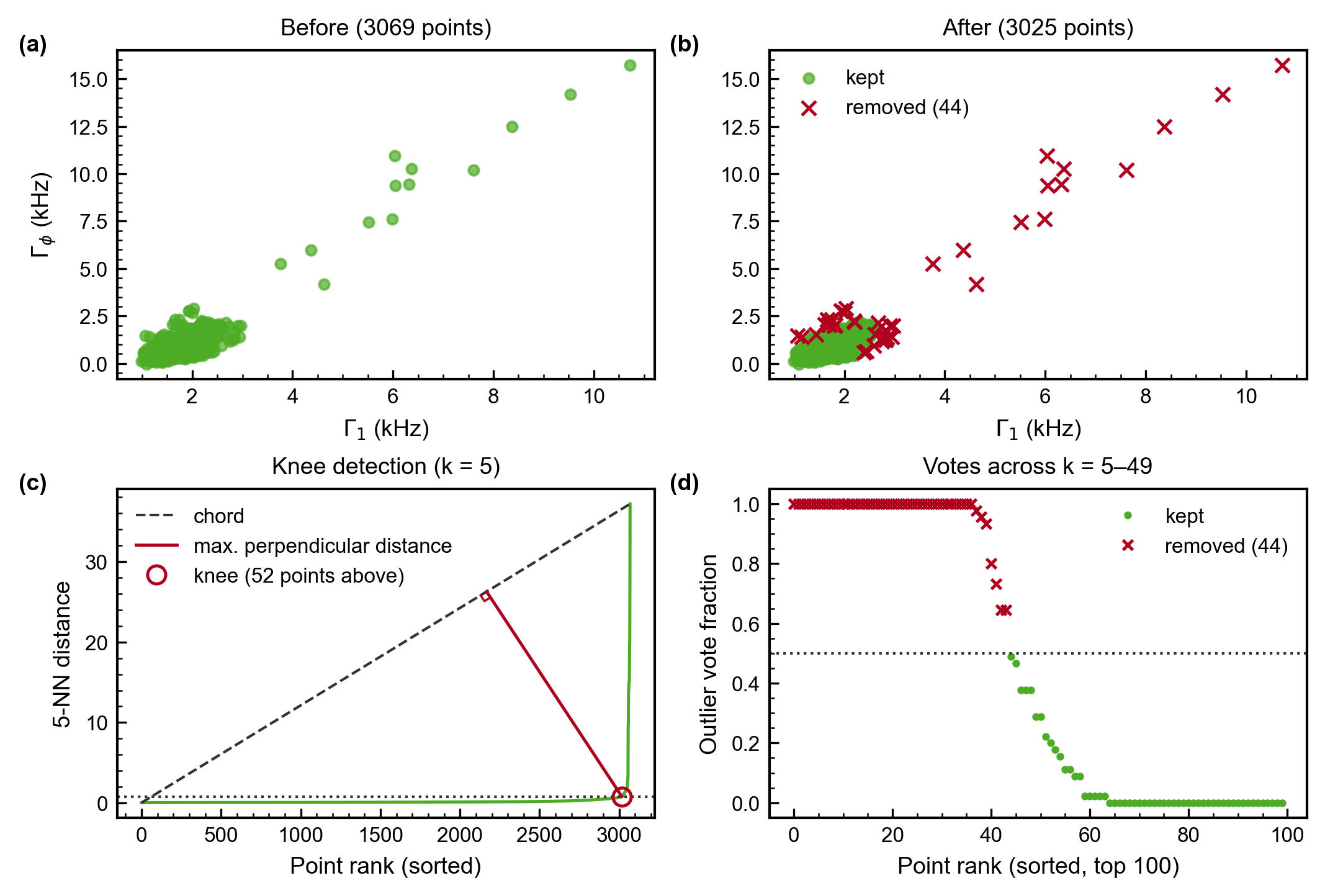}
  \caption{\textbf{Outlier Removal}: (a) Relaxation and dephasing rate scatter plot from dataset 1. (b) Relaxation and dephasing rate scatter plot with outliers indicated with a red $\times$. (c) Outlier threshold using $k = 5$ nearest neighbors for each data point. (d) Sorted outlier vote fraction for $k = 5$ to $k = 49$ for the 100 points with the highest outlier vote fraction.
  }
   \label{fig_si:outlier_removal}
\end{figure*}

To prevent sparse outliers from skewing the correlation metrics and data fitting, we use a density-based approach to remove outliers in the $\Gamma_\phi$ vs. $\Gamma_1$ scatter plot (Fig. \ref{fig_si:outlier_removal}(a), (b)). Each point $(\Gamma_{1, j}, \Gamma_{\phi, j})$ (where the data points are individually indexed by $j$) is converted to standardized values $(\Gamma_{1, j}^{s}, \Gamma_{\phi, j}^{s})$. The standardized values $\Gamma_{1, j}^s$ (and analogously for $\Gamma_{\phi, j}^s$) are defined as follows:
\begin{equation} 
\Gamma_{1, j}^s = \frac{\Gamma_{1, j} - \textrm{median}(\Gamma_1)}{\textrm{MAD}(\Gamma_1)}
\end{equation}
\begin{equation} 
\textrm{MAD}(\Gamma_1) = 1.4826 \times \textrm{median}(|\Gamma_{1, j} - \textrm{median}(\Gamma_1)|)
\end{equation}
MAD is the \textit{median absolute deviation}, which is an outlier-robust version of the standard deviation. The factor of 1.4826 is a normalizing constant such that the MAD of a Gaussian distribution equals its standard deviation.

For a fixed value $k$, we calculate for each point $(\Gamma_{1, j}^s, \Gamma_{\phi, j}^s)$ the Euclidean distance to its $k$th nearest neighbor (kNN-distance). We classify all points whose kNN-distance is greater than some threshold $t_k$ as an outlier. To determine the threshold $t_k$, we plot the sorted kNN-distance and find the point on the curve which is farthest from the chord connecting the left and right endpoints of the curve (Fig. \ref{fig_si:outlier_removal}(c)).

We perform this procedure for each $k$ between 5 and 49, and for each $k$ label each data point $(\Gamma_{1, j}^s, \Gamma_{\phi, j}^s)$ as an outlier or non-outlier. All points which are labeled as an outlier by a majority vote (i.e. labeled as an outlier by more than 50\% of $k$ values between 5 and 49) are removed. With this majority-vote procedure, the outlier removal procedure is robust to the value of $k$. This procedure is also robust to the majority vote threshold of 50\% (Fig. \ref{fig_si:outlier_removal}(d)).

\subsection{Piecewise Linear Fitting of $\Gamma_\phi$ vs. $\Gamma_1$}
\label{ReLUFit}

\begin{figure*}
   \centering
  \includegraphics{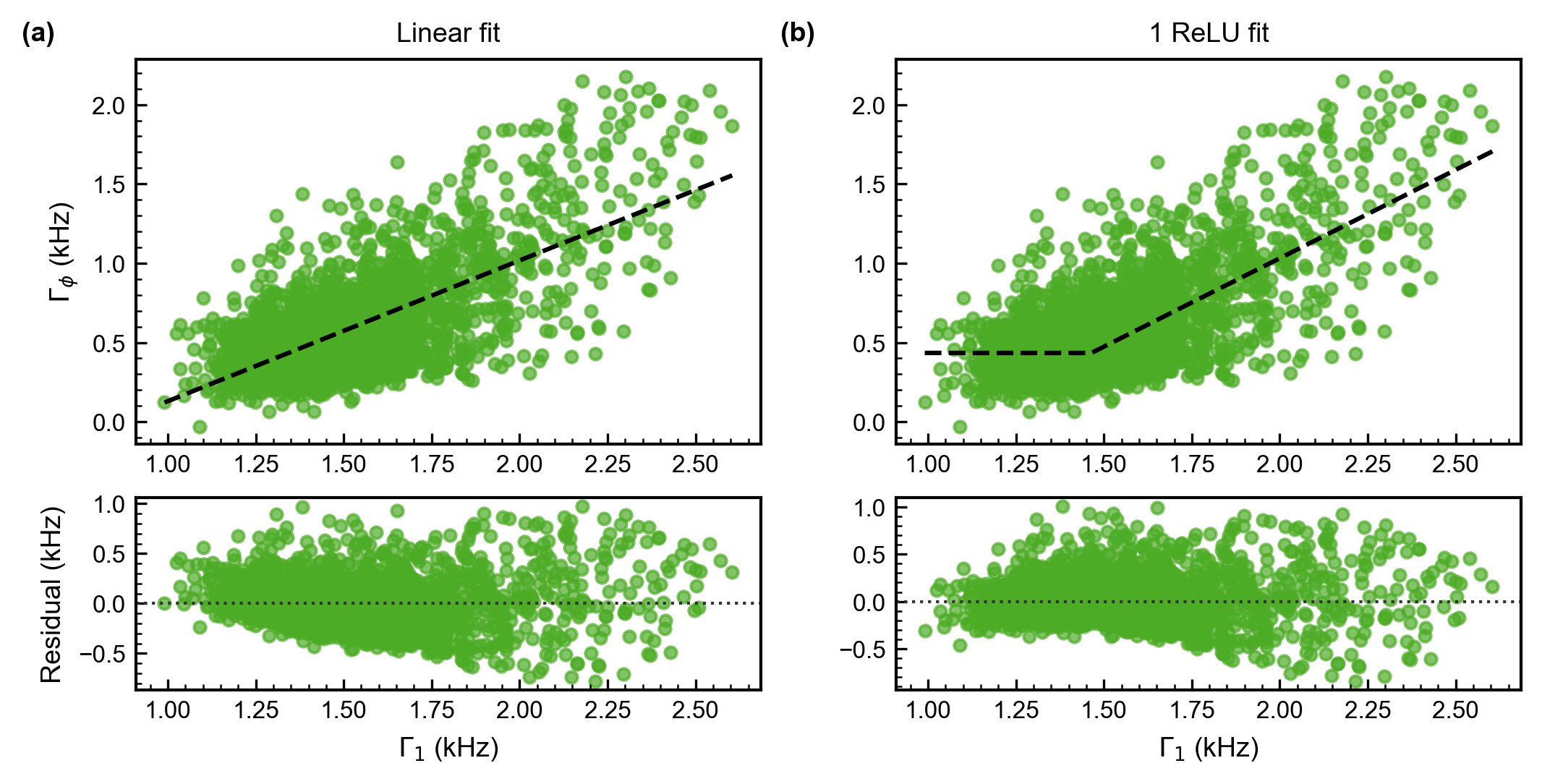}
  \caption{\textbf{Residual Analysis}: Comparison of residuals for a linear fit (a) and a piecewise linear fit with 1 ReLU (b). The residuals in (a) show a distinct downward trend, which is not apparent in (b).
  }
   \label{fig_si:residual_analysis}
\end{figure*}

Analysis of residuals suggests that the correlation between $\Gamma_\phi$ and $\Gamma_1$ is nonlinear, as shown by a decreasing trend in the residuals in Fig. \ref{fig_si:residual_analysis}. In order to investigate how the fitted slope between $\Gamma_\phi$ and $\Gamma_1$ changes based on the value of $\Gamma_1$, we fit the data to a functional form given by:
\begin{equation}
f(x) = \sum_{i=1}^k a_i \textrm{ReLU}(w_i x + b_i) + C \qquad \textrm{ReLU}(x) = \max(x, 0)
\label{eq:sum_relus}
\end{equation}
The fitted parameters are $a_i$, $w_i$, $b_i$, and $C$. The rectified linear unit (ReLU) is a nonlinear function widely used in neural networks for nonlinear data fitting to construct machine learning models. Assuming that the data deviates from Eq. \eqref{eq:sum_relus} with Gaussian noise $\mathcal{N}(0, \sigma^2)$, the fitting uses maximum likelihood estimation (MLE) by selecting the fitted parameters which maximize the weighted log-likelihood $\log L$ on the data $(\Gamma_{1, j}, \Gamma_{\phi, j})$:
\begin{equation}
    \log L = - \frac12 \sum_j W_j \left[\log 2 \pi \sigma^2 - \frac{(\Gamma_{\phi, j} - f(\Gamma_{1, j}))^2}{2 \sigma^2}\right]
\end{equation}
We use density-based weights $W_j$ for each point to prevent sparse points in the scatter from over-influencing the fit. The density-based weight for each point is $1/d_j$, where $d_j$ is the Euclidean distance to the 5th nearest neighbor. The actual weights $W_j$ are normalized such that $\sum_j W_j = n$, where $n$ is the number of data points. The minimization is performed using the L-BFGS-B optimizer with 100 random restarts.

We perform this fitting for $k = 1$ up to $k = 5$. To select the best model while preventing overfitting, we calculate the Bayes Information Criterion (BIC) for the fitted model at each $k$:
\begin{equation} 
\textrm{BIC} = -2 \log L + (3k + 2) \log n_{\textrm{eff}}
\end{equation}
where $\log L$is the log-likelihood as before and $n_{\textrm{eff}}$ is the effective number of data points using Kish's effective sample size \cite{kish1965survey} due to the weights: 
\begin{equation} 
n_{\textrm{eff}} \equiv \frac{n}{1 + \textrm{Var}(W_j)}
\end{equation}
We select the model with $k$ ReLUs such that BIC is minimized. Note that the BIC metric penalizes increased model complexity (additional ReLUs) through the $(3k + 2) \log n_{\textrm{eff}}$ term. 

\begin{figure*}
   \centering
  \includegraphics{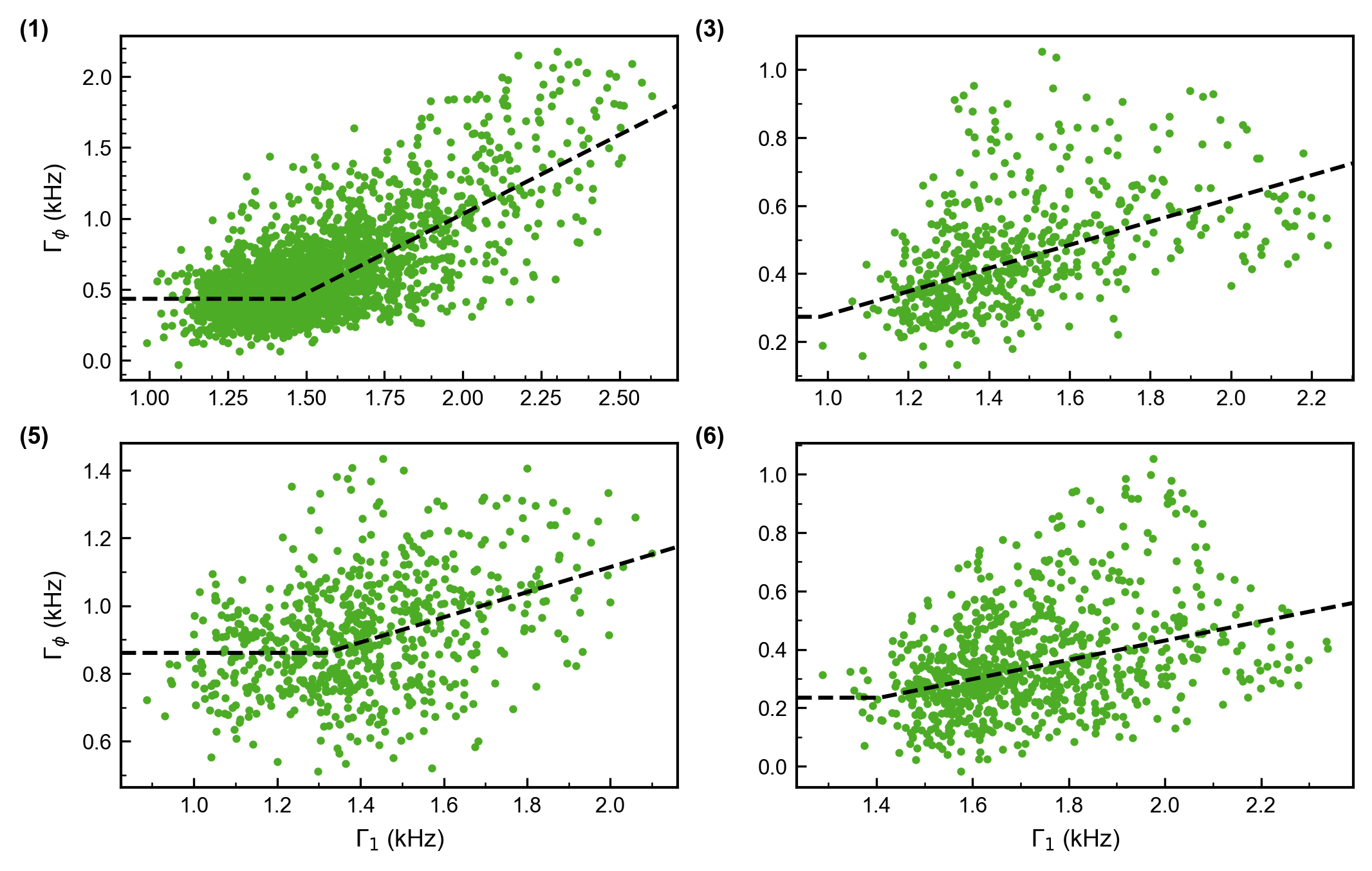}
  \caption{\textbf{Correlation Analysis for Various Datasets}: Piecewise linear fitting performed on four datasets as indicated in parentheses at the top left of each subpanel.
  }
   \label{fig_si:corr_datasets}
\end{figure*}

\subsection{Correlation Analysis on Various Datasets}

We have performed this piecewise linear fitting procedure on four datasets where the average $T_1$ and $T_{2E}$ are comparable (Fig. \ref{fig_si:corr_datasets}). In datasets 1 and 5, we clearly observe uncorrelated and correlated regimes of relaxation and dephasing. However, in datasets 3 and 6 we only observe correlated relaxation and dephasing.

\section{Noise Spectral Density Calculations from CPMG Dynamical Decoupling Traces}
\label{SwCalc}

\subsection{Deriving the Coherence Integral}
Let $\varphi(t)$ be the excess phase acquired during free evolution in a $T_2$ experiment. The decay curve due to qubit dephasing $|\langle e^{i \varphi(t)} \rangle|$ can be expanded in cumulants $\kappa_n[\varphi(t)]$ of the random variable $\varphi(t)$:
\begin{equation}
    \langle e^{i \varphi(t)} \rangle = \exp\left(\sum_{n=1}^\infty \frac{i^n}{n!} \kappa_n[\varphi(t)] \right) 
\end{equation}
We assume zero-mean Gaussian noise, so $\kappa_n[\varphi(t)] = 0$ except for $n = 2$. Thus $\langle e^{i \varphi(t)} \rangle = \exp(-\frac12 \langle \varphi(t)^2\rangle)$. Let $\delta\omega(t)$ represent qubit frequency fluctuations during the free evolution time and let $g(t) = \pm 1$ represent the effect of refocusing pulses. The phase acquired can be expressed as:
\begin{equation} 
\varphi(t) = \int_0^t dt^\prime \, \delta\omega(t^\prime) g(t^\prime)
\label{eq:accum_phase}
\end{equation}
We compute the second moment of the accumulated phase as:
\begin{align}
\langle \varphi(t)^2 \rangle &= \left\langle \int_0^t dt_1 \, \delta\omega(t_1) g(t_1) \int_0^t dt_2 \, \delta\omega(t_2) g(t_2) \right\rangle \\
&= \int_0^t dt_1 \int_0^t dt_2 \, \langle \delta \omega(t_1) \delta \omega(t_2) \rangle g(t_1) g(t_2)
\end{align}
Assuming that the noise is stationary, the Wiener-Khinchin theorem states that:
\begin{equation}
\langle \delta \omega(t_1) \delta \omega(t_2) \rangle = \int_{-\infty}^{\infty} \frac{d \omega}{2 \pi} S(\omega) e^{i \omega (t_1 - t_2)}
\end{equation}
We use the two-sided noise spectral density $S(\omega)$. The second moment of the accumulated phase becomes:
\begin{equation}
\begin{aligned} 
\langle \varphi(t)^2 \rangle &= \int_{-\infty}^{\infty} \frac{d \omega}{2 \pi} S(\omega) \\
&\left[\int_0^t dt_1 \, g(t_1) e^{i \omega t_1} \int_0^t dt_2 \, g(t_2) e^{i \omega t_2} \right]
\end{aligned}
\end{equation}
The term in brackets is the squared magnitude of the Fourier transform of the pulse sequence $g(t)$. We define the term in brackets to be $t^2 g_N(\omega, t)$, where $g_N(\omega, t)$ is the filter function of the pulse sequence (the factor of $t^2$ arises so that $g_N(\omega, t)$ is dimensionless). Noting that the two-sided noise spectral density $S(\omega)$ and the filter function $g_N(\omega, t)$ are even in $\omega$, we have that:
\begin{equation} 
\langle \varphi(t)^2 \rangle = 2 \int_0^\infty \frac{d \omega}{2 \pi} S(\omega) t^2 g_N(\omega, t)
\end{equation}
Thus the decay curve due to qubit dephasing is:
\begin{equation} 
\langle e^{i \varphi(t)} \rangle = \exp\left(-t^2 \int_0^\infty \frac{d \omega}{2 \pi} S(\omega) g_N(\omega, t)\right)
\end{equation}
The expression in the exponential is defined as the coherence integral $\chi_N(t)$:
\begin{equation}
    \chi_N(t) = t^2 \int_0^\infty \frac{d \omega}{2 \pi} S(\omega) g_N(\omega, t)
    \label{eq:coherence_integral}
\end{equation}
For a CPMG pulse sequence, the filter function $g_N(\omega, t)$ is given by:
\begin{equation}
\begin{split}
g_N(\omega, \tau) &= \frac{1}{(\omega \tau)^2} \Big|1 + (-1)^{1 + N} \exp(i \omega \tau) \\
  &+ 2 \sum_{j=1}^N (-1)^j \exp(i \omega \delta_j \tau) \cos(\omega t_\pi / 2) \Big|^2
\end{split}
\label{eq:filter_SI}
\end{equation}
where $\delta_j \in [0, 1]$ is the normalized position of the center of the $j$th pulse between the $X_{\pi/2}$ pulses, $\tau$ is the total free evolution time, and $t_\pi$ is the length of the $X_\pi$ pulse. 

\subsection{Converting CPMG TO $S(\omega)$}
This calculation is similar to that in \cite{bland_millisecond_2025, bylander_noise_2011}. The CPMG pulse sequence consists of two $X_{\pi/2}$ pulses with a series of $(\tau / (2N) - Y_\pi - \tau / (2N))^N$ pulses between them. The excited state population as a function of total delay time $\tau$ is modeled as:
\begin{equation} 
P(\tau) = A e^{-\tau/(2T_1)}e^{-\Gamma_p N} e^{-\chi_N(\tau)} + C
\label{eq:cpmg_decay}
\end{equation} 
where $A$ is the total contrast of the decay curve, $C$ is the excited state population at long delay times, $\Gamma_p$ is the pulse dephasing rate, and $\chi_N(\tau)$ is the coherence integral defined in equation \eqref{eq:coherence_integral}.
Solving for $\chi_N(\tau)$ in equation \eqref{eq:cpmg_decay}, we find 
\begin{equation} \chi_N(\tau) = -\ln \left(\frac{P(\tau) - C}{A} e^{\tau/(2T_1)}e^{\Gamma_p N}\right) \end{equation}
In equation \eqref{eq:coherence_integral} we can approximate $g_N(\omega, t)$ as a delta function centered at $\omega^\prime = N / 2 \tau$ with area $\Delta \omega \approx 1/\tau$. Thus equation \eqref{eq:coherence_integral} becomes: 
\begin{equation} \chi_N(\tau) = \tau^2 S(\omega^\prime, \tau) \Delta \omega \approx \tau S(\omega^\prime) g_N(\omega^\prime, \tau)\end{equation}
Equating the two expressions for $\chi_N(\tau)$: 
\begin{equation} 
\label{eq:SwConv}
S\left(\omega^\prime = \frac{N}{2 \tau}\right) = -\frac{1}{\tau g_N(\omega^\prime, \tau)} \ln \left( \frac{P(\tau) - C}{A} e^{\tau/2T_1} e^{\Gamma_p N}\right)
\end{equation}
The right-hand side can be extracted entirely from measured values. To estimate the pulse dephasing rate $\Gamma_p$, we define the following function:
\begin{equation} f(N, \Gamma_p) = \frac{P(\tau_\varepsilon) - 0.5}{0.48} e^{\tau_\varepsilon / 2T_1} e^{\Gamma_p N} \end{equation}
The function $f(N, \Gamma_p)$ estimates the pure dephasing at $e^{-\chi_N(\tau = 0)}$ for a given $\Gamma_p$. Ideally we have that $\tau_\varepsilon = 0$, but in practice this is often a small value due to finite latency between pulses. The definition of $f(N, \Gamma_p)$ also uses a guess of $C = 0.5$ and $A = 0.48$, which are close to their actual values. We estimate $\Gamma_p$ by selecting the value of $\Gamma_p$ which minimizes the variance of $f(N, \Gamma_p)$ across all $N$. We note that since we are minimizing variance, the guess of $A = 0.48$ is innocuous since it scales all variances by the same constant. Typical values of $\Gamma_p$ are $\sim 10^{-5}$. After estimating $\Gamma_p$, we extract $A$ and $C$ by fitting to all the CPMG decay curves of various $N$ in a given timestep simultaneously.

\subsection{Data Cutoffs for $S(\omega)$ Calculation}
\label{SwCutoff}

In converting CPMG population measurements $P(\tau)$ to $S(\omega)$, we must exclude data from small $\tau \approx 0$ due to numerical instabilities and large $\tau$ due to quantum projection noise dominating the exponentially small decay curve signal. In this section we derive bounds for the values of $\tau$ that should be excluded when converting between $P(\tau)$ and $S(\omega)$.

For long delay times, suppose the measured decay curve follows a functional form $P(\tau) = A_{\textrm{eff}} e^{-(\tau / T_2)^n} + C$, where $A_{\textrm{eff}} \equiv A e^{-\Gamma_p N}$ from \eqref{eq:cpmg_decay}. If the true excited state population is $p$, then the measured population $P(\tau)$ is the sample mean of a Bernoulli random variable with probability of success $p$ averaged across $N$ measurements. Thus $P(\tau)$ has mean $p$ and standard deviation $\sqrt{p(1-p) / N}$. Hence the decay curve $P(\tau)$ is no longer reliable when:
\begin{equation} A_{\textrm{eff}} e^{-(\tau / T_2)^n} < \sqrt{\frac{p(1-p)}{N}} \end{equation}
This is true for:
\begin{equation} \tau > T_2 \left[\ln \left(A_{\textrm{eff}} \sqrt{\frac{N}{p(1-p)}}\right)\right]^{1/n} \end{equation}
This bound is most conservative for $p = 0.5$, and assuming pulse dephasing $e^{-\Gamma_p N}$ is negligible, we have that $A_{\textrm{eff}} \approx 0.5$. Thus the bound becomes:
\begin{equation} \tau > T_2 (\ln \sqrt{N})^{1/n} \end{equation}
For pure exponential decay $n = 1$ and averaging over $N = 300$ shots (as is typical for data in this paper), we must exclude $\tau > 2.8 T_2$. In practice we usually exclude $\tau > 2 T_2$ for improved data quality.

For short delay times, $P(\tau) - P(0)$ tends to zero which causes the logarithm factor in equation \eqref{eq:SwConv} to diverge, while for small $\tau$ the $1/\tau$ factor diverges also, rendering \eqref{eq:SwConv} highly unstable numerically. Suppose we have some measurement noise $\sigma$ so that the measurement of $P(\tau)$ is $P(\tau) + \sigma$. Furthermore for notational convenience we define $L$ and $\tilde{L}$ to be the following quantity with and without measurement error, respectively:
\begin{equation}
L = \frac{P(\tau) - C}{A e^{-\Gamma_p N}} e^{\tau / 2T_1} \qquad \tilde{L} = \frac{P(\tau) + \sigma - C}{A e^{-\Gamma_p N}} e^{\tau / 2T_1}
\label{eq:LDef}
\end{equation}
We note that we can write \eqref{eq:SwConv} as:
\begin{equation}
S\left(\omega^\prime = \frac{N}{2\tau}\right) = -\frac{\ln L}{\tau g_N(\omega^\prime, \tau)}
\end{equation}
Let $\sigma_S$ be the error in $S(\omega)$ due to the measurement error $\sigma$. The relative error in $S(\omega)$ is thus:
\begin{equation} 
\frac{\sigma_S}{S(\omega)} = -\frac{|\ln \tilde{L} - \ln L|}{\ln L} = \frac{|\ln(\tilde{L} / L)|}{\ln L}
\end{equation}
Using the functional form $P(\tau) = A_{\textrm{eff}} e^{-(\tau / T_2)^n}$ as before and \eqref{eq:LDef}, we have that:
\begin{equation}
\ln\left(\frac{\tilde{L}}{L}\right) = \ln\left(1 + \frac{\sigma}{P(\tau) - C}\right) = \ln\left(1 + \frac{\sigma}{A_{\textrm{eff}}} e^{-(\tau / T_2)^n}\right)
\end{equation}
\begin{equation}
\ln L = -\left(\frac{\tau}{T_2}\right)^n + \frac{\tau}{2T_1}
\end{equation}
Approximating to first-order $\ln(1 + x) \approx x$ and excluding data points where the relative error in $S(\omega)$ is greater than some threshold $\varepsilon$, we have that:
\begin{equation}
\frac{\sigma_S}{S(\omega)} \approx \frac{\frac{\sigma}{A_{\textrm{eff}}} e^{-(\tau / T_2)^n}}{\left(\frac{\tau}{T_2}\right)^n - \frac{\tau}{2T_1}} > \varepsilon
\end{equation}
To obtain a closed-form solution, we neglect the $T_1$ contribution $\tau / (2T_1)$ and approximate $e^{-(\tau / T_2)^n} \approx 1$ for small $\tau$. Making these approximations, we have that the values of $\tau$ to exclude due to numerical instability are:
\begin{equation}
\tau < \left(\frac{\sigma}{A_{\textrm{eff}} \varepsilon}\right)^{1/n} T_2
\end{equation}
We substitute some estimated typical values $\sigma \approx 0.05$ (which is estimated from a rough estimate from readout fidelity fluctuations), $n=1$ (pure exponential decay), $A_{\textrm{eff}} \approx 0.5$, $\varepsilon = 0.5$, which suggests that $\tau < 0.2 T_2$ should be excluded. In this work, we exclude $\tau < 0.24 T_2$ for improved data quality, though we note that due to the many approximations such values should only be treated as rough estimates.

\subsection{Additional $S(\omega)$ Examples}
\label{SwExamples}

\begin{figure*}
   \centering
  \includegraphics{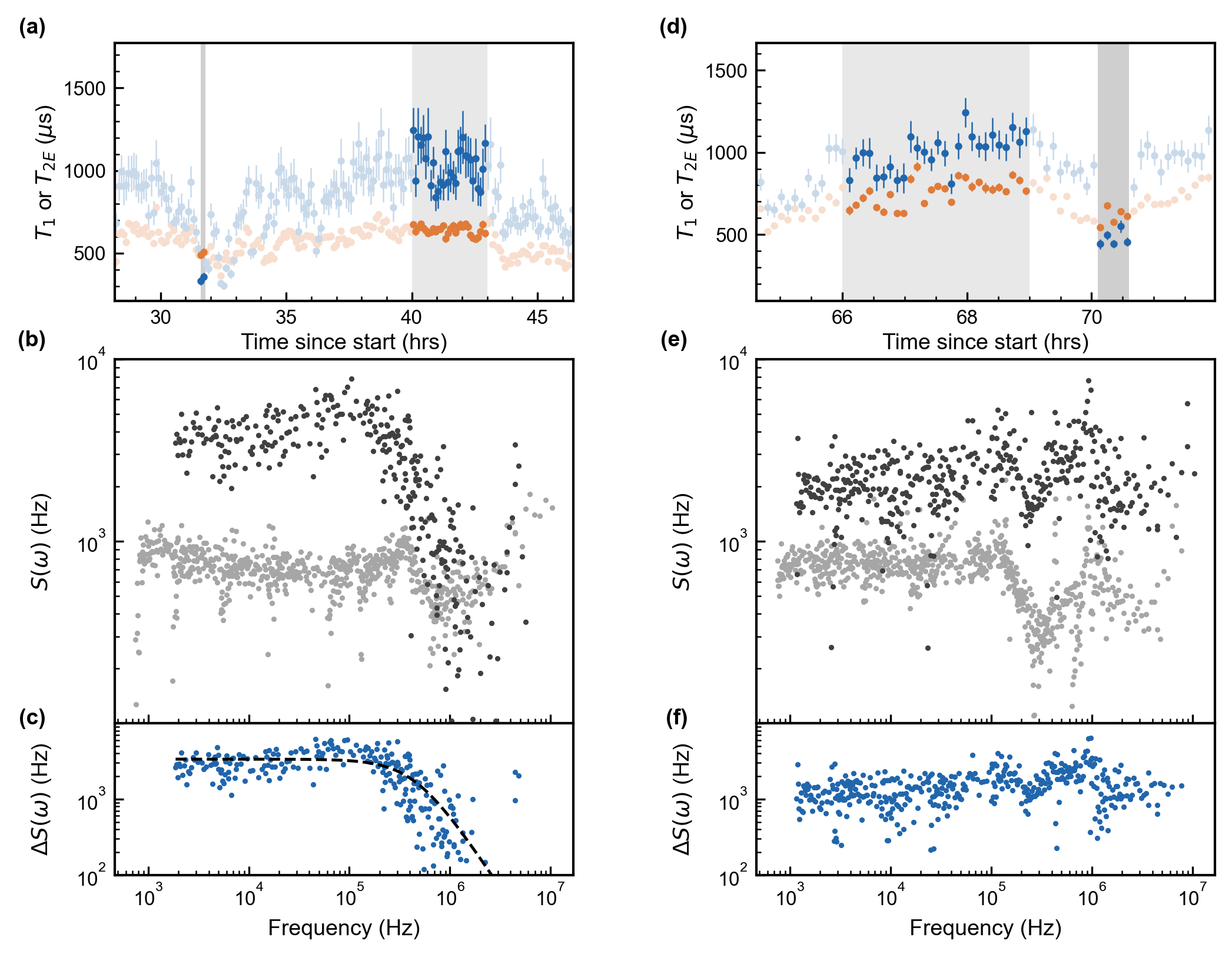}
  \caption{\textbf{Additional $S(\omega)$ Comparisons}: (a)--(c) from dataset 6, (d)--(f) from dataset 3. (a, d) Time trace of $T_1$ (orange) and $T_{2\text{E}}$ (blue) with data selected inside (dark green) and outside (light green) an excursion. (b, e) $S(\omega)$ extracted from CPMG decay traces, averaged over the time points as highlighted in (a, d). (c) $S(\omega)$ difference inside and outside the excursion in (a), which shows a Lorentzian profile with a fitted roll-off frequency at $454 \pm 36 \textrm{ kHz}$. (f) $S(\omega)$ difference inside and outside the excursion in (d), showing white noise with no apparent roll-off. 
  }
   \label{fig_si:sw_examples}
\end{figure*}

Additional examples of $S(\omega)$ measurements inside and outside an excursion are shown in Fig. \ref{fig_si:sw_examples}. We observe excess noise with a roll-off frequency of $454 \pm 36 \textrm{ kHz}$ in Fig. \ref{fig_si:sw_examples}(a) -- (c). This differs from the qubit's readout resonator linewidth $\kappa = 326 \textrm{ kHz}$ as well as the roll-off frequency in Fig. \ref{fig:cpmg}. We also observe excess noise with no apparent roll-off frequency in Fig. \ref{fig_si:sw_examples}(d) -- (f). The roll-off frequency is likely outside the bandwidth of measured frequencies.

\section{Non-Gaussian and Non-Stationary Noise Features}
\label{NonStandardSw}

\begin{figure}
   \centering
  \includegraphics{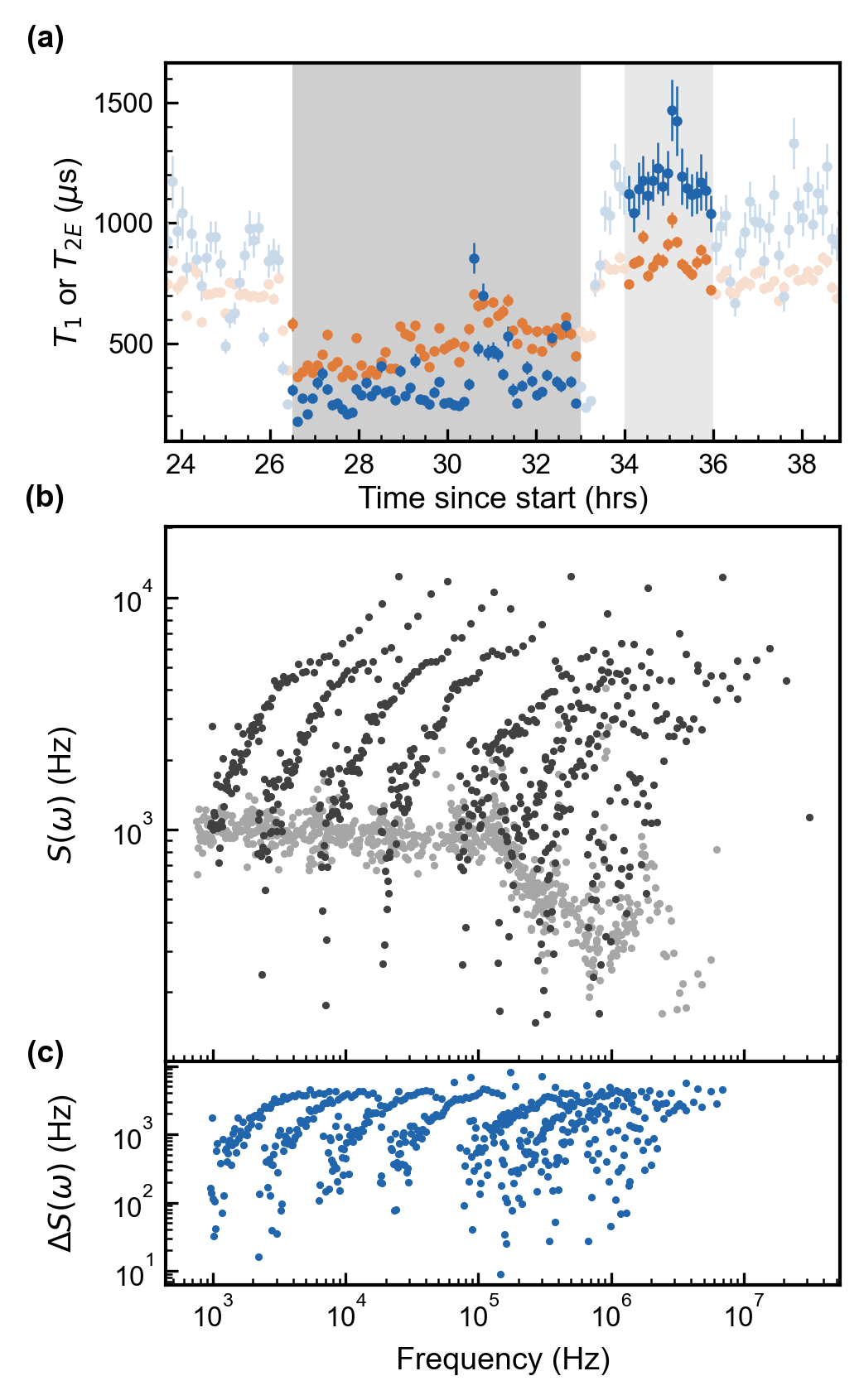}
  \caption{\textbf{Upward-Sloping and Discontinuous $S(\omega)$}: 
  Data extracted from dataset 3. Averaged $S(\omega)$ extracted from CPMG decay curves inside (dark green) and outside (light green) the excursion.}
   \label{fig_si:sw_stripes}
\end{figure}

In some cases we observe upward-sloping and discontinuous $S(\omega)$ inside an excursion (Fig. \ref{fig_si:sw_stripes}). While upward-sloping $S(\omega)$ is anomalous, the discontinuities are unphysical. The derivation of the coherence integral used for the $S(\omega)$ calculations assumes that the noise source is Gaussian and stationary. However, since excursions are caused by single TLSs coupling resonantly with the qubit, this noise source can be non-Gaussian \cite{sung2019nongaussian, cywinski2008enhance} or non-stationary. 

\subsection{Stretched-Exponent Analysis of CPMG Decay Curves}

We fit each of the CPMG decay curves to a stretched exponential function $P(t) = A e^{-\Gamma_p N} e^{-(t/ T_2)^n} + C$, where $A$, $C$, $T_2$, and $n$ are fit parameters. Converting the CPMG decay data to a noise spectral density $S(\omega)$ following Appendix \ref{SwCalc}, the stretched exponent $n$ determines the shape of the noise spectral density $S(\omega)$ using equation \eqref{eq:SwConv}:
\begin{align*} 
S\left(\omega = \frac{N}{2 \tau}\right) &= -\frac{1}{\tau g_N(\omega, \tau)} \ln\left(e^{-(\tau/T_2)^n} e^{\tau / 2T_1} \right) \\
&= \frac{(\tau / T_2)^n - \tau / 2T_1}{\tau g_N(\omega, \tau)} \\
&= \frac{\tau^{n-1}}{T_2^n g_N(\omega, \tau)} - \frac{1}{2T_1 g_N(\omega, \tau)}
\end{align*}
We find that $g_N(\omega, \frac{N}{2 \omega})$ can be well-approximated as a constant independent of $\omega$. Writing this as a function of $\omega = \frac{N}{2 \tau}$:
\[S(\omega) \propto  \left(\frac{N}{2T_2}\right)^{n-1} \frac{1}{T_2} \frac{1}{\omega^{n-1}} - \frac{1}{2T_1}\]
In general, a CPMG decay curve with stretched exponent $n$ yields an $S(\omega)$ with functional form $1/\omega^{n-1}$. Thus a stretched exponent of $n = 1$ yields a white noise spectrum while a stretched exponent of $n = 2$ yields a $1/f$ noise spectrum. 

\begin{figure}
   \centering
  \includegraphics{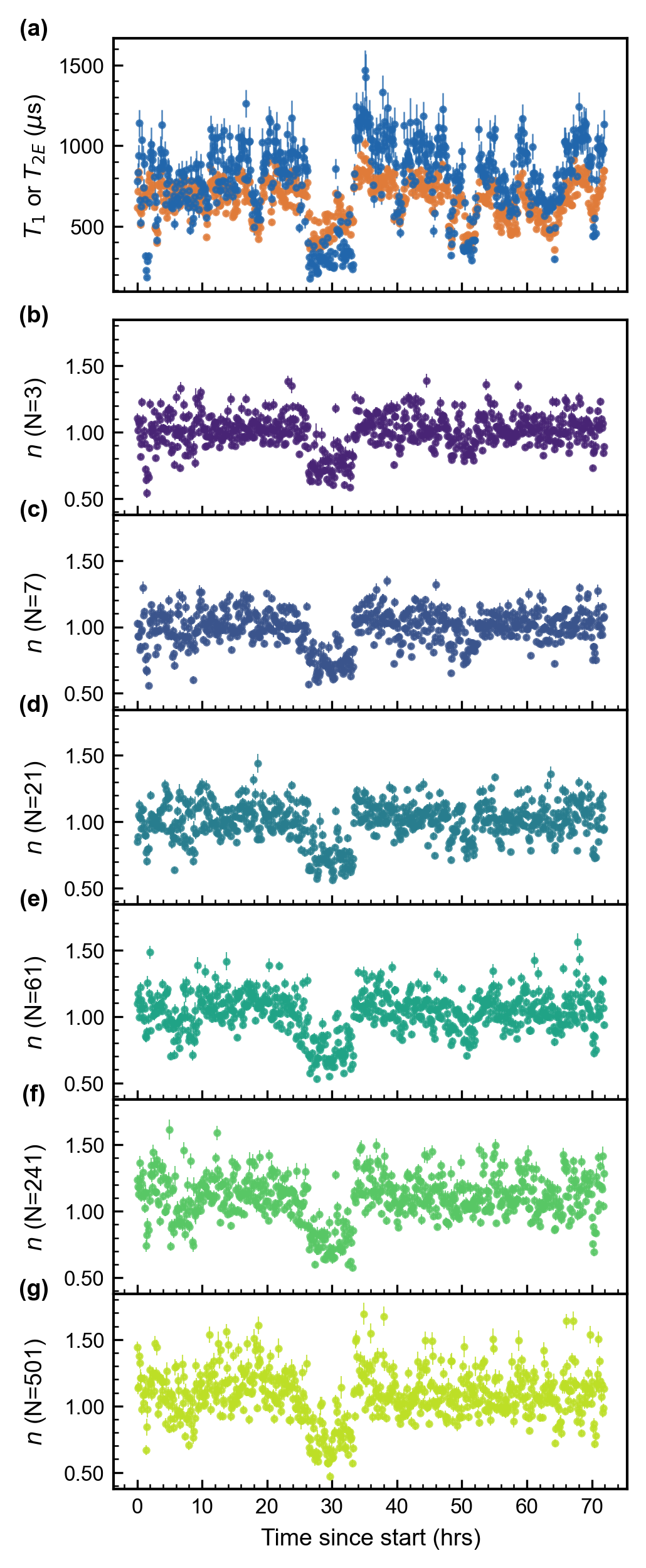}
  \caption{\textbf{Stretched exponent over time}: (a) $T_1$ (orange) and $T_{2\text{E}}$ (blue) measured over time. (b) -- (g) Fitted stretched exponent $n$ over time for CPMG decay curves with various numbers of $\pi$-pulses: (b) $N = 3$, (c) $N = 7$, (d) $N = 21$, (e) $N = 61$, (f) $N = 241$, (g) $N = 501$.} 
   \label{fig_si:n_vs_time}
\end{figure}

The upward-sloping $S(\omega)$ suggests decay curves with stretched exponent $n < 1$. Examining the decay curves inside and outside the excursion in Fig. \ref{fig_si:sw_stripes}, we can track the temporal fluctuations of the stretched exponent $n$ (Fig. \ref{fig_si:n_vs_time}). Excursions coincide with $n < 1$, which is surprising, while outside an excursion, the decay curves follow a pure exponential $n \approx 1$. In order to investigate potential causes of a stretched exponent $n < 1$, we perform numerics to simulate qubit dephasing where the qubit noise environment is dominated by coupling to a single TLS.

\subsection{Numerical Simulations of TLS-Induced Qubit Dephasing}

\begin{figure*}
   \centering
  \includegraphics{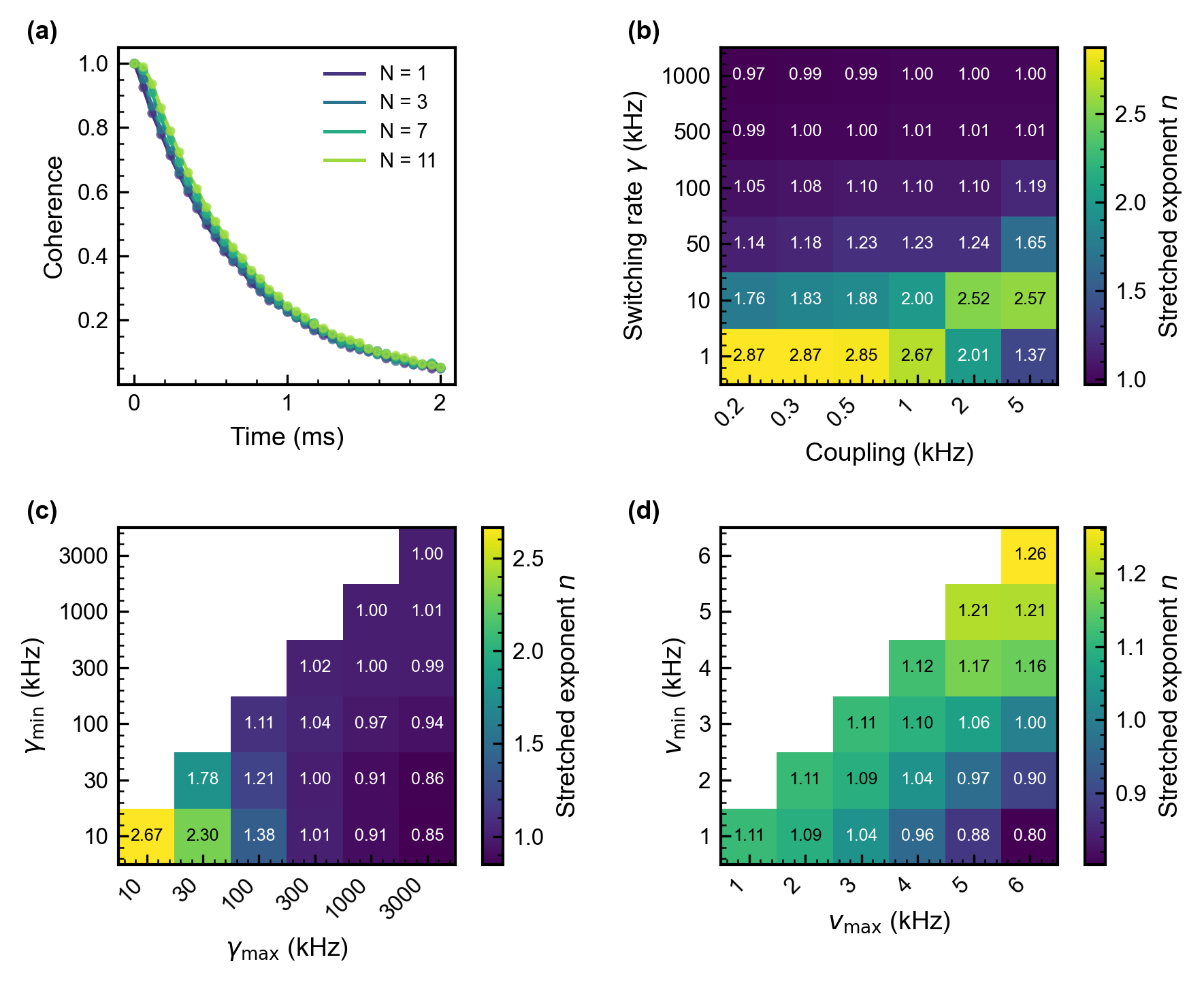}
  \caption{\textbf{Numerical Simulations of Random Telegraph Noise Dephasing}: (a) Synthetic CPMG decay curves with various numbers of $\pi$-pulses $N$ assuming a random telegraph noise model with coupling frequency 3 kHz and switching rate 100 kHz. (b) Fitted stretched exponent $n$ of CPMG decay curves with $N = 3$ $\pi$-pulses for various switching rates and coupling strengths. (c) Fitted stretched exponent for a fixed coupling $A = 3 \textrm{ kHz}$ and non-stationary switching rate varying shot-to-shot between $\gamma_{\textrm{min}}$ and $\gamma_{\textrm{max}}$. (d) Fitted stretched exponent for a fixed switching rate $\gamma = 100 \textrm{ kHz}$ and non-stationary coupling strength varying shot-to-shot between $\nu_{\textrm{min}}$ and $\nu_{\textrm{max}}$.}
   \label{fig_si:rtn_models}
\end{figure*}

We model qubit dephasing due to a single TLS as a time-dependent perturbation to the qubit frequency $\delta \omega(t) = \pm \nu$, where $\delta \omega(t)$ exhibits random telegraph switching between $\pm \nu$ with a switching rate $\gamma$. The magnitude of $\nu$ represents the qubit-TLS coupling strength. Using this model of $\delta\omega(t)$, we perform Monte Carlo simulations and generate synthetic CPMG decay curves $\langle e^{i \varphi(t)} \rangle$ where $\varphi(t)$ is defined as in \eqref{eq:accum_phase} (Fig. \ref{fig_si:rtn_models}(a)). Performing these simulations sweeping the qubit-TLS coupling strength $|\nu|$ and switching rate $\gamma$ only yields decay curves with $n > 1$ (Fig. \ref{fig_si:rtn_models}(b)), while we observe in experiment decay curves with $n < 1$ during an excursion (Fig. \ref{fig_si:n_vs_time}(b)). However, if we allow $\nu$ or $\gamma$ to vary over the course of the experiment, we find that decay curves of $n < 1$ are possible (Fig. \ref{fig_si:rtn_models}(c), (d)). Stretched exponents $n < 1$ have also been observed in NV centers due to configurational averaging of surface spins \cite{dwyer2022probing}.

\section{AC Stark Shift Calibration}
\label{ACStarkCal}

\begin{figure}
   \centering
  \includegraphics{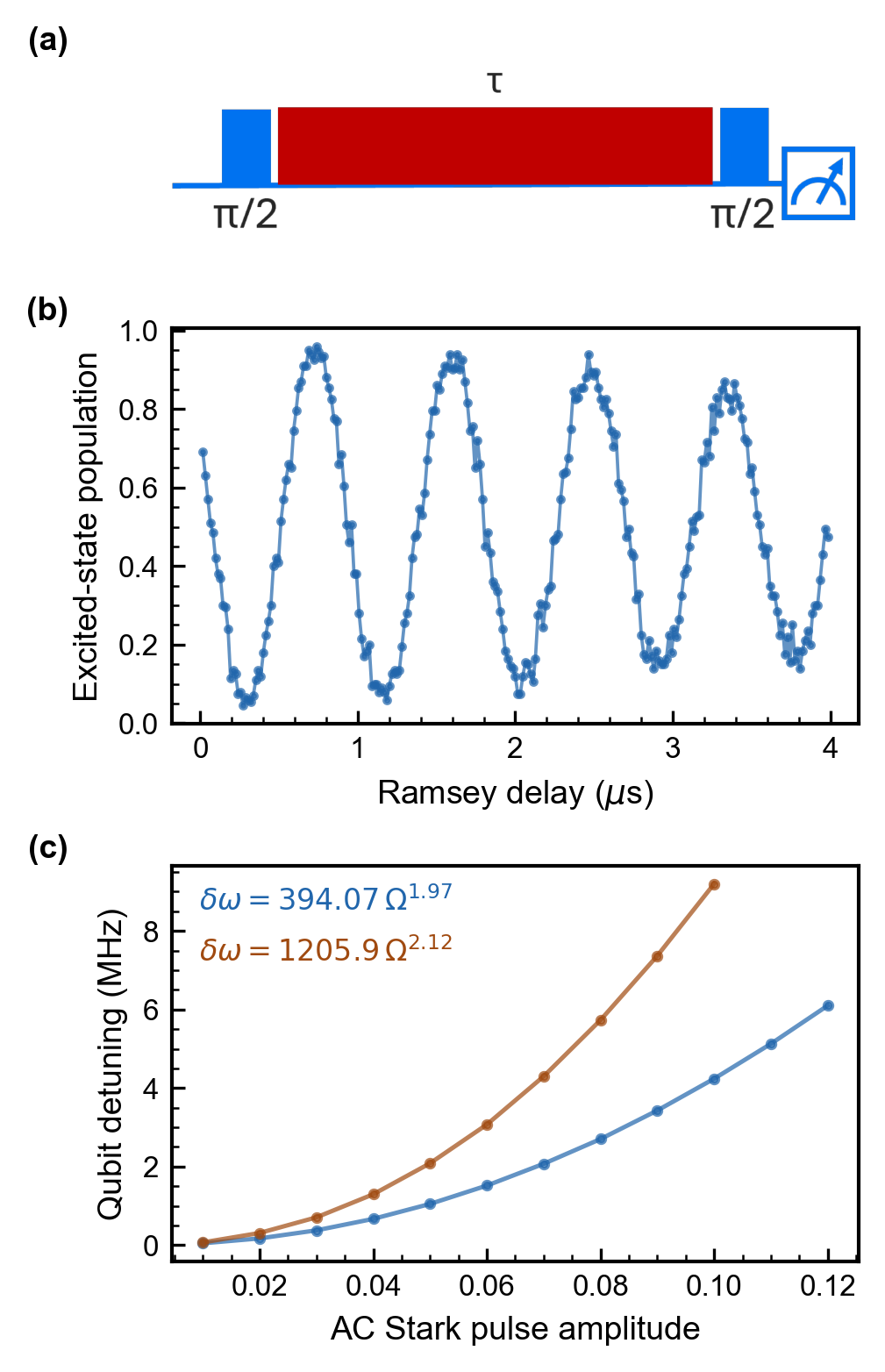}
  \caption{\textbf{AC Stark shift calibration}: (a) Ramsey pulse sequence with AC Stark shift pulse during the free evolution time. (b) Sample Ramsey trace showing oscillatory features used to measure the qubit frequency detuning. (c) Qubit detuning $\delta \omega$ (MHz) vs. AC Stark shift pulse amplitude $\Omega$ for an AC Stark shift pulse with positive detuning $\Delta = 50 \textrm{ MHz}$ (blue) and negative detuning $\Delta = -50 \textrm{ MHz}$ (brown). Measured values (data points) with a power-law fit (continuous curve) show a nearly quadratic dependence between qubit detuning and AC Stark pulse amplitude.
  }
   \label{fig_si:stark_calibration}
\end{figure}

We use a method similar to \cite{carroll_dynamics_2022} to calibrate the correspondence between Stark shift pulse amplitude and qubit frequency detuning. All calibration is performed with a fixed Stark shift detuning frequency $\Delta$. We measure the qubit frequency at various Stark shift pulse amplitudes using a Ramsey pulse sequence with an AC Stark shift pulse during the free evolution time (Fig. \ref{fig_si:stark_calibration}(a)). We sweep the delay time only for very short $\tau$ much less than the Ramsey coherence time $T_{2R}$, so the Ramsey trace is almost purely oscillatory and has negligible exponential decay. We measure the qubit frequency detuning by fitting the Ramsey trace to a sinusoidal function (Fig. \ref{fig_si:stark_calibration}(b)). 

The measured qubit frequency detuning $\delta \omega$ is plotted versus Stark shift pulse amplitude $\Omega$ (Fig. \ref{fig_si:stark_calibration}(c)). A power-law fit shows that the qubit frequency detuning has a nearly quadratic dependence on the Stark shift pulse amplitude, which agrees with the theoretical prediction $\delta \omega \propto \Omega^2$. We perform separate calibrations for positive and negative Stark shift pulse detuning to allow for TLS spectroscopy both above and below the qubit frequency.

\section{TLS Spectroscopy Fitting}
\label{LorentzianFit}

\subsection{Data Collection and Processing}
We perform TLS spectroscopy by applying an AC Stark shift to the qubit frequency. The pulse sequence is a standard $T_1$ pulse sequence with an additional detuned AC Stark pulse during the delay time; the calibration of this procedure is described in Appendix \ref{ACStarkCal}. The delay time is held constant (100 $\mu$s for dataset 7 in the main text), and we average over a fixed number of shots to estimate the excited state population $P_e(f, t)$ over time $t$ for each Stark-shifted qubit frequency $f$ (the populations are averaged over 500 shots for the dataset 7 in the main text, corresponding to a temporal resolution of 3 to 4 minutes per measurement). Note that we use $t$ to denote wall-clock time from repeated measurements rather than the delay time in the $T_1$ pulse sequence.

We then convert from $P_e(f, t)$ to $\Gamma_1(f, t)$ with $\Gamma_1(f, t) = -\frac{1}{\tau}\ln(P_e(f, t) / P_{\text{init}}(t))$ where $P_{\text{init}}(t)$ is the initial excited state population (as measured by the interleaved full $T_1$ sweeps). 

\subsection{Error Bars}
Next, we calculate the error bars of $\Gamma_1(f,t)$ at each  qubit frequency and time. We include two sources of error: quantum projection noise (QPN) and fit error.

First, we can calculate the error due to QPN:
\begin{equation} \sigma_{P_e} = \sqrt{\frac{P_e(1-P_e)}{n}}\end{equation}
where $n$ is the total number of shots averaged to estimate the population. We can then propagate this to $\Gamma_1(f, t) = -\frac{1}{\tau}\ln(P_e(f, t) / P_{\text{init}}(t))$: 
\begin{equation} \sigma_{\Gamma_1(f,t), \text{QPN}} = \frac{\sigma_{P_e}}{\tau |P_e|}
\end{equation}

We can also calculate uncertainty due to fit error. Because $\Gamma_1(f,t)$ is calculated by measuring the population at a single fixed delay time $\tau$, the fit error cannot be calculated directly from fitting to a single data point. Instead, we estimate the fit error using the interleaved full $T_1$ sweeps. Specifically, the full $T_1$ sweeps contain data given by $P_e(\tau) = P_0 + Ae^{-\Gamma_1 \tau}$ for many delay times $\tau$. Across all repeated full $T_1$ measurements, we estimate $\sigma^2_{P_0}$, $\sigma^2_{A}$, and $\text{Cov}(P_0, A)$. Using $\sigma^2_{P_0}$, $\sigma^2_{A}$, and $\text{Cov}(P_0, A)$, we can then estimate $\sigma_{\Gamma_1(f,t), \text{\text{fit error}}}$ as:
\begin{equation} \begin{aligned}
\sigma_{\Gamma_1(f,t), \text{\text{fit error}}}^2 &= \left|\frac{\partial \Gamma_1}{\partial P_0}\right|^2\sigma^2_{P_0} + \left|\frac{\partial \Gamma_1}{\partial A}\right|^2\sigma^2_{A} \\
&+ 2 \frac{\partial \Gamma_1}{\partial P_0}\frac{\partial \Gamma_1}{\partial A}\text{Cov}(P_0, A) 
\end{aligned} \end{equation}

\begin{figure}[H]
   \centering
  \includegraphics[width=\columnwidth]{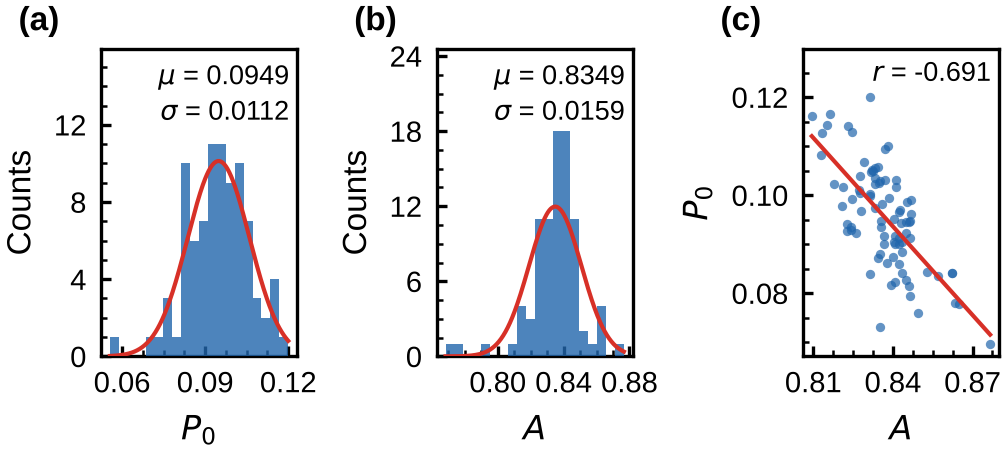}
  \caption{\textbf{Fit Error from full $T_1$ Sweeps}: Analysis of full sweeps of $T_1$ which are interleaved into the AC Stark measurement; together, these are used to calculate fit error. Histograms of $P_0$ (a) and $A$ (b) in $T_1$ decay curve as modeled by $P_e(\tau) = P_0 + Ae^{-\Gamma_1 \tau}$; resulting histograms are fitted to normal distributions. Scatterplot of fitted $P_0$ and $A$ values (c); resulting slope is used to estimate $\text{Cov}(P_0,A)$.}
   \label{fig:gscatter}
\end{figure}

We then add $\sigma_{\Gamma_1(f,t), \text{QPN}}$ and $\sigma_{\Gamma_1(f,t), \text{\text{fit error}}}$ in quadrature: \begin{equation} 
\sigma_{\Gamma(f,t), \text{total}} = \sqrt{\sigma^2_{\Gamma_1(f,t), \text{QPN}} + \sigma^2_{\Gamma_1(f,t), \text{\text{fit error}}}}
\end{equation}

\subsection{Time-Independent Background}

We first find the time-independent background $\Gamma_{1, \textrm{ Background}}(f)$ in equation \eqref{eq:TLSbudget}, which we decompose into a sum $B_0 + B(f)$, where $B_0$ is frequency-independent and $B(f)$ is frequency-dependent. We define the time- and frequency-independent background $B_0$ to be the 1st percentile of all measured $\Gamma_1$ values from AC Stark shift measurements across time and frequency in a given dataset.

\begin{figure}[H]
   \centering
  \includegraphics[width=\columnwidth]{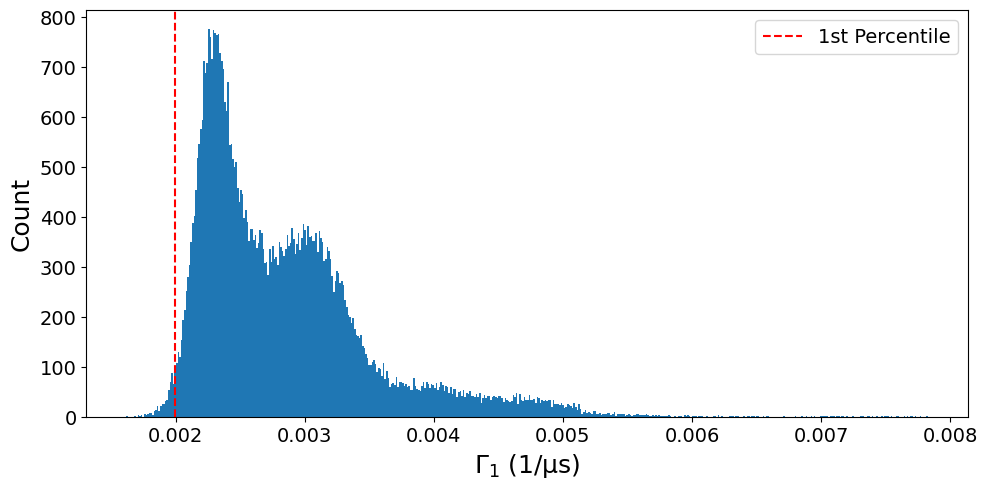}
  \caption{\textbf{Time and Frequency Independent Background}: Histogram of all $\Gamma_1$ values across time and frequency measured using AC Stark shift at a fixed $\tau = 100 \; \mu \text{s}$. Red vertical dashed line shows the $\Gamma_1$ at the 1st percentile, which defines the value of the time and frequency independent background $B_0$.}
   \label{fig:gscatter}
\end{figure}

Then, for each dataset, we calculate a time-independent, frequency-dependent background $B(f)$. 
To do this, we divide the data into four equally-sized epochs. In each epoch, we take the 20th percentile estimate of $\Gamma_1$. For each frequency, we then select the minimum of the 20th percentile estimates across the four epochs, which gives an unsmoothed frequency-dependent background. We then apply a Gaussian smoothing with kernel size $\sigma_f = 0.06 + (0.45 - 0.06)\text{min}(\frac{|f|}{1.5}, 1)$. Finally, we choose a constant offset and normalize such that $B(f)$ integrates to zero in a region that is flat and close to the minimum; in the main text dataset, this was $f < -15 \text{ MHz}$. We subtract $B_0 + B(f)$ from $\Gamma_1(f, t)$ and fit the remaining model to this difference.

\begin{figure}[H]
   \centering
  \includegraphics[width=\columnwidth]{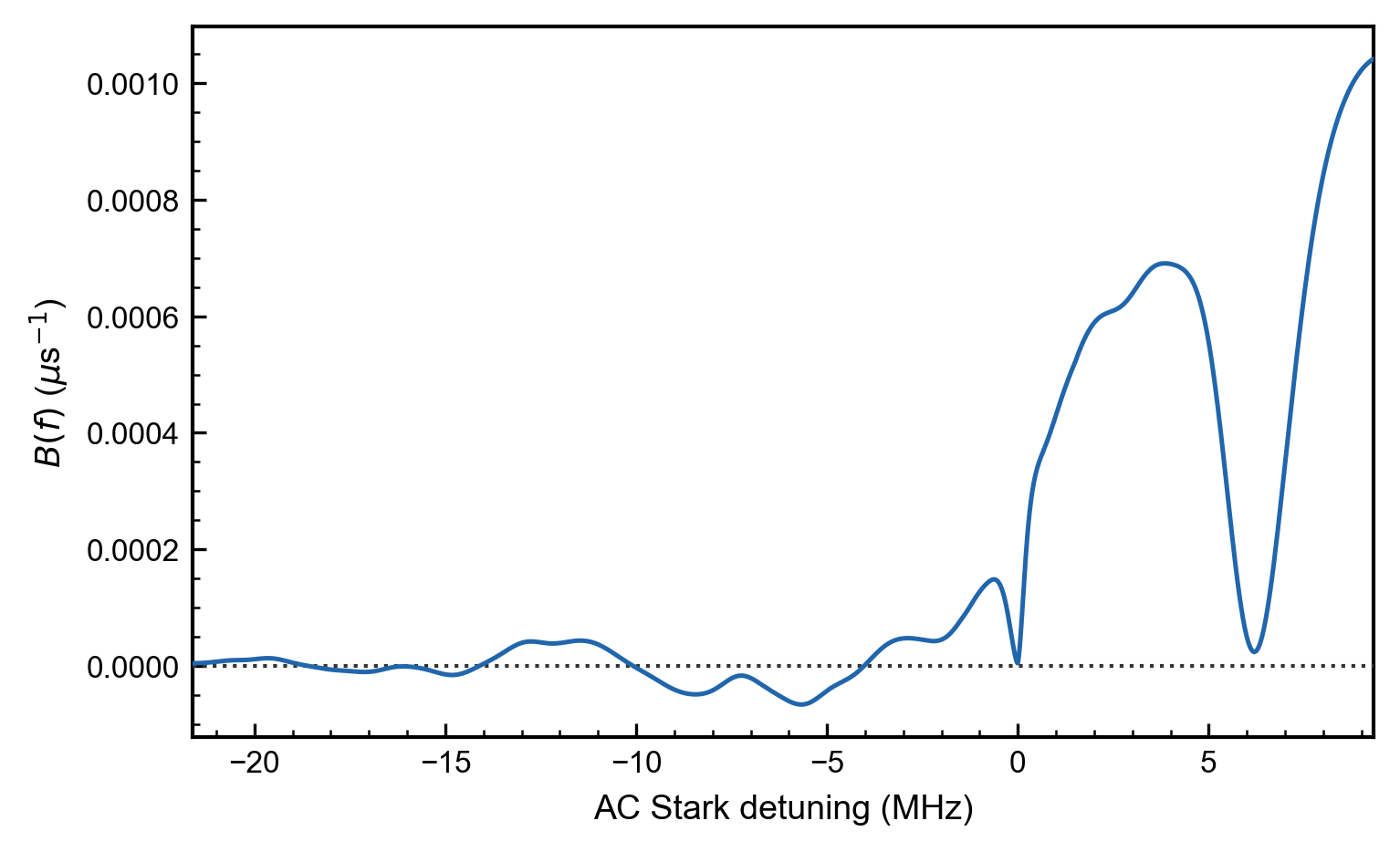}
  \caption{\textbf{Frequency-dependent background}: Frequency dependent background $B(f)$ as calculated from the 2D color plot $\Gamma_1(f, t)$. This is subtracted from the raw data, thus removing any features which do not spectrally diffuse over the course of the measurement.
  }
   \label{B(f)}
\end{figure}

\begin{figure*}[t]
    \centering
    \includegraphics[width=\textwidth]{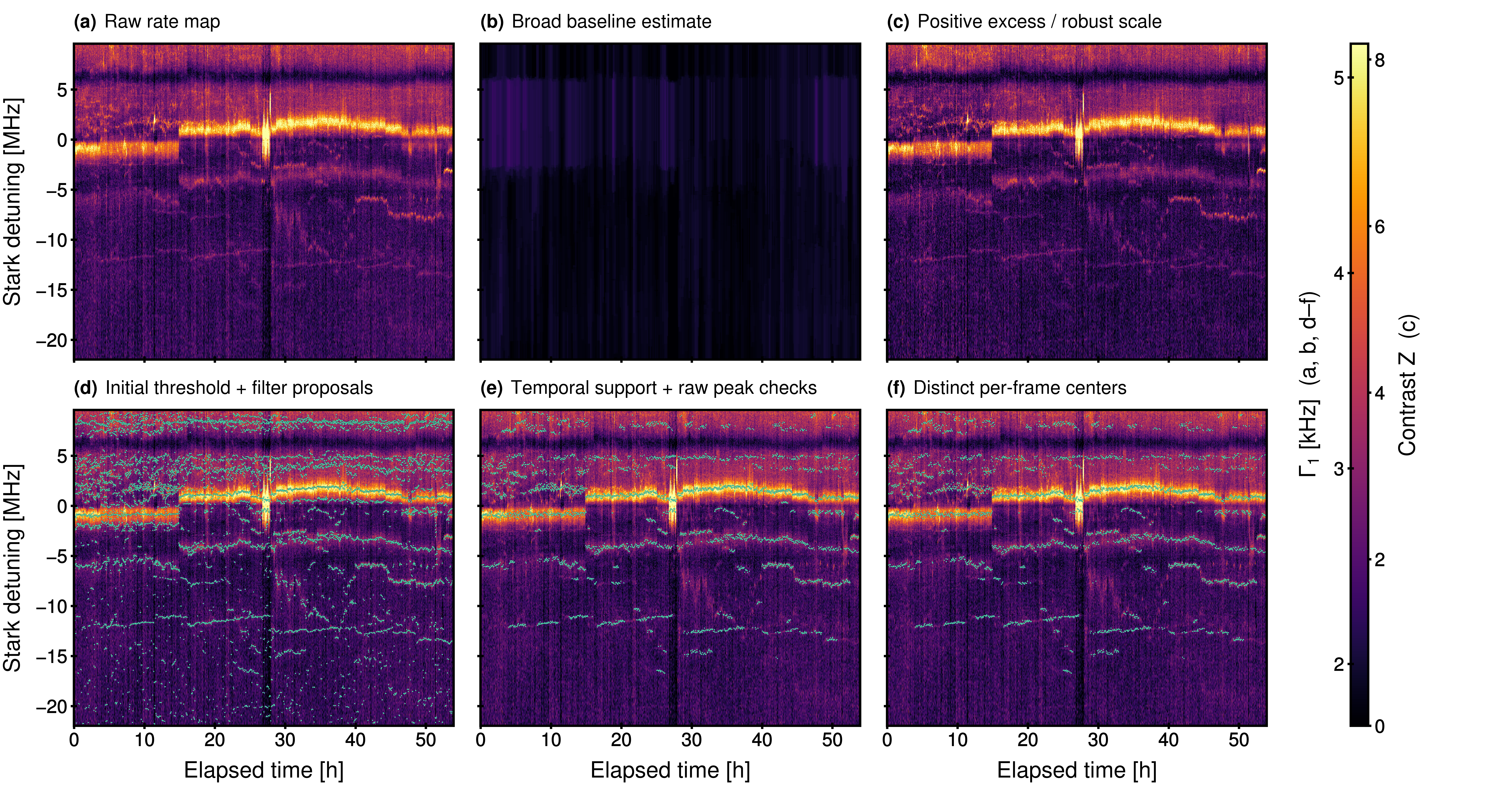}
    \caption{
    \textbf{Resonance identification algorithm.}
    (a) Measured relaxation-rate map $\Gamma_1(f,t)$ used as the input to the identification algorithm.
    (b) Slowly varying background obtained using a morphological opening along frequency followed by a median over neighboring spectra.
    (c) Normalized positive excess $Z(f,t)$ used for the contrast-based search.
    (d) Initial resonance proposals generated by the contrast-based search and the complementary Lorentzian-template search. The template search operates in parallel on a temporally filtered version of the measured relaxation-rate map.
    (e) Candidate centers remaining after localization to a resolved maximum in the spectrum at the current time point and tests of support in neighboring spectra.
    (f) Final candidate resonance centers after consolidation of duplicate nearby proposals, overlaid on the measured relaxation-rate map.
    }
    \label{fig:figSI_identification_stages}
\end{figure*}

\subsection{Ridge-Based Peak Fitting}

Having corrected the data and calculated the uncertainty on each estimate $\Gamma_1(f,t)$, we next identify candidate resonance centers in the resulting relaxation-rate map. This is not equivalent to simply finding local maxima of $\Gamma_1(f)$ at each time $t$. The spectroscopy maps contain features with a wide range of amplitudes and linewidths, superimposed on a background that varies with frequency and slowly changes over time. A weak but well-resolved peak can therefore be difficult to distinguish from fluctuations in a single spectrum, while small variations on top of a broad feature can appear as additional local maxima.

To address these issues, we identify resonances using both their spectral structure within a given spectrum and supporting information from neighboring spectra. The procedure is illustrated in Fig.~\ref{fig:figSI_identification_stages}. We first remove the slowly varying background and search for regions of enhanced relaxation [Fig.~\ref{fig:figSI_identification_stages}(a)--(c)]. We then perform a complementary search for weaker features with a localized resonance-like lineshape [Fig.~\ref{fig:figSI_identification_stages}(d)]. Proposals from either search must correspond to a resolved maximum in the measured spectrum and have supporting structure in neighboring spectra [Fig.~\ref{fig:figSI_identification_stages}(e)]. Finally, overlapping proposals are consolidated to produce the candidate resonance centers shown in Fig.~\ref{fig:figSI_identification_stages}(f).

We denote the measured relaxation-rate map by
$M_{ij}=\Gamma_1(f_i,t_j)$. Since the resonances appear as relatively narrow
enhancements on top of a frequency-dependent background, we estimate this
background using a morphological opening along frequency followed by a median
over nine neighboring spectra. A morphological opening consists of taking a
local minimum followed by a local maximum over the same frequency window,
suppressing narrow positive features while retaining broader variations in the
spectrum. Using a 160-bin window, we define
$b_{ij}=\mathrm{med}_{|v|\leq4}[\mathcal{O}_{160}M]_{i,j+v}$ and the excess
relaxation $S_{ij}=\max(M_{ij}-b_{ij},0)$. We then normalize $S$ by its robust
median-absolute-deviation scale in each spectrum to obtain a contrast map
$Z_{ij}=S_{ij}/\widetilde{s}_j$, shown in
Fig.~\ref{fig:figSI_identification_stages}(c). This allows the search to depend
on relaxation relative to the local background rather than on the absolute
value of $\Gamma_1$.

We search $Z$ from strong to weak contrast using a two-threshold procedure:
pixels above a high threshold seed a candidate region, while connected pixels
above a lower threshold are allowed to join it. The three seed/growth pairs are
$(7.0,4.5)$, $(5.0,3.0)$, and $(4.0,2.5)$. Once a strong feature is identified,
its surrounding region is excluded from subsequent searches, allowing weaker
nearby structure to be resolved without repeatedly identifying the same
resonance. Within each region, distinct local maxima are proposed separately
when they have sufficient prominence and separation; otherwise the region is
represented by its excess-weighted center. Broad features are additionally
compared across neighboring spectra before being split, reducing the tendency
to identify small fluctuations on a broad peak as separate resonances.

The contrast-based search can miss weak peaks that nevertheless have a clear
spectral shape. We therefore perform a second search by median-filtering the
measured map over five neighboring spectra and convolving along frequency with
zero-mean Lorentzian-shaped templates of several widths. The maximum response
over the template widths defines a complementary detection map, whose local
maxima provide additional resonance proposals
[Fig.~\ref{fig:figSI_identification_stages}(d)]. The templates are used only to
identify localized spectral structure; their widths and amplitudes are not
interpreted as fitted TLS parameters.

Finally, all proposals are checked against the measured spectrum at the
corresponding time point. A candidate is retained only if it lies near a local
maximum with sufficient prominence and two-sided spectral contrast, which
suppresses detections on the shoulders of stronger features. We additionally
require supporting structure in neighboring spectra to reject isolated
fluctuations. For contrast-based candidates this support is determined from
repeated local contrast and brightness, while template-based candidates are
required to show a recurring filter response and similar local spectral
structure. Nearby proposals are then consolidated unless the measured spectrum
contains distinct maxima separated by a clear minimum. The resulting points in
Fig.~\ref{fig:figSI_identification_stages}(f) are therefore candidate resonance
centers at each time point.

This algorithm returns a set of peak centers for each timestep $t$. We then fit each frame of the data to a vertical offset and a sum of Lorentzians. The Lorentzians are constrained such that a peak must be fit within 0.5 MHz of each peak identified by the tracking algorithm. 
\[\Gamma_{1, \text{Broadband}} + \sum_i \frac{2g_i^2 \Gamma_{d, i}}{\Gamma_{d, i} + 4\pi^2(f-f_d)^2}\] 
These fits also have the following bounds: $\Gamma_{1, \text{Broadband}} \in [0, \infty)$, peak width $\in [\text{frequency distance between points, 5}] \text{ MHz}$, and peak height $\in [0, \infty) \text{ Hz}$. 

\subsection{BIC-Based Peak Fitting}
The tracking algorithm identifies the clearest features and is designed to suppress type I error (i.e. identifying peaks that do not exist). Therefore, there are additional features that we can identify. To do this, we consider the residuals after the ridge-based fitting. For each timestep, we choose eight candidate peaks (from the local maxima of the residuals); we chose eight candidates because we found that the BIC was minimized for fewer than eight additional peaks. Then, there are $2^8 = 256$ possible combinations of whether each peaks is included in the fit or not. We try all $256$ possibilities and choose the best one in terms of $\text{BIC} = -2 \log L + 3k\log n$ where $L$ is the likelihood, $k$ is the number of additional peaks, and $n$ is the number of data points in each frame.

\begin{figure}[H]
   \centering
  \includegraphics[width=\columnwidth]{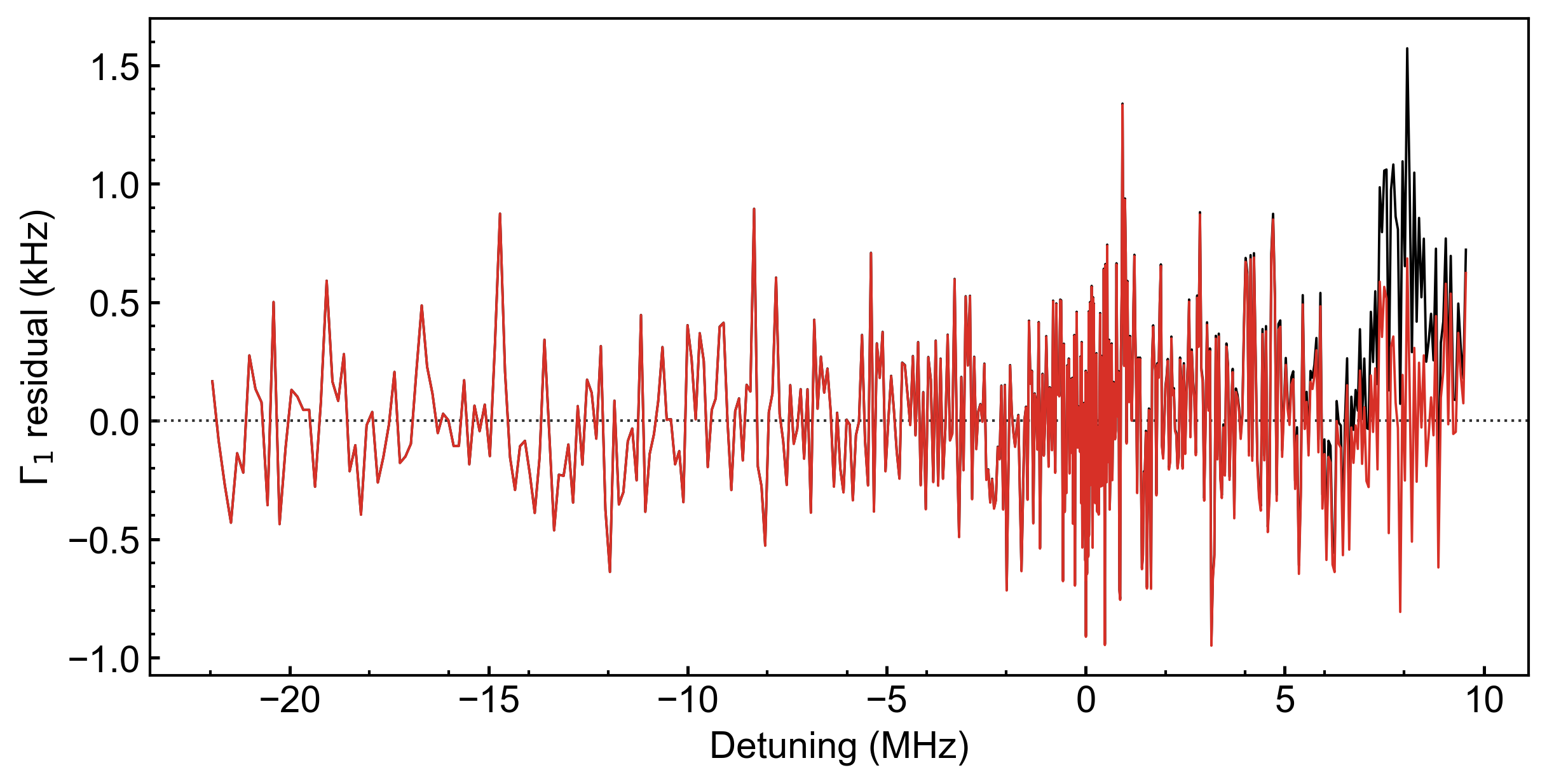}
  \caption{\textbf{Residuals}: Residuals after ridge-based peak fitting (black) and both ridge- and BIC-based peak fitting (red) for $t=29.04$. BIC-based fitting is done on residuals after ridge-based peak fitting (black) and is designed to identify peaks that were initially missed; peak around 7 MHz in black is removed after BIC-based fitting. 
  }
   \label{B(f)}
\end{figure}

Additionally, these fits have the following bounds: peak center $\in [\text{minimum detuning} - 0.01 \textrm{ MHz}, 12 \textrm{ MHz}]$, peak width $\in [0.15 \textrm{ MHz}, 5 \textrm{ MHz}]$, and peak height $\in [0, \infty]$. Additionally, fits with BIC peaks that are $< 0.1 \text{ MHz}$ from a ridge-based detected peak or another BIC peak are disqualified. 

\begin{figure}
   \centering
  \includegraphics[width=\columnwidth]{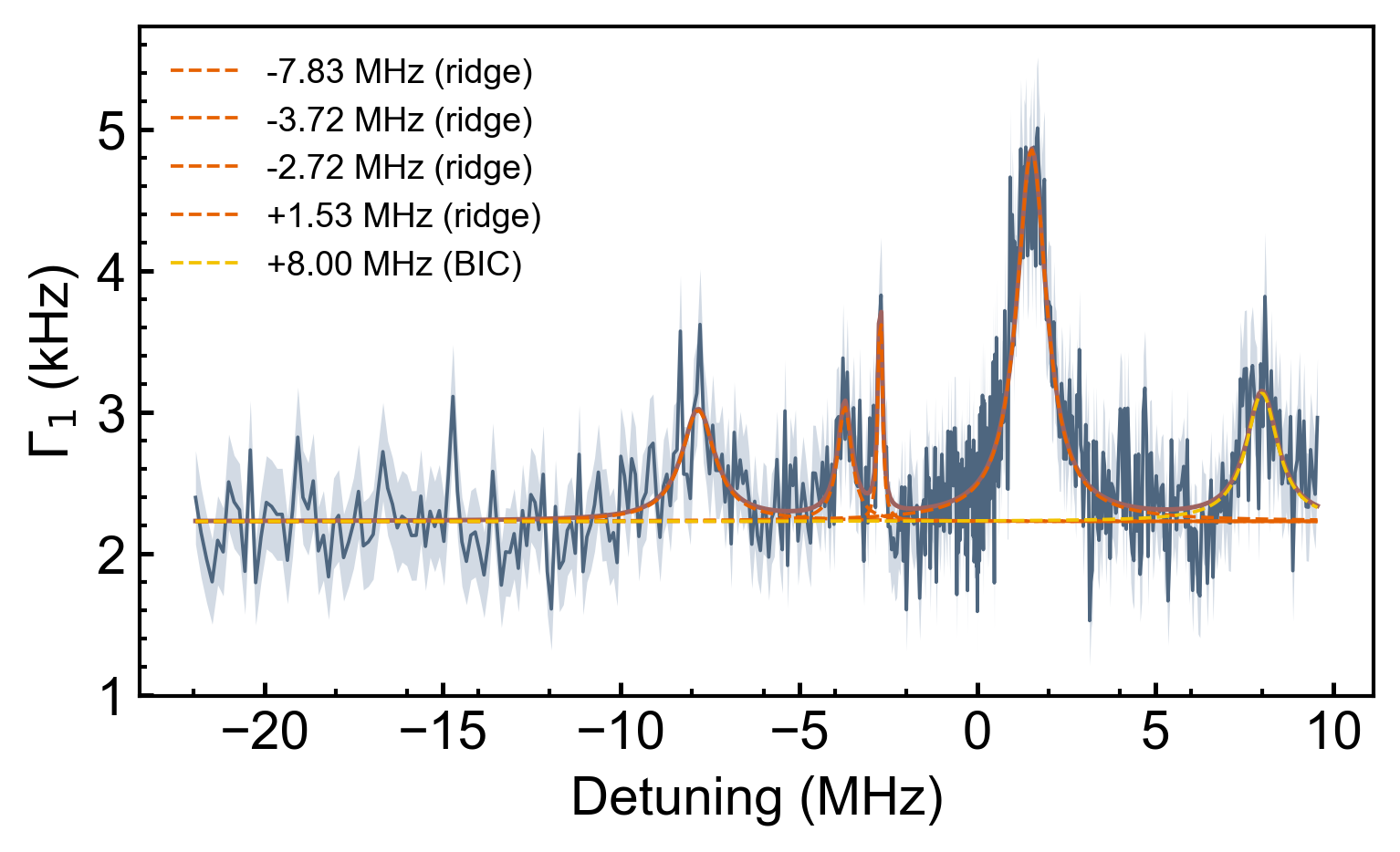}
  \caption{\textbf{Lorentzian Fitting}: The data (grey) is fit to a sum of Lorentzians and a vertical offset for $t=29.04$ hours in dataset 7. Lorentzians are fit in two stages: the first stage is ridge-based fitting (orange), the second stage is BIC-based fitting (yellow) applied to the residuals. 
  }
   \label{B(f)}
\end{figure}

\subsection{Full Reconstruction}
We now have a fully specified fit $\hat{\Gamma}_1(f, t)$, which we evaluate at $f = 0$ to model fluctuations in qubit $\Gamma_1$ at the qubit frequency: 
\begin{equation}
\begin{aligned}
    \hat{\Gamma}_{1}(f = 0, t) &= \underbrace{\Gamma_{1, \text{Background}}}_{1^{\text{st}} \text{ percentile of color plot}} \\
    &+ \underbrace{B(f=0)}_{\text{frequency dependent background}} \\
    &+ \underbrace{\Gamma_{1, \text{Broadband}}(t)}_{\text{vertical shift fitted in each frame}} \\
    &+ \underbrace{\sum_i \Gamma_{1, \text{TLS}}^{(i)}(f=0,t)}_{\text{Lorentzians evaluated at $f=0$}}
\end{aligned}
\end{equation}
Physically, $\Gamma_{1, \text{Background}}$ is background due to frequency- and time-independent sources. $B(f = 0)$ is the frequency-dependent background evaluated at zero detuning. $\Gamma_{1, \text{Broadband}}(t)$ are contributions from frequency-dependent but time-independent sources, such as unresolved TLSs, equilibrium and non-equilibrium quasiparticles, or electronics. It is estimated from the fitted vertical offset from the ridge-based fit. Finally, $\sum_i \Gamma_{1, \text{TLS}}^{(i)}(f=0,t)$ is the contribution of individually resolvable TLSs. It is estimated by evaluating all Lorentzians (both ridge and BIC) and zero detuning. 

\subsection{Summary Statistics}
We assess the quality of the fitted model by comparing our measured $\Gamma_{1}$ (from full $T_1$ sweeps) to modeled $\hat{\Gamma}_{1}$ using Lin's Concordance Correlation Coefficient $r_c$ (Appendix \ref{LinsCCC}). We find good agreement between $\Gamma_1$ and $\hat{\Gamma}_1$ with $r_c = 0.831$.

\begin{figure}
   \centering
  \includegraphics[width=\columnwidth]{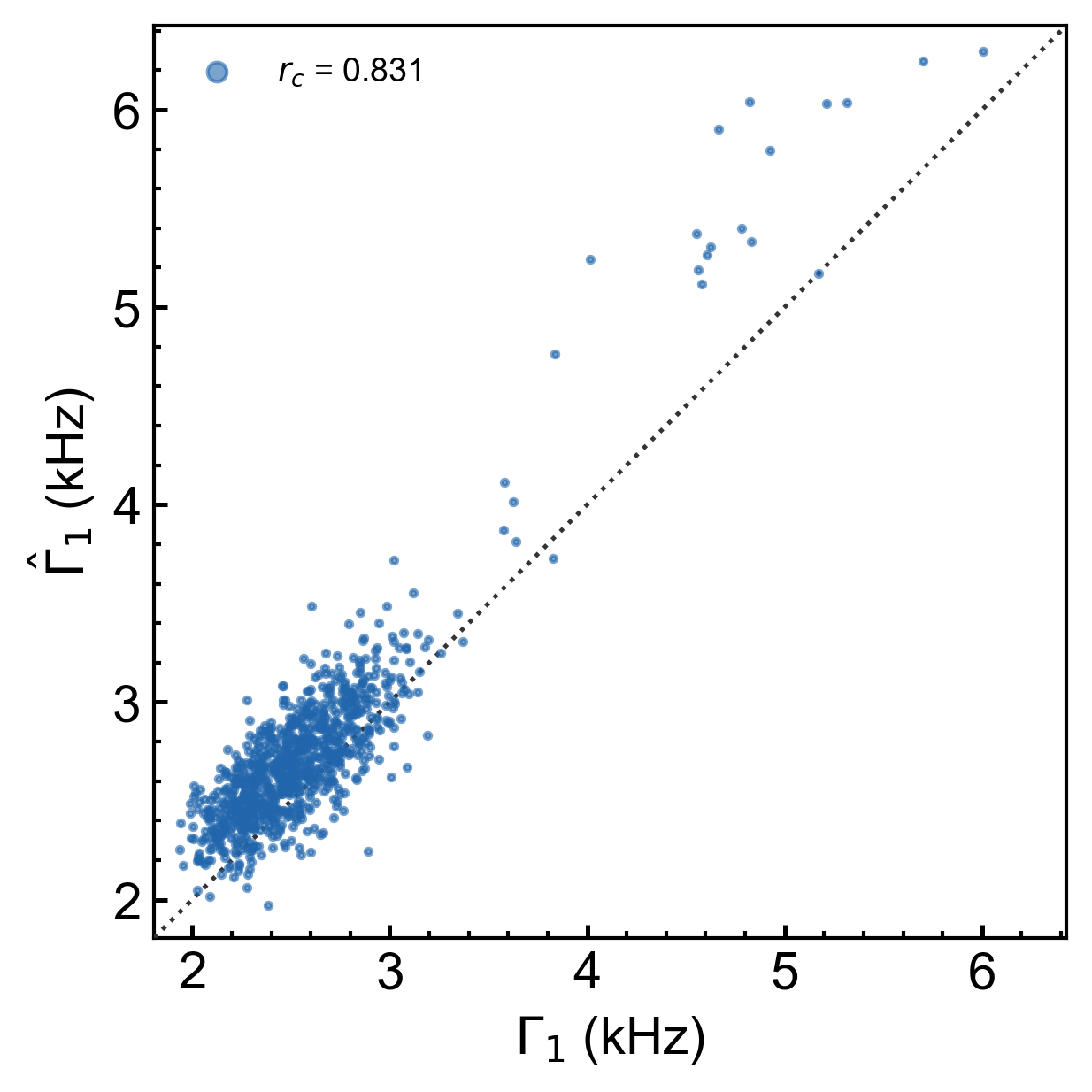}
  \caption{\textbf{Scatterplot of $\Gamma_1$ and $\hat{\Gamma}_1$}: Scatterplot of $\Gamma_1(t)$ as directly measured from interleaved full $T_1$ sweeps and $\hat{\Gamma}_1(t)$ as estimated from TLS fitting, with $\hat{\Gamma}_1 = \Gamma_1$ (dotted line) as a reference. $\Gamma_1(t)$ and $\hat{\Gamma}_1(t)$ show good agreement with $r_c = 0.831$, though $\hat{\Gamma}_1(t)$ contains a slight systematic overestimate. 
  }
   \label{B(f)}
\end{figure}

To quantify the relative contribution of TLSs and broadband fluctuations in the temporal fluctuations of $\hat{\Gamma}_1$, we calculate the following summary statistics (values are from the main text dataset):  
\begin{align}
    10.8 &= \text{Var}(\sum_i \Gamma_{1, \text{TLS}, i}(f = 0, t)) / \text{Var}(\Gamma_{1, \text{Broadband}}(t))\\
    0.902 &= \text{Corr}(\Gamma_{1, \text{full sweep}}(t), \sum_i \Gamma_{1, \text{TLS}, i}(f = 0, t))\\
    -0.056 &= \text{Corr}(\Gamma_{1, \text{full sweep}}(t), \Gamma_{1, \text{Broadband}}(t))
\end{align}

\section{Lin's Concordance Correlation Coefficient}
\label{LinsCCC}

Lin's Concordance Correlation Coefficient (CCC) $r_c$ \cite{lin_concordance_1989} is a statistical metric which is commonly used to evaluate the quality of a new measurement as compared to a gold standard ($\hat{\Gamma}_1(t)$ and $\Gamma_1(t)$ respectively in this case). It penalizes, scatter around the best-fit line (i.e. Pearson correlation $r$), systematic error (i.e. points are far from the $y=x$ line).

For an arbitrary situation, where a variable is measured by one scheme to produce $\vec{x}$ and another to produce $\vec{y}$, the sample form can be written as:
\begin{equation}
    r_c = \frac{2r s_x s_y}{s_x^2 + s_y^2 + (\bar{x} - \bar{y})^2}\\
\end{equation}
where $r$ is the Pearson correlation, $s_x = \sqrt{\frac{1}{n}\sum_i (x_i - \bar{x})^2}, s_y = \sqrt{\frac{1}{n}\sum_i (y_i - \bar{y})^2}$, and $\bar{x}, \bar{y}$ are the respective sample means. Additionally, if $\bar{x} = \bar{y}$ and $s_x = s_y$, then $r_c = r$. 


\bibliography{bib}

@PREAMBLE{
 "\providecommand{\noopsort}[1]{}" 
 # "\providecommand{\singleletter}[1]{#1}%" 
}

@article{google_surface_2025,
	title = {Quantum error correction below the surface code threshold},
	volume = {638},
    url = {https://www.nature.com/articles/s41586-024-08449-y},
	urldate = {2024-12-09},
	journal = {Nature},
	publisher = {Nature Publishing Group},
	author = {{Google Quantum AI and Collaborators}},
	month = feb,
	year = {2025},
	pages = {920--926},
}

@article{krinner2022realizing,
  title={Realizing repeated quantum error correction in a distance-three surface code},
  author={Krinner, Sebastian and Lacroix, Nathan and Remm, Ants and Di Paolo, Agustin and Genois, Elie and Leroux, Catherine and Hellings, Christoph and Lazar, Stefania and Swiadek, Francois and Herrmann, Johannes and others},
  journal={Nature},
  volume={605},
  number={7911},
  pages={669--674},
  year={2022},
  publisher={Nature Publishing Group UK London},
  doi={10.1038/s41586-022-04566-8},
  url={https://doi.org/10.1038/s41586-022-04566-8}
}

@article{Gong2021Experimental,
    author = {Gong, Ming and Yuan, Xiao and Wang, Shiyu and Wu, Yulin and Zhao, Youwei and Zha, Chen and Li, Shaowei and Zhang, Zhen and Zhao, Qi and Liu, Yunchao and Liang, Futian and Lin, Jin and Xu, Yu and Deng, Hui and Rong, Hao and Lu, He and Benjamin, Simon C and Peng, Cheng-Zhi and Ma, Xiongfeng and Chen, Yu-Ao and Zhu, Xiaobo and Pan, Jian-Wei},
    title = "{Experimental exploration of five-qubit quantum error-correcting code with superconducting qubits}",
    journal = {National Science Review},
    volume = {9},
    number = {1},
    pages = {nwab011},
    year = {2021},
    month = {01},
    issn = {2095-5138},
    doi = {10.1093/nsr/nwab011},
    url = {https://doi.org/10.1093/nsr/nwab011},
}

@article{sivak2023realtime,
  title={Real-time quantum error correction beyond break-even},
  author={Sivak, VV and Eickbusch, Alec and Royer, Baptiste and Singh, Shraddha and Tsioutsios, Ioannis and Ganjam, Suhas and Miano, Alessandro and Brock, BL and Ding, AZ and Frunzio, Luigi and others},
  journal={Nature},
  volume={616},
  number={7955},
  pages={50--55},
  year={2023},
  publisher={Nature Publishing Group UK London},
  doi = {10.1038/s41586-023-05782-6},
  url = {https://doi.org/10.1038/s41586-023-05782-6},
}

@article{ofek2016extending,
  title={Extending the lifetime of a quantum bit with error correction in superconducting circuits},
  author={Ofek, Nissim and Petrenko, Andrei and Heeres, Reinier and Reinhold, Philip and Leghtas, Zaki and Vlastakis, Brian and Liu, Yehan and Frunzio, Luigi and Girvin, Steven M and Jiang, Liang and others},
  journal={Nature},
  volume={536},
  number={7617},
  pages={441--445},
  year={2016},
  publisher={Nature Publishing Group UK London},
  doi = {10.1038/nature18949},
  url = {https://doi.org/10.1038/nature18949}
}

@article{barends2015digital,
  title={Digital quantum simulation of fermionic models with a superconducting circuit},
  author={Barends, Rami and Lamata, Lucas and Kelly, Julian and Garc{\'\i}a-{\'A}lvarez, L and Fowler, Austin G and Megrant, Anthony and Jeffrey, Evan and White, Ted C and Sank, Daniel and Mutus, Josh Y and others},
  journal={Nature communications},
  volume={6},
  number={1},
  pages={7654},
  year={2015},
  publisher={Nature Publishing Group UK London},
  doi = {10.1038/ncomms8654},
  url = {https://doi.org/10.1038/ncomms8654}
}

@article{marcos2013superconducting,
  title = {Superconducting Circuits for Quantum Simulation of Dynamical Gauge Fields},
  author = {Marcos, D. and Rabl, P. and Rico, E. and Zoller, P.},
  journal = {Phys. Rev. Lett.},
  volume = {111},
  issue = {11},
  pages = {110504},
  numpages = {6},
  year = {2013},
  month = {Sep},
  publisher = {American Physical Society},
  doi = {10.1103/PhysRevLett.111.110504},
  url = {https://link.aps.org/doi/10.1103/PhysRevLett.111.110504}
}

@article{kandala2017hardware,
  title={Hardware-efficient variational quantum eigensolver for small molecules and quantum magnets},
  author={Kandala, Abhinav and Mezzacapo, Antonio and Temme, Kristan and Takita, Maika and Brink, Markus and Chow, Jerry M and Gambetta, Jay M},
  journal={Nature},
  volume={549},
  number={7671},
  pages={242--246},
  year={2017},
  publisher={Nature Publishing Group},
  doi = {10.1038/nature23879},
  url = {https://doi.org/10.1038/nature23879}
}

@article{zhang2022synthesizing,
  title = {Synthesizing Five-Body Interaction in a Superconducting Quantum Circuit},
  author = {Zhang, Ke and Li, Hekang and Zhang, Pengfei and Yuan, Jiale and Chen, Jinyan and Ren, Wenhui and Wang, Zhen and Song, Chao and Wang, Da-Wei and Wang, H. and Zhu, Shiyao and Agarwal, Girish S. and Scully, Marlan O.},
  journal = {Phys. Rev. Lett.},
  volume = {128},
  issue = {19},
  pages = {190502},
  numpages = {6},
  year = {2022},
  month = {May},
  publisher = {American Physical Society},
  doi = {10.1103/PhysRevLett.128.190502},
  url = {https://link.aps.org/doi/10.1103/PhysRevLett.128.190502}
}

@article{mi2022time,
  title={Time-crystalline eigenstate order on a quantum processor},
  author={Mi, Xiao and Ippoliti, Matteo and Quintana, Chris and Greene, Ami and Chen, Zijun and Gross, Jonathan and Arute, Frank and Arya, Kunal and Atalaya, Juan and Babbush, Ryan and others},
  journal={Nature},
  volume={601},
  number={7894},
  pages={531--536},
  year={2022},
  publisher={Nature Publishing Group UK London},
  doi={10.1038/s41586-021-04257-w},
  url={https://doi.org/10.1038/s41586-021-04257-w}
}

@article{andersen2025thermalization,
  author  = {Andersen, T. I. and Astrakhantsev, N. and Karamlou, A. H. and Berndtsson, J. and Motruk, J. and Szasz, A. and Gross, J. A. and Schuckert, A. and Westerhout, T. and Zhang, Y. and Forati, E. and Rossi, D. and Kobrin, B. and Di Paolo, A. and Klots, A. R. and Drozdov, I. and Kurilovich, V. and Petukhov, A. and Ioffe, L. B. and Elben, A. and Rath, A. and Vitale, V. and Vermersch, B. and others},
  title   = {Thermalization and criticality on an analogue--digital quantum simulator},
  journal = {Nature},
  volume  = {638},
  number  = {8049},
  pages   = {79--85},
  year    = {2025},
  month   = feb,
  doi     = {10.1038/s41586-024-08460-3},
  url     = {https://doi.org/10.1038/s41586-024-08460-3}
}

@article{kollar2019hyperbolic,
  title={Hyperbolic lattices in circuit quantum electrodynamics},
  author={Koll{\'a}r, Alicia J and Fitzpatrick, Mattias and Houck, Andrew A},
  journal={Nature},
  volume={571},
  number={7763},
  pages={45--50},
  year={2019},
  publisher={Nature Publishing Group UK London},
  doi = {doi.org/10.1038/s41586-019-1348-3},
  url = {https://doi.org/10.1038/s41586-019-1348-3}
}

@article{google_otoc_2025,
  title = {Observation of Constructive Interference at the Edge of Quantum Ergodicity},
  author = {{Google Quantum AI and Collaborators}},
  journal = {Nature},
  volume = {646},
  number = {8086},
  pages = {825--830},
  year = {2025},
  month = oct,
  doi = {10.1038/s41586-025-09526-6},
  publisher = {Springer Nature},
  url = {https://doi.org/10.1038/s41586-025-09526-6}
}

@article{bland_millisecond_2025,
	title = {Millisecond lifetimes and coherence times in {2D} transmon qubits},
	volume = {647},
	copyright = {2025 The Author(s), under exclusive licence to Springer Nature Limited},
	issn = {1476-4687},
	url = {https://www.nature.com/articles/s41586-025-09687-4},
	doi = {10.1038/s41586-025-09687-4},
	number = {8089},
	urldate = {2026-02-10},
	journal = {Nature},
	publisher = {Nature Publishing Group},
	author = {Bland, Matthew P. and Bahrami, Faranak and Martinez, Jeronimo G. C. and Prestegaard, Paal H. and Smitham, Basil M. and Joshi, Atharv and Hedrick, Elizabeth and Kumar, Shashwat and Yang, Ambrose and Pakpour-Tabrizi, Alexander C. and Jindal, Apoorv and Chang, Ray D. and Cheng, Guangming and Yao, Nan and Cava, Robert J. and de Leon, Nathalie P. and Houck, Andrew A.},
	month = nov,
	year = {2025},
	pages = {343--348}
}

@article{olszewski_krypton_2026,
  author  = {Olszewski, Maciej W. and Kong, Lingda and Reinhardt, Simon and Tong, Daniel and Du, Xinyi and Di Gianluca, Gabriele and Lu, Haoran and Roy, Saswata and Zhang, Luojia and Biedron, Aleksandra B. and Muller, David A. and Fatemi, Valla},
  title   = {Krypton-sputtered tantalum films for scalable high-performance quantum devices},
  journal = {Nature Materials},
  year    = {2026},
  doi     = {10.1038/s41563-026-02718-z},
  url     = {https://doi.org/10.1038/s41563-026-02718-z}
}

@misc{acharya_ultrahigh_2026,
  title         = {Ultra-high {Q-factor} superconducting tantalum resonators on 300 mm {Si} wafers},
  author        = {R. Acharya and D. Perez Lozano and Ts. Ivanov and S. Massar and C. Vrancken and Y. Canvel and Y. Li and A. M. Vadiraj and J. Van Damme and S. Aghaeimeibodi and A. Khalajhedayati and M. Mongillo and O. Painter and D. Wan and A. Poto{\v{c}}nik and K. De Greve},
  year          = {2026},
  eprint        = {2606.10719},
  archivePrefix = {arXiv},
  primaryClass  = {quant-ph},
  url           = {https://arxiv.org/abs/2606.10719}
}

@article{carroll_dynamics_2022,
	title = {Dynamics of superconducting qubit relaxation times},
	volume = {8},
	copyright = {2022 The Author(s)},
	issn = {2056-6387},
	url = {https://www.nature.com/articles/s41534-022-00643-y},
	doi = {10.1038/s41534-022-00643-y},
	number = {1},
	urldate = {2026-03-05},
	journal = {npj Quantum Information},
	publisher = {Nature Publishing Group},
	author = {Carroll, M. and Rosenblatt, S. and Jurcevic, P. and Lauer, I. and Kandala, A.},
	month = nov,
	year = {2022},
	pages = {132},
}

@article{burnett_decoherence_2019,
	title = {Decoherence benchmarking of superconducting qubits},
	volume = {5},
	copyright = {2019 The Author(s)},
	issn = {2056-6387},
	url = {https://www.nature.com/articles/s41534-019-0168-5},
	doi = {10.1038/s41534-019-0168-5},
	number = {1},
	urldate = {2025-12-27},
	journal = {npj Quantum Information},
	publisher = {Nature Publishing Group},
	author = {Burnett, Jonathan J. and Bengtsson, Andreas and Scigliuzzo, Marco and Niepce, David and Kudra, Marina and Delsing, Per and Bylander, Jonas},
	month = jun,
	year = {2019},
	pages = {54},
}

@article{berritta_real-time_2026,
	title = {Real-{Time} {Adaptive} {Tracking} of {Fluctuating} {Relaxation} {Rates} in {Superconducting} {Qubits}},
	volume = {16},
	url = {https://link.aps.org/doi/10.1103/gk1b-stl3},
	doi = {10.1103/gk1b-stl3},
	number = {1},
	urldate = {2026-03-04},
	journal = {Physical Review X},
	publisher = {American Physical Society},
	author = {Berritta, Fabrizio and Benestad, Jacob and Krzywda, Jan A. and Krause, Oswin and Marciniak, Malthe A. and Krøjer, Svend and Warren, Christopher W. and Hogedal, Emil and Nylander, Andreas and Ahmad, Irshad and Osman, Amr and Biznárová, Janka and Rommel, Marcus and Roudsari, Anita Fadavi and Bylander, Jonas and Tancredi, Giovanna and Danon, Jeroen and Hastrup, Jacob and Kuemmeth, Ferdinand and Kjaergaard, Morten},
	month = feb,
	year = {2026},
	pages = {011025},
}

@article{klimov_fluctuations_2018,
	title = {Fluctuations of {Energy}-{Relaxation} {Times} in {Superconducting} {Qubits}},
	volume = {121},
	issn = {0031-9007, 1079-7114},
	url = {https://link.aps.org/doi/10.1103/PhysRevLett.121.090502},
	doi = {10.1103/PhysRevLett.121.090502},
	number = {9},
	urldate = {2026-03-28},
	journal = {Physical Review Letters},
	author = {Klimov, P. V. and Kelly, J. and Chen, Z. and Neeley, M. and Megrant, A. and Burkett, B. and Barends, R. and Arya, K. and Chiaro, B. and Chen, Yu and Dunsworth, A. and Fowler, A. and Foxen, B. and Gidney, C. and Giustina, M. and Graff, R. and Huang, T. and Jeffrey, E. and Lucero, Erik and Mutus, J. Y. and Naaman, O. and Neill, C. and Quintana, C. and Roushan, P. and Sank, Daniel and Vainsencher, A. and Wenner, J. and White, T. C. and Boixo, S. and Babbush, R. and Smelyanskiy, V. N. and Neven, H. and Martinis, John M.},
	month = aug,
	year = {2018},
	pages = {090502},
}

@article{li_gate_2023,
  author = {Li, Zhiyuan and Liu, Pei and Zhao, Peng and Mi, Zhenyu and
            Xu, Huikai and Liang, Xuehui and Su, Tang and Sun, Weijie and
            Xue, Guangming and Zhang, Jing-Ning and Liu, Weiyang and
            Jin, Yirong and Yu, Haifeng},
  title = {Error per single-qubit gate below $10^{-4}$ in a superconducting qubit},
  journal = {npj Quantum Information},
  volume = {9},
  pages = {111},
  year = {2023},
  doi = {10.1038/s41534-023-00781-x},
  url = {https://doi.org/10.1038/s41534-023-00781-x}
}

@article{schlor_correlating_2019,
  title = {Correlating Decoherence in Transmon Qubits: Low Frequency Noise by Single Fluctuators},
  author = {Schl{\"o}r, Steffen and Lisenfeld, J{\"u}rgen and M{\"u}ller, Clemens and
            Bilmes, Alexander and Schneider, Andre and Pappas, David P. and
            Ustinov, Alexey V. and Weides, Martin},
  journal = {Phys. Rev. Lett.},
  volume = {123},
  issue = {19},
  pages = {190502},
  numpages = {6},
  year = {2019},
  month = {Nov},
  publisher = {American Physical Society},
  doi = {10.1103/PhysRevLett.123.190502},
  url = {https://link.aps.org/doi/10.1103/PhysRevLett.123.190502}
}

@article{zhao_combating_2022,
  title = {Combating fluctuations in relaxation times of fixed-frequency transmon qubits with microwave-dressed states},
  author = {Zhao, Peng and Ma, Teng and Jin, Yirong and Yu, Haifeng},
  journal = {Phys. Rev. A},
  volume = {105},
  issue = {6},
  pages = {062605},
  numpages = {12},
  year = {2022},
  month = {Jun},
  publisher = {American Physical Society},
  doi = {10.1103/PhysRevA.105.062605},
  url = {https://link.aps.org/doi/10.1103/PhysRevA.105.062605}
}

@article{zhu_disentangling_2025,
  title = {Disentangling the Impact of Quasiparticles and Two-Level Systems on the Statistics of Superconducting-Qubit Lifetime},
  author = {Zhu, S. and You, Xinyuan and Alyanak, Ugur and Bal, Mustafa and
            Crisa, Francesco and Garattoni, Sabrina and Lunin, A. and
            Pilipenko, Roman and Murthy, Akshay and Romanenko, Alexander and
            Grassellino, Anna},
  journal = {Physical Review Applied},
  volume = {24},
  issue = {3},
  pages = {034002},
  year = {2025},
  publisher = {American Physical Society},
  doi = {10.1103/x4ny-17lk},
  url = {https://link.aps.org/doi/10.1103/x4ny-17lk}
}

@article{muller_interacting_2015,
  title = {Interacting two-level defects as sources of fluctuating high-frequency noise in superconducting circuits},
  author = {M{\"u}ller, Clemens and Lisenfeld, J{\"u}rgen and Shnirman, Alexander and Poletto, Stefano},
  journal = {Phys. Rev. B},
  volume = {92},
  issue = {3},
  pages = {035442},
  numpages = {13},
  year = {2015},
  month = {Jul},
  publisher = {American Physical Society},
  doi = {10.1103/PhysRevB.92.035442},
  url = {https://link.aps.org/doi/10.1103/PhysRevB.92.035442}
}

@article{you_stabilizing_2022,
  title = {Stabilizing and Improving Qubit Coherence by Engineering the Noise Spectrum of Two-Level Systems},
  author = {You, Xinyuan and Huang, Ziwen and Alyanak, Ugur and
            Romanenko, Alexander and Grassellino, Anna and Zhu, Shaojiang},
  journal = {Physical Review Applied},
  volume = {18},
  issue = {4},
  pages = {044026},
  year = {2022},
  month = oct,
  publisher = {American Physical Society},
  doi = {10.1103/PhysRevApplied.18.044026},
  url = {https://link.aps.org/doi/10.1103/PhysRevApplied.18.044026}
}

@article{etxezarreta_martinez_time-varying_2021,
	title = {Time-varying quantum channel models for superconducting qubits},
	volume = {7},
	copyright = {2021 The Author(s)},
	issn = {2056-6387},
	url = {https://www.nature.com/articles/s41534-021-00448-5},
	doi = {10.1038/s41534-021-00448-5},
	number = {1},
	urldate = {2026-04-20},
	journal = {npj Quantum Information},
	publisher = {Nature Publishing Group},
	author = {Etxezarreta Martinez, Josu and Fuentes, Patricio and Crespo, Pedro and Garcia-Frias, Javier},
	month = jul,
	year = {2021},
	pages = {115}
}

@article{weeden_statistics_2026,
  title = {Statistics of Strongly Coupled Defects in Superconducting Qubits},
  author = {Weeden, S. and Harrison, D. C. and Patel, S. and Snyder, M. and
            Blackwell, E. J. and Spahn, G. and Abdullah, S. and Takeda, Y. and
            Plourde, B. L. T. and Martinis, J. M. and McDermott, R.},
  journal = {Physical Review Applied},
  volume = {25},
  issue = {4},
  pages = {044050},
  year = {2026},
  month = apr,
  publisher = {American Physical Society},
  doi = {10.1103/4ssz-6ctb},
  url = {https://link.aps.org/doi/10.1103/4ssz-6ctb}
}

@article{gustavsson_suppressing_2016,
  author  = {Gustavsson, Simon and
             Yan, Fei and
             Catelani, Gianluigi and
             Bylander, Jonas and
             Kamal, Archana and
             Birenbaum, Jeffrey and
             Hover, David and
             Rosenberg, Danna and
             Samach, Gabriel and
             Sears, Adam P. and
             Weber, Steven J. and
             Yoder, Jonilyn L. and
             Clarke, John and
             Kerman, Andrew J. and
             Yoshihara, Fumiki and
             Nakamura, Yasunobu and
             Orlando, Terry P. and
             Oliver, William D.},
  title   = {Suppressing relaxation in superconducting qubits by quasiparticle pumping},
  journal = {Science},
  year    = {2016},
  volume  = {354},
  number  = {6319},
  pages   = {1573--1577},
  doi     = {10.1126/science.aah5844},
  url     = {https://doi.org/10.1126/science.aah5844}
}

@article{bylander_noise_2011,
	title = {Noise spectroscopy through dynamical decoupling with a superconducting flux qubit},
	volume = {7},
	copyright = {2011 Springer Nature Limited},
	issn = {1745-2481},
	url = {https://www.nature.com/articles/nphys1994},
	doi = {10.1038/nphys1994},
	number = {7},
	urldate = {2026-02-10},
	journal = {Nature Physics},
	publisher = {Nature Publishing Group},
	author = {Bylander, Jonas and Gustavsson, Simon and Yan, Fei and Yoshihara, Fumiki and Harrabi, Khalil and Fitch, George and Cory, David G. and Nakamura, Yasunobu and Tsai, Jaw-Shen and Oliver, William D.},
	month = jul,
	year = {2011},
	pages = {565--570}
}

@article{viola_dynamical_1998,
  author  = {Lorenza Viola and Seth Lloyd},
  title   = {Dynamical suppression of decoherence in two-state quantum systems},
  journal = {Physical Review A},
  volume  = {58},
  number  = {4},
  pages   = {2733--2744},
  year    = {1998},
  month   = oct,
  doi     = {10.1103/PhysRevA.58.2733},
  url     = {https://link.aps.org/doi/10.1103/PhysRevA.58.2733},
  publisher = {American Physical Society}
}

@article{koch_model_2007,
	title = {Model for 1/f {Flux} {Noise} in {SQUIDs} and {Qubits}},
	volume = {98},
	copyright = {http://link.aps.org/licenses/aps-default-license},
	issn = {0031-9007, 1079-7114},
	url = {https://link.aps.org/doi/10.1103/PhysRevLett.98.267003},
	doi = {10.1103/PhysRevLett.98.267003},
	number = {26},
	urldate = {2026-05-11},
	journal = {Physical Review Letters},
	author = {Koch, Roger H. and DiVincenzo, David P. and Clarke, John},
	month = jun,
	year = {2007},
	pages = {267003}
}

@article{schreier_suppressing_2008,
	title = {Suppressing charge noise decoherence in superconducting charge qubits},
	volume = {77},
	copyright = {http://link.aps.org/licenses/aps-default-license},
	issn = {1098-0121, 1550-235X},
	url = {https://link.aps.org/doi/10.1103/PhysRevB.77.180502},
	doi = {10.1103/PhysRevB.77.180502},
	number = {18},
	urldate = {2026-05-11},
	journal = {Physical Review B},
	author = {Schreier, J. A. and Houck, A. A. and Koch, Jens and Schuster, D. I. and Johnson, B. R. and Chow, J. M. and Gambetta, J. M. and Majer, J. and Frunzio, L. and Devoret, M. H. and Girvin, S. M. and Schoelkopf, R. J.},
	month = may,
	year = {2008},
	pages = {180502}
}

@article{christensen_anomalous_2019,
	title = {Anomalous charge noise in superconducting qubits},
	volume = {100},
	url = {https://link.aps.org/doi/10.1103/PhysRevB.100.140503},
	doi = {10.1103/PhysRevB.100.140503},
	number = {14},
	urldate = {2026-04-23},
	journal = {Physical Review B},
	publisher = {American Physical Society},
	author = {Christensen, B. G. and Wilen, C. D. and Opremcak, A. and Nelson, J. and Schlenker, F. and Zimonick, C. H. and Faoro, L. and Ioffe, L. B. and Rosen, Y. J. and DuBois, J. L. and Plourde, B. L. T. and McDermott, R.},
	month = oct,
	year = {2019},
	pages = {140503}
}

@article{yan_spectroscopy_2012,
	title = {Spectroscopy of low-frequency noise and its temperature dependence in a superconducting qubit},
	volume = {85},
	issn = {1098-0121, 1550-235X},
	url = {http://arxiv.org/abs/1201.5665},
	doi = {10.1103/PhysRevB.85.174521},
	number = {17},
	urldate = {2026-04-18},
	journal = {Physical Review B},
	author = {Yan, Fei and Bylander, Jonas and Gustavsson, Simon and Yoshihara, Fumiki and Harrabi, Khalil and Cory, David G. and Orlando, Terry P. and Nakamura, Yasunobu and Tsai, Jaw-Shen and Oliver, William D.},
	month = may,
	year = {2012},
	pages = {174521}
}

@article{mcewen_gapEngineering_2024,
  author = {McEwen, Matt and Miao, Kevin C. and Atalaya, Juan and Bilmes, Alexander and Crook, Alex and Bovaird, Jenna and Kreikebaum, John Mark and Zobrist, Nicholas and Jeffrey, Evan and Ying, Bicheng and Bengtsson, Andreas and Chang, Hung-Shen and Dunsworth, Andrew and Kelly, Julian and Zhang, Yaxing and Forati, Ebrahim and Acharya, Rajeev and Iveland, Justin and Liu, Wayne and Kim, Seon and Burkett, Brian and Megrant, Anthony and Chen, Yu and Neill, Charles and Sank, Daniel and Devoret, Michel and Opremcak, Alex},
  title = {Resisting High-Energy Impact Events through Gap Engineering in Superconducting Qubit Arrays},
  journal = {Physical Review Letters},
  volume = {133},
  number = {24},
  pages = {240601},
  year = {2024},
  doi = {10.1103/PhysRevLett.133.240601},
  publisher = {American Physical Society}
}

@article{kurilovich_correlated_2026,
  author = {Kurilovich, Vladislav D. and Roberts, Gabrielle and Martin, Leigh S. and McEwen, Matt and others},
  title = {Correlated Phase Error Bursts in a Gap-Engineered Superconducting Qubit Array},
  journal = {Physical Review X},
  volume = {16},
  number = {2},
  year = {2026},
  doi = {10.1103/1bl4-b2f7},
  publisher = {American Physical Society}
}

@article{riste_millisecond_2013,
	title = {Millisecond charge-parity fluctuations and induced decoherence in a superconducting transmon qubit},
	volume = {4},
	copyright = {2013 The Author(s)},
	issn = {2041-1723},
	url = {https://www.nature.com/articles/ncomms2936},
	doi = {10.1038/ncomms2936},
	number = {1},
	urldate = {2026-05-25},
	journal = {Nature Communications},
	publisher = {Nature Publishing Group},
	author = {Ristè, D. and Bultink, C. C. and Tiggelman, M. J. and Schouten, R. N. and Lehnert, K. W. and DiCarlo, L.},
	month = may,
	year = {2013},
	pages = {1913}
}

@article{blais_circuit_2021,
	title = {Circuit {Quantum} {Electrodynamics}},
	volume = {93},
	issn = {0034-6861, 1539-0756},
	url = {http://arxiv.org/abs/2005.12667},
	doi = {10.1103/RevModPhys.93.025005},
	number = {2},
	urldate = {2025-10-30},
	journal = {Reviews of Modern Physics},
	author = {Blais, Alexandre and Grimsmo, Arne L. and Girvin, S. M. and Wallraff, Andreas},
	month = may,
	year = {2021},
	pages = {025005},
}

@article{krantz_quantum_2019,
	title = {A quantum engineer's guide to superconducting qubits},
	volume = {6},
	issn = {1931-9401},
	url = {https://pubs.aip.org/apr/article/6/2/021318/570326/A-quantum-engineer-s-guide-to-superconducting},
	doi = {10.1063/1.5089550},
	number = {2},
	urldate = {2025-10-30},
	journal = {Applied Physics Reviews},
	author = {Krantz, P. and Kjaergaard, M. and Yan, F. and Orlando, T. P. and Gustavsson, S. and Oliver, W. D.},
	month = jun,
	year = {2019},
	pages = {021318}
}

@misc{chen_scalable_2025,
	title = {Scalable and {Site}-{Specific} {Frequency} {Tuning} of {Two}-{Level} {System} {Defects} in {Superconducting} {Qubit} {Arrays}},
	url = {http://arxiv.org/abs/2503.04702},
	doi = {10.48550/arXiv.2503.04702},
	urldate = {2026-03-26},
	publisher = {arXiv},
	author = {Chen, Larry and Lee, Kan-Heng and Liu, Chuan-Hong and Marinelli, Brian and Naik, Ravi K. and Kang, Ziqi and Goss, Noah and Kim, Hyunseong and Santiago, David I. and Siddiqi, Irfan},
	month = mar,
	year = {2025},
	note = {arXiv:2503.04702 [quant-ph]}
}

@article{rower_evolution_2023,
	title = {Evolution of 1/f {Flux} {Noise} in {Superconducting} {Qubits} with {Weak} {Magnetic} {Fields}},
	volume = {130},
	url = {https://link.aps.org/doi/10.1103/PhysRevLett.130.220602},
	doi = {10.1103/PhysRevLett.130.220602},
	number = {22},
	urldate = {2025-12-27},
	journal = {Physical Review Letters},
	publisher = {American Physical Society},
	author = {Rower, David A. and Ateshian, Lamia and Li, Lauren H. and Hays, Max and Bluvstein, Dolev and Ding, Leon and Kannan, Bharath and Almanakly, Aziza and Braumüller, Jochen and Kim, David K. and Melville, Alexander and Niedzielski, Bethany M. and Schwartz, Mollie E. and Yoder, Jonilyn L. and Orlando, Terry P. and Wang, Joel I-Jan and Gustavsson, Simon and Grover, Jeffrey A. and Serniak, Kyle and Comin, Riccardo and Oliver, William D.},
	month = may,
	year = {2023},
	pages = {220602}
}

@article{lisenfeld_electric_2019,
	title = {Electric field spectroscopy of material defects in transmon qubits},
	volume = {5},
	copyright = {2019 The Author(s)},
	issn = {2056-6387},
	url = {https://www.nature.com/articles/s41534-019-0224-1},
	doi = {10.1038/s41534-019-0224-1},
	number = {1},
	urldate = {2026-05-19},
	journal = {npj Quantum Information},
	publisher = {Nature Publishing Group},
	author = {Lisenfeld, Jürgen and Bilmes, Alexander and Megrant, Anthony and Barends, Rami and Kelly, Julian and Klimov, Paul and Weiss, Georg and Martinis, John M. and Ustinov, Alexey V.},
	month = nov,
	year = {2019},
	pages = {105}
}

@article{colao_zanuz_mitigating_2025,
  title = {Mitigating losses of superconducting qubits strongly coupled to defect modes},
  author = {Colao Zanuz, Dante and Ficheux, Quentin and Michaud, Laurent and
            Orekhov, Alexei and Hanke, Kilian and Flasby, Alexander and
            Bahrami Panah, Mohsen and Norris, Graham J. and
            Kerschbaum, Michael and Remm, Ants and Swiadek, Fran{\c c}ois and
            Hellings, Christoph and Laz{\u a}r, Stefania and Scarato, Colin and
            Lacroix, Nathan and Krinner, Sebastian and Eichler, Christopher and
            Wallraff, Andreas and Besse, Jean-Claude},
  journal = {Phys. Rev. Applied},
  volume = {23},
  issue = {4},
  pages = {044054},
  numpages = {14},
  year = {2025},
  month = {Apr},
  publisher = {American Physical Society},
  doi = {10.1103/PhysRevApplied.23.044054},
  url = {https://link.aps.org/doi/10.1103/PhysRevApplied.23.044054}
}

@article{lisenfeld_mappingTLS_2026,
  author  = {Jürgen Lisenfeld and Alexander K. Händel and Etienne Daum and Benedikt Berlitz and Alexander Bilmes and Alexey V. Ustinov},
  title   = {Mapping the positions of Two-Level-Systems on the surface of a superconducting transmon qubit},
  journal = {npj Quantum Information},
  volume  = {12},
  pages   = {80},
  year    = {2026},
  doi     = {10.1038/s41534-026-01272-5},
  url     = {https://doi.org/10.1038/s41534-026-01272-5}
}

@article{MadrasSokal1988,
  author    = {Neal Madras and Alan D. Sokal},
  title     = {The Pivot Algorithm: A Highly Efficient Monte Carlo Method for the Self-Avoiding Walk},
  journal   = {Journal of Statistical Physics},
  year      = {1988},
  volume    = {50},
  number    = {1--2},
  pages      = {109--186},
  doi       = {10.1007/BF01022990}
}

@article{sung2019nongaussian,
  author  = {Sung, Y. and Beaudoin, F. and Norris, L. M. and
             Sung, K. J. and Kubo, Y. and Ladd, T. D. and
             Blais, A. and Cory, D. G.},
  title   = {Non-Gaussian noise spectroscopy with a superconducting qubit sensor},
  journal = {Nature Communications},
  volume  = {10},
  pages   = {3715},
  year    = {2019},
  doi     = {10.1038/s41467-019-11699-4},
  url     = {https://doi.org/10.1038/s41467-019-11699-4}
}

@article{cywinski2008enhance,
  author  = {Cywi{\'n}ski, {\L}ukasz and
             Lutchyn, Roman M. and
             Nave, C. P. and
             Das Sarma, S.},
  title   = {How to enhance dephasing time in superconducting qubits},
  journal = {Physical Review B},
  volume  = {77},
  number  = {17},
  pages   = {174509},
  year    = {2008},
  doi     = {10.1103/PhysRevB.77.174509},
  url     = {https://doi.org/10.1103/PhysRevB.77.174509}
}

@article{dwyer2022probing,
  author  = {Dwyer, Bo L. and Rodgers, L. V. H. and Urbach, E. K. and
             Bluvstein, Dolev and Sangtawesin, S. and Zhou, H. and
             Nassab, Y. and Fitzpatrick, M. and Yuan, Z. and
             De Greve, K. and Peterson, E. L. and Knowles, H. and
             Sumarac, T. and Chou, J. P. and Gali, A. and
             Dobrovitski, V. V. and Lukin, M. D. and De Leon, N. P.},
  title   = {Probing Spin Dynamics on Diamond Surfaces Using a Single Quantum Sensor},
  journal = {PRX Quantum},
  volume  = {3},
  number  = {4},
  pages   = {040328},
  year    = {2022},
  doi     = {10.1103/PRXQuantum.3.040328},
  url     = {https://doi.org/10.1103/PRXQuantum.3.040328}
}

@book{kish1965survey,
  author    = {Kish, Leslie},
  title     = {Survey Sampling},
  publisher = {John Wiley \& Sons},
  address   = {New York},
  year      = {1965}
}

@article{thorbeck_two-level-system_2023-2,
	title = {Two-{Level}-{System} {Dynamics} in a {Superconducting} {Qubit} {Due} to {Background} {Ionizing} {Radiation}},
	volume = {4},
	issn = {2691-3399},
	url = {https://link.aps.org/doi/10.1103/PRXQuantum.4.020356},
	doi = {10.1103/PRXQuantum.4.020356},
	number = {2},
	urldate = {2026-08-25},
	journal = {PRX Quantum},
	author = {Thorbeck, Ted and Eddins, Andrew and Lauer, Isaac and McClure, Douglas T. and Carroll, Malcolm},
	month = jun,
	year = {2023},
	pages = {020356},
}

@article{lin_concordance_1989,
	title = {A concordance correlation coefficient to evaluate reproducibility},
	volume = {45},
	issn = {0006-341X},
	number = {1},
	journal = {Biometrics},
	author = {Lin, L. I.},
	month = mar,
	year = {1989},
	pages = {255--268},
}

\end{document}